\documentclass[11pt]{article}
\usepackage[left=1in,right =1in, top=1in, bottom=1in]{geometry}
\usepackage[linktocpage]{hyperref}
\usepackage[noadjust]{cite}
\usepackage{float}
\usepackage{amsfonts}
\usepackage{comment}
\usepackage{csquotes}
\usepackage{subcaption}
\usepackage{amsmath}
\usepackage{amssymb}
\usepackage{latexsym}
\usepackage{graphicx}
\usepackage[english]{babel}
\usepackage[font={small}]{caption}

\usepackage{amsfonts}
\usepackage{latexsym}
\usepackage{graphicx}
\usepackage[english]{babel}
\usepackage{url}
\usepackage{enumerate}
\usepackage[shortlabels]{enumitem}
\usepackage{ulem}
\usepackage{tikz}
\DeclareMathOperator{\Tr}{Tr}

\begin{document}
\input epsf

\def\p{\partial}
\def\h{{1\over 2}}
\def\be{\begin{equation}}
\def\bea{\begin{eqnarray}}
\def\ee{\end{equation}}
\def\eea{\end{eqnarray}}
\def\d{\partial}
\def\la{\lambda}
\def\eps{\epsilon}
\def\b{\bigskip}
\def\m{\medskip}

\newcommand{\newsection}[1]{\section{#1} \setcounter{equation}{0}}

\def\q{\quad}

\newcommand{\MyRed}{\color [rgb]{0.8,0,0}}
\newcommand{\MyGreen}{\color [rgb]{0,0.7,0}}
\newcommand{\MyBlue}{\color [rgb]{0,0,0.8}}
\newcommand{\MyBrown}{\color [rgb]{0.8,0.4,0.1}}
\newcommand{\MyPurple}{\color [rgb]{0.6,0.0,0.6}}
\def\MH#1{{\MyRed [MH: #1]}}
\def\SM#1{{\MyPurple [SM: #1]}}   
\def\BG#1{{\MyBlue [BG: #1]}}
\def\MM#1{{\MyGreen [MM: #1]}}

\def\t{\tilde}
\def\r{\rightarrow}
\def\nn{\nonumber\\}

\newcommand\blfootnote[1]{%
  \begingroup
  \renewcommand\thefootnote{}\footnote{#1}%
  \addtocounter{footnote}{-1}%
  \endgroup
}

\begin{flushright}
\end{flushright}
\vspace{20mm}
\begin{center}
{\LARGE Contrasting the fuzzball and wormhole paradigms\\
 for black holes}
\\
\vspace{18mm}
{\bf Bin Guo$^1$\blfootnote{$^{1}$guo.1281@osu.edu}, Marcel R. R. Hughes$^{2}$\blfootnote{$^{2}$hughes.2059@osu.edu}, Samir D. Mathur$^3$\blfootnote{$^{3}$mathur.16@osu.edu } and Madhur Mehta$^4$\blfootnote{$^{4}$mehta.493@osu.edu}  
\\}

\vspace{10mm}
Department of Physics,\\ The Ohio State University,\\ Columbus,
OH 43210, USA

\vspace{4mm}
\end{center}


\thispagestyle{empty}
\begin{abstract}
\vspace{2mm}

We examine an interesting set of recent proposals describing a `wormhole paradigm' for black holes. These proposals require that in some effective variables, semiclassical low-energy  dynamics emerges at the horizon. We prove the `effective small corrections theorem' to show that such an effective horizon behavior is not compatible with the requirement that the black hole radiate like a piece of coal as seen from outside. This theorem thus concretizes the fact that the proposals within the wormhole paradigm require some nonlocality linking the hole and its distant radiation. We try to illustrate various proposals for nonlocality by making simple bit models to encode the nonlocal effects. In each case, we find either nonunitarity of evolution in the black hole interior or a nonlocal Hamiltonian interaction between the hole and infinity; such an interaction is not present for burning coal. We examine recent arguments about the Page curve and observe that the quantity that is argued to follow the Page curve of a normal body is not the entanglement entropy but a different quantity. It has been suggested that this replacement of the quantity to be computed arises from the possibility of topology change in gravity which can generate replica wormholes. We examine the role of topology change in quantum gravity but do not find any source of connections between different replica copies in the path integral for the R\'{e}nyi entropy. We also contrast the wormhole paradigm with the fuzzball paradigm, where the fuzzball does radiate like a piece of coal. Just as in the case of a piece of coal, the fuzzball does not have low-energy semiclassical dynamics at its surface at energies $E\sim T$ (effective dynamics at energies $E\gg T$ is possible under the conjecture of fuzzball complementarity, but these $E\gg T$ modes have no relevance to the Page curve or the information paradox).

\end{abstract}
\newpage

\tableofcontents

\setcounter{page}{1}

\numberwithin{equation}{section}

\section{Introduction}\label{secintro}
\normalem

In 1974 Hawking discovered that black holes evaporate by producing entangled pairs at the horizon \cite{Hawking:1975vcx,Hawking:1976ra}. The two members of the pair $\{ b,c\}$ are in an entangled state which can be schematically written as 
\be
|\psi\rangle_{pair}={1\over \sqrt{2}}\Big (|0\rangle_b|0\rangle_c+|1\rangle_b|1\rangle_c\Big ) +O(\epsilon) \ .
\label{pair}
\ee
Here the $O(\epsilon)$ correction takes into account any small quantum gravity corrections not captured by the leading order `quantum fields on curved space' computation done by Hawking. This pair creation process leads to a monotonically rising entanglement entropy $S_{ent}$ of the hole with its radiation over time. We then get a violation of quantum unitarity when the hole evaporates away; an aspect of a problem known as the black hole information paradox. The graph of $S_{ent}$ with time is called the Page curve; thus another way of stating the problem is that the Page curve for Hawking's computation does not come down to zero at the end of the evaporation process.

Over the past two decades a resolution to this puzzle has emerged in string theory; this resolution is called the fuzzball paradigm \cite{Lunin:2001jy,Mathur:2003hj,Lunin:2002iz,Kanitscheider:2007wq,Mathur:2005zp,Bena:2007kg,Chowdhury:2010ct,Bena:2015bea,Bena:2016agb,Bena:2016ypk}. In string theory, we must make a black hole by taking a bound state of the strings and branes in the theory. In each case where such a bound state has been made, it has been found that the bound state is a `fuzzball': a horizon-sized quantum object with no horizon. If all black hole microstates are assumed to have this behavior, then the black hole is an object no different from a planet or a star; it radiates from its surface like any normal body, not by the production of entangled pairs. In other words, there is \emph{no} analogue of (\ref{pair}). In this way the fuzzball paradigm resolves the information paradox (for an overview of the fuzzball paradigm, see appendix~\ref{appA}). 

Recently, there has been interest in looking for an alternative resolution of the information paradox; we will call this attempt the `wormhole paradigm'. The central aspect of the wormhole paradigm is the requirement that, in some effective variables, we \emph{do} have an approximation to low-energy semiclassical dynamics at the horizon. This low-energy effective dynamics will lead to the creation of entangled pairs in the effective variables
\be
|\psi_{ef\!f}\rangle_{pair}={1\over \sqrt{2}}\Big (|0\rangle_{b,ef\!f}|0\rangle_{c,ef\!f}+|1\rangle_{b,ef\!f}|1\rangle_{c,ef\!f}\Big ) +O(\epsilon) \ .
\label{twopp}
\ee
This pair creation again leads to a monotonically rising entanglement of the hole with its radiation. The problem is then argued to be resolved by nonlocal effects in the gravity theory that connect the hole to its radiation. In a rough picture we can imagine this connection to be in the form of a wormhole extending from the hole to each radiation quantum \cite{Maldacena:2013xja}; hence the term `wormhole paradigm' for this class of models.

There has been quite some confusion about the wormhole paradigm. What does it assume and what does it show? One reason for this confusion is that there are several different lines of thought that come under the rough umbrella of the wormhole paradigm. The common feature shared by all these approaches is that there should be an emergence of effective semiclassical behavior at the horizon, i.e. we should get vacuum pair production in the state (\ref{twopp}). In this manner the wormhole approaches all differ from the fuzzball paradigm, where we do not have any such pair production. However, the arguments for why the Page curve should come down (in spite of having the pair creation (\ref{twopp})) differ between different formulations of the wormhole paradigm. As noted above, the Page curve will be argued to come down using some postulate involving nonlocality. This postulate, however, is not explicit in many of the approaches. One of our main goals will be to make such nonlocality explicit.

Some confusion is also caused by the fact that there are several different aspects of these arguments, whose roles are sometimes not sufficiently clarified:

\begin{enumerate}[start=1,
    labelindent=\parindent,
    leftmargin =2.5\parindent,
    label=(A\arabic*)]
\item \label{A1}
The existence of nonlocal Hamiltonian interactions between the hole and its radiation.
\item \label{A2}
The idea that the effective bits appearing in (\ref{twopp}) can be made as combinations of bits in the hole as well as bits  in the radiation at infinity.
\item \label{A3}
The idea that if nonlocal effects are `small' then they could be  somehow  consistent with the notion that physics far from the hole should be `normal'.
\item \label{A4}
The idea that degrees of freedom (bits) describing quanta at infinity are not independent of  the degrees of freedom in the hole.
\end{enumerate}
%
Note that the nonlocality needed in the wormhole paradigm is not the `nonlocality' of entanglement. In ordinary quantum mechanics two quanta with an arbitrarily large separation can be in an entangled state. Such an entanglement is present between the hole and its radiation in Hawking's original computation of radiation. But this entanglement does not create any interaction between the hole and its radiation. This entanglement will not by itself bring the Page curve down; rather, this entanglement is the basic cause of the information paradox.

Our goal in this article is to try to clarify the above confusions by detailing what we understand about the various proposals. In seeking this clarity we make simple bit models to explain what we think the different proposals are saying. Our hope is that these models will prompt a discussion in the field that will make precise what is assumed, what is claimed and what is proved in the different approaches to the wormhole paradigm. It may be that  proponents of the wormhole paradigm have other models in mind; in that case it would be very useful to have these other models expressed in the same  bit model language that we use here so that the underlying ideas behind the paradigm become clear.

Let us now summarize the various issues arising in the fuzzball and wormhole paradigms; in the process we will keep in mind how points \ref{A1}-\ref{A4} above appear in the various arguments.

\subsection{Different models for radiating objects}

Let us list the various kinds of radiating objects that appear in the discussion of the fuzzball and wormhole paradigms. In what follows, the reader should assume the following: (a) the exact theory has a unitary evolution, unless stated otherwise, (b) the bits being described are bits of the exact theory, unless they are explicitly termed effective bits and, (c) the radiation quanta at infinity are always the exact bits that are observed by an apparatus placed at infinity.

\subsubsection{Burning coal}\label{seccoal}

First consider how a normal object like a piece of coal burns away. The state of the emitted quanta depend on the state of the coal they are emitted from, since the emitted quanta are produced by interactions between the quanta making up the coal. But these interactions are short ranged; once a radiated quantum has left the vicinity of the coal, \emph{its state is no longer affected by the coal}. The Page curve for the coal at first rises and then falls back down to zero.

One might say that there could be some small long-ranged interactions between the emitted quanta and the remaining coal. But the point is that such interactions are not the reason that the Page curve drops down to zero {for a piece of burning coal}. We could perfectly well take a model of the coal where the interactions between the radiated quantum and the remaining coal fall to zero outside some radius $R_{max}$ and in such a model we will still find that the Page curve comes down to zero at the end of evaporation. 

A third fact that is of relevance in view of point \ref{A4} above is that degrees of freedom at infinity are independent of the degrees of freedom in the coal. The `bits' at infinity are made by excitation of say, a scalar field $\phi(x)$ near infinity, while bits in the coal are made from fields in the region of the coal. These are independent degrees of freedom. To summarize, the properties of burning coal are:

\begin{enumerate}[start=1,
    labelindent=\parindent,
    leftmargin =2.5\parindent,
    label=(C\arabic*)]
\item \label{C1}
There are no relevant interactions between the radiated quanta and the remaining coal once these radiated quanta have left the vicinity of the coal.
\item \label{C2}
The bits at infinity which describe the radiation are independent of the bits that make up the remaining coal.
\item \label{C3}
The Page curve first rises and then falls back to zero at the end of the burning process.
\end{enumerate}

\subsubsection{The semiclassical black hole}

The semiclassical black hole radiates quanta by pulling pairs out of the \emph{vacuum}. Thus these quanta have no information about the details of the matter which made the hole. The state of the created pair $\{b,c\}$ can be modelled as
\be
|\psi\rangle_{pair}={1\over \sqrt{2}}\Big (|0\rangle_b|0\rangle_c+|1\rangle_b|1\rangle_c\Big ) \ .
\label{pairw}
\ee
One can imagine that there may well be small quantum gravity effects that modify the state of the created pair by the $O(\epsilon)$ corrections noted in (\ref{pair}); here the correction to any pair can depend on the matter which made the hole and the state of the quanta that fell into the hole at earlier steps of pair creation. Some people had originally hoped that small $O(\epsilon)$ corrections to (\ref{pair}) would somehow introduce suitable correlations among the radiated quanta and bring the entanglement down to zero. However, the small corrections theorem \cite{Mathur:2009hf} showed that this is not possible; the entanglement entropy $S_{ent}(N)$ after $N$ emissions will continue to rise as
\be
 S_{ent}(N+1)>S_{ent}(N)+\ln 2-2\epsilon \ .
 \label{three}
 \ee
Thus we need an order \emph{unity} correction to the low-energy dynamics at the horizon in order to resolve the information paradox. 

To summarize, the semiclassical hole is defined as one where we study quantum fields on the fixed geometry of the classical black hole and include the possibility of small corrections arising out of nonperturbative quantum gravity processes. For this semiclassical hole, we have the following properties:

\begin{enumerate}[start=1,
    labelindent=\parindent,
    leftmargin =2.5\parindent,
    label=(SC\arabic*)]
\item \label{SC1}
The horizon is a vacuum region to leading order. Thus semiclassical dynamics holds to leading order in this region. The metric can be taken as
\be
g_{\mu\nu}=\bar g_{\mu\nu}+ h_{\mu\nu} \ ,
\label{mfour}
\ee
with $\bar g_{\mu\nu}$ being the classical black hole metric and $h_{\mu\nu}$ is small. For a scalar field on this background, we will have 
\be
\square \phi\approx 0 \ ,
\label{mone}
\ee
around the horizon. This dynamics will give rise to the creation of entangled pairs of the form (\ref{pair}).

\item \label{SC2}
There are no relevant interactions between the radiated quanta and the remaining hole once these radiated quanta have left the vicinity of the hole.

\item \label{SC3}
The bits at infinity which describe the radiation are independent of the bits that make up the remaining hole.

\item \label{SC4}
The Page curve will keep rising monotonically in the form (\ref{three}).

\end{enumerate}

\subsubsection{Fuzzballs}\label{secfuzzball}

A fuzzball behaves just like a piece of coal. The nontrivial step here is the demonstration that brane bound states in string theory do not generate the geometry of the semiclassical hole; instead they generate extended objects that have no horizon or singularity. In \cite{Gibbons:2013tqa} it was shown how the traditional no-hair theorems are violated by specific features of string theory.

A fuzzball radiates from its surface just like a piece of coal radiates photons from its surface. For a piece of coal we do not have any effective pair creation of the states (\ref{pair}); similarly, for a fuzzball we will not have any effective variables where we get (\ref{pair}). This issue is very important and will be discussed in more detail below. Thus for a fuzzball we have the following behavior:

\begin{enumerate}[start=1,
    labelindent=\parindent,
    leftmargin =2.5\parindent,
    label=(F\arabic*)]
\item \label{F1}
There are no relevant interactions between the radiated quanta and the remaining fuzzball once these radiated quanta have left the vicinity of the fuzzball.
\item \label{F2}
The bits at infinity which describe the radiation are independent of the bits that make up the remaining fuzzball.
\item \label{F3}
The Page curve first rises and then falls back to zero at the end of the burning process.
\item \label{F4}
There are no effective variables in which we get (\ref{twopp}).
\item \label{F5}
The full structure of string theory is required to obtain `fuzzballs'; i.e. to obtain objects that do not collapse to the traditional semiclassical hole. Thus a simple theory like (1+1)-dimensional JT gravity will not have fuzzballs; in such a theory we will just get the traditional semiclassical hole.
\end{enumerate}

\subsubsection{An impossibility}\label{seceff}

The discovery that brane bound states in string theory generate fuzzballs with no horizon gives a simple resolution to the information paradox. In the initial days of the fuzzball paradigm, some people thought that this change to the geometry of the hole was too radical; they hoped that small corrections to the traditional black hole could somehow encode enough correlations in the radiated quanta to bring the Page curve down to zero. The small corrections theorem (\ref{three}) showed that this hope could not be realized; one needs an order unity correction to horizon dynamics, so the fuzzball paradigm was a \emph{natural} resolution to the puzzle rather than a radical one. 

At this point some people felt that the following might be possible. Suppose it was true that in the exact quantum gravity theory the microstates of the hole behaved like pieces of coal. But this description of the microstates would be very complicated. Suppose that there was some choice of effective variables describing the complicated degrees of freedom of the black hole, in which an approximation to the semiclassical dynamics emerged. In that case getting an exact description of the hole in terms of fuzzballs would be correct, but it may be that the effective variables would give a simpler and more useful description of the dynamics. To make this suggestion more precise, suppose we require the effective description to have the following properties:
\begin{enumerate}[start=1,
    labelindent=\parindent,
    leftmargin =3\parindent,
    label=(EFF\arabic*)]
\item \label{EFF1}
There are no relevant interactions in the exact theory between the radiated quanta  and the remaining hole once these radiated quanta have left the vicinity of the coal. (This is just like \ref{C1} for burning coal.)

\item \label{EFF2}
The bits at infinity which describe the radiation are independent of the bits that make up the remaining hole. (This is just like \ref{C2} for burning coal.)

\item \label{EFF3}
The effective degrees of freedom describing the hole are obtained (in some possibly very complicated way) from all the degrees of freedom in the region of the hole; say in the region $r<10\,r_h$, where $r_h$ is the radius of the hole.\footnote{In the rest of this paper we will use this region $r<10\,r_h$ as describing the region up to a few horizon radii from the center of the hole; the reader could of course substitute the number $10$ by any other number of his choice.} Note that from property \ref{EFF2}, these degrees of freedom making the effective bits are independent of the degrees of freedom near infinity.

\item \label{EFF4}
We will be quite generous in how little we demand from these effective variables:

\begin{enumerate}[start=1,
    labelindent=\parindent,
    leftmargin =0.6\parindent,
    label=(\roman*)]
\item \label{i}
The semiclassical dynamics of the hole has to emerge only approximately with these variables. Thus for a scalar field the equation $\square \phi=0$ in the vicinity of the horizon can be relaxed to
\be
\square \phi_{ef\!f}=O(\epsilon) \ , \quad \epsilon\ll 1 \ .
\label{mtwo}
\ee

\item \label{ii}
This effective semiclassical dynamics has to only describe low-energy physics. This low-energy physics includes  modes with wavelengths $r_h/100\lesssim \lambda \lesssim 10\,r_h$ since it is the stretching of modes in this range which gives the Hawking pair production that we are interested in. But the effective dynamics does not have to work for wavelengths down to string  length or Planck length.

\item \label{iii}
The evolution (\ref{mtwo})  will yield the creation of approximate entangled pairs (\ref{twopp}) in the region of the hole. We do not require that the above effective description describe the creation of \emph{all} the pairs emitted by the hole. We merely ask that it describes the emission of a few pairs, after which we may have to choose a new set of effective variables $\phi_{ef\!f}(x)$ to continue getting an effective semiclassical dynamics around the horizon.

\end{enumerate}
\hspace{-1.45cm}Given the above conditions one can prove the following:
\item \label{EFF5}
The Page curve for the \emph{exact} theory will keep rising monotonically; i.e. we cannot get the analogue of property \ref{C3} of burning coal.
\end{enumerate}
This statement can be proved by a simple adaptation of the proof of the small corrections theorem (\ref{three}). In this adaptation we will use the pair creation (\ref{twopp}) for \emph{effective} variables in place of the pair creation (\ref{pair}) for semiclassical excitations around the traditional black hole geometry. The result \ref{EFF5} will be termed the effective small corrections theorem; we will see its derivation in section~\ref{sec small}.

To summarize the content of \ref{EFF1}-\ref{EFF5}: \emph{we cannot require that the black hole behaves like a piece of coal as seen from outside and also require that the variables in the region $r<10\,r_h$ give rise to effective semiclassical dynamics around the horizon.}

\subsubsection{Wormhole model - I: Nonlocal construction of effective variables}  

We have seen above that in the fuzzball paradigm the black hole behaves like burning coal as seen from outside, thus properties \ref{C1}-\ref{C3} are reflected in \ref{F1}-\ref{F3}. Furthermore, for a piece of coal we do not have any effective dynamics where we see pair creation at the horizon; likewise, for fuzzballs we have \ref{F4} which says that there is no such effective description of pair creation. Thus if we did not wish to insist on getting (\ref{mtwo}) for low-energy modes, then fuzzballs already provide a resolution of the information paradox.

As we have noted, the wormhole paradigm \emph{does} ask for effective low-energy semiclassical dynamics around the horizon and so will have the effective pair creation (\ref{twopp}). As discussed in section~\ref{seceff} above, this effective pair creation cannot be achieved while, (i) keeping all the properties of burning coal \ref{C1}-\ref{C3} and, (ii) making the effective degrees of freedom only out of degrees of freedom in the region $r<10\,r_h$. 

The wormhole paradigm tries to get around this difficulty by a variety of ways, which we will now discuss. We start with the idea that the effective variables can be made using both the exact bits in the region of the hole and the exact bits at infinity. We have the following postulates:

\begin{enumerate}[start=1,
    labelindent=\parindent,
    leftmargin =2.5\parindent,
    label=(WI-\arabic*)]

\item \label{WI-1}
There are no relevant interactions between the radiated quanta and the remaining hole once these radiated quanta have left the vicinity of the coal. Here we are talking about the \emph{exact} bits of the theory (just like \ref{C1} for burning coal).

\item \label{WI-2}
The bits at infinity which describe the radiation are independent of the bits that make up the remaining hole. Again we are talking about the \emph{exact} bits of the theory (just like \ref{C2} for burning coal).

\item \label{WI-3}
The effective degrees of freedom describing the hole are obtained (in some possibly very complicated way) from the exact bits in the region of the hole ($r<10\,r_h$), \emph{as well as} the exact bits making up the radiation $R$ that has been emitted from the hole.

\item \label{WI-4}
We ask for the same effective dynamics that was listed in (EFF4) in section~\ref{seceff} above.

\item \label{WI-5}
We ask that the Page curve for the \emph{exact} bits making up the radiation $R$ come down to zero at the end of the evaporation process.
\end{enumerate}
This may look like an appealing set of properties to ask for, since we have asked for the exact dynamics to be like that of coal and have also asked for effective semiclassical dynamics around the horizon. However, the construction \ref{WI-3} runs into immediate difficulty with the postulate \ref{WI-1} as follows. Suppose we do have a set of effective bits which make the horizon region an approximation to the semiclassical hole, say in the region $r<10\,r_h$. Place an apparatus at $r=5\,r_h$ that sends a beam into the hole and another at $r=5\,r_h$ that checks for a reflected beam. Since the effective variables yield semiclassical dynamics to a first approximation, there will be very little reflected beam; the incoming beam will fall into the hole. Now take the exact bits in the radiation $R$ and change their state; for example if they were spins, rotate the spins using a magnetic field applied in the radiation region near infinity. Since the effective bits near the horizon involved the exact bits at infinity, the state of the effective bits near the horizon will change; for example by
\be
{1\over \sqrt{2}}\Big (|0\rangle_{b,ef\!f}|0\rangle_{c,ef\!f}+|1\rangle_{b,ef\!f}|1\rangle_{c,ef\!f}\Big ) ~\r~{1\over \sqrt{2}}\Big (|0\rangle_{b,ef\!f}|1\rangle_{c,ef\!f}+|1\rangle_{b,ef\!f}|0\rangle_{c,ef\!f}\Big ) \ .
\label{mthree}
\ee
But if the state on the left-hand side of (\ref{mthree}) was the local vacuum at the horizon, the state on the right-hand side will \emph{not} be the vacuum. Thus the beam will now reflect off the hole and be observed in the detection apparatus.\footnote{For an explicit example of how waves reflect off a fuzzball see \cite{Chua:2021sew}.} Thus we see that manipulating the radiation bits $R$ at infinity in a suitable way can change the structure of the hole in the region $r<10\,r_h$. As a consequence, we should modify \ref{WI-1} to read
 
 \begin{enumerate}[start=1,
    labelindent=\parindent,
    leftmargin =2.5\parindent,
    label=(WI-\arabic*')]
 
\item \label{WI-1'} Modifying the radiation quanta at infinity will change the dynamics that is observed by experimenters in the vicinity of the hole. This is different from the behavior of a piece of burning coal, where manipulating the radiation quanta at infinity does not change the observations of an experimenter examining the coal. We will investigate models of this type in section~\ref{secnonlocaldef}.
 
\end{enumerate}
Maldacena \cite{MaldacenaSaclayComment1} has suggested a model where manipulating the bits at infinity will lead to a modification of the bits in the \emph{interior} of the hole.\footnote{We could regard this interior as the region inside of the QES, in the cases where the QES is within the horizon.} In particular, he has conjectured that if the hole has evaporated away completely and its contents have pinched off into a baby universe, then manipulating the radiation bits $R$ at infinity can extract information from this baby universe.
 
As we will observe in section~\ref{requirement} however, manipulating the bits at infinity will also force to a modification of the bits in the vicinity of the horizon and this latter modification will lead to the above noted change in the results of an experimenter who seeks to scatter light off the hole. It is very important to understand why in any situation where the interior bits can be manipulated from infinity, the horizon region must \emph{also} change under such modifications. We have the following situation:
 
\begin{enumerate}[start=1,
    labelindent=\parindent,
    leftmargin =2\parindent,
    label=(\alph*)]
 
\item \label{(a10)}
Suppose that we take a bit model where the bits $b_{ef\!f}, c_{ef\!f}$ emerging in the Hawking process are made by using the bits at infinity as well as the bits in the region $r<10\, r_h$. In this case manipulating the bits at infinity will change the observations of an experimenter outside the hole who is trying to reflect a beam off the hole.
 
\item \label{(b10)}
Suppose we say that the effective bits around the horizon do \emph{not} involve the bits at infinity. Then we find that the $b_{ef\!f}, c_{ef\!f}$ are entangled with each other just as in Hawking's original computation. In this case the effective small corrections theorem will tell us that the Page curve of the exact theory will have to keep rising monotonically. 
 
\item \label{(c10)}

Suppose we proceed as in case \ref{(b10)}, but try to get the Page curve to come down by requiring that the $c_{ef\!f}$ quanta in the interior of the hole (i.e. in the `island' region) are made as some combinations of the bits at infinity as well as the bits in the region $r<10\,r_h$. Such an attempt will lead to nonunitarity of evolution in the region $r<10\,r_h$. The reason is the we have required approximate semiclassical evolution in the region around the horizon, and this semiclassical evolution takes the $c_{ef\!f}$ created around the horizon in (b) above and deposits them on the island. So we do not have any freedom to choose what bits the $c_{ef\!f}$ in the island are made of; in particular, the $c_{ef\!f}$ that end up on the island are maximally entangled with the $b_{ef\!f}$ the escape to infinity. If we try to annihilate the $c_{ef\!f}$ that are created at the horizon so that they do {\it not} reach the island, then we get a nonunitarity of evolution in the region $r<10\,r_h$. 
\end{enumerate}
In short, we have not been able to find a unitary bit model where, (i) the horizon exhibits an effective semiclassical dynamics, (ii) The Page curve comes down and, (iii) an experimenter outside the hole cannot see modifications of horizon behavior when bits at infinity are manipulated.

\subsubsection{Wormhole model - II: Nonlocal Hamiltonian interactions}\label{nonlocal}

Some people have argued that quantum gravity can have nonlocal Hamiltonian interactions and that it is these interactions that resolve the puzzle. One could then ask: why do we not see these nonlocal interactions in everyday experiments? The idea of these proposals would then be that the nonlocal effects are not large, in a sense that we explain below with the following postulates:

\begin{enumerate}[start=1,
    labelindent=\parindent,
    leftmargin =2.5\parindent,
    label=(WII-\arabic*)]

\item \label{WII-1} 
There are nonlocal interactions between the radiation $R$ and the remaining degrees of freedom in the region $r<10\,r_h$. Thus this is different from postulate \ref{C1} for burning coal.

\item \label{WII-2}
These nonlocal effects can change, for example, the spin of any radiation quantum at infinity through an interaction that depends on the state in the region $r<10\,r_h$. But this change in the spin would be small if we look at, (i) just a few radiation quanta and, (ii) look at these quanta over timescales much shorter than the Hawking evaporation time.

\item \label{WII-3}
With these interactions, the Page curve comes down to zero by the end of the evaporation process.

\end{enumerate}
We will give a toy model for such nonlocal interactions in section~\ref{secnonlocal}. We do not believe that such nonlocal interactions actually arise in string theory. Here we just observe that if one \emph{does} postulate such effects, then a model along the lines of \ref{WII-1}-\ref{WII-3} is possible. It is important to note that \ref{WII-1} is different from \ref{C1} for burning coal, so we cannot say that in such models the hole behaves like a piece of burning coal as seen from outside.

\subsubsection{Wormhole model - III: Identifying bits between the hole and infinity}\label{secintroidentify}

Another set of arguments has taken the track of altering the condition \ref{C2} that we had for coal. That is, we postulate that the degrees of freedom far from the hole are not independent of degrees of freedom in the region of the hole $r<10\,r_h$. Thus the postulates would have the form

\begin{enumerate}[start=1,
    labelindent=\parindent,
    leftmargin =3\parindent,
    label=(WIII-\arabic*)]

\item \label{WIII-1}
There are no Hamiltonian terms giving an interaction between the bits in the region $r<10\,r_h$ and the radiation region $R$.

\item \label{WIII-2}
The degrees of freedom at infinity are not independent of the degrees of freedom in the region $r<10\,r_h$. (This is different from \ref{C2} for burning coal.)

\item \label{WIII-3}
We require the horizon have the semiclassical behavior of the traditional hole; i.e. we have the creation of entangled pairs (\ref{pair}) at the horizon. (We do not talk about effective bits, since the identification of bits is supposed to resolve the puzzle while maintaining semiclassical dynamics (\ref{mfour}) at the horizon.)

\item \label{WIII-4}
We require that the Page curve comes down to zero at the end of the evaporation process.

\end{enumerate}
However, there is an immediate issue with such a model. By \ref{WIII-3}, we create entangled pairs at the horizon. Each of the excitations $\{b, c\}$ have two states $\{0,1\}$, so we have a $4$-dimensional Hilbert space from these excitations. The excitation $b$ moves off to infinity, while $c$ falls into the hole; at this stage we still have a $4$-dimensional space of states for this pair of quanta. Now suppose we wish to make an identification of bits, so that the bit representing $b$ is identified with the bit representing $c$. We can try this in two ways:

\begin{enumerate}[start=1,
    labelindent=\parindent,
    leftmargin =2\parindent,
    label=(\roman*)]

\item \label{(i)}
We require that the state of the bit $b$ become the state of the bit $c$. Since $b$ and $c$ have the same state, the Hilbert space spanned by them is now $2$-dimensional. The reduction from $4$ to $2$ dimensions is a nonunitary evolution of the system.

\item \label{(ii)}
We keep all $4$ states of the $b,c$ pair but introduce a nonlocal Hamiltonian between the $b,c$ quanta so that the states where $b,c$ are not identified rise in energy to a level where they are not part of the low-energy space of excitations. With this model we find that the $b$ quantum at infinity does not behave like a normal quantum: it costs energy to change its state between $|0\rangle$ and $|1\rangle$, while a similar bit radiated by a piece of coal does not have such an energy increase.
\end{enumerate}
The above are two very crude models of what happens if we try to identify bits. One can try to include more complicated identifications, however, the essential difficulty will remain the same: the continuous production of new entangled pairs gives an enlargement of the Hilbert space of excitations and if we try to bring the Page curve down by introducing identifications between bits, then we either have nonunitarity or find that the bits at infinity behave differently from bits radiated from a piece of coal. We will consider models of this type in section~\ref{identify}.

\subsubsection{The Page curve}

It has been argued that one can obtain a Page curve for the black hole that comes down to zero at the end of evaporation using general arguments of semiclassical gravity (e.g. (1+1)-dimensional JT gravity) without knowing the details of the quantum gravity theory. 
We will argue that one cannot get the Page curve in this manner. In more detail, we do the following:

\begin{enumerate}[start=1,
    labelindent=\parindent,
    leftmargin =2\parindent,
    label=(\alph*)]

\item \label{(a)}
It has been stated that such a semiclassical argument for the Page curve is  similar to the Gibbons-Hawking computation of black hole entropy. We will see, however, that there is an important difference. With the Gibbons-Hawking computation we start with a Euclidean path integral with time period $\beta$ which correctly counts the states for \emph{any} system with Hamiltonian $H$ and spectrum $\{E_i\}$, through the partition function
\be
Z(\beta)=\Tr \big[e^{-\beta H}\big]=\sum_i e^{-\beta E_i} \ .
\label{dthtwo2}
\ee
One then observes that there is a plausible classical saddle for this path integral, with the only assumption made being that this saddle should give a good approximation to the integral. For the Page curve computation we are not able to cast the the starting path integral as a quantity that gives the entanglement for an arbitrary system. For example, in computing the second R\'{e}nyi entropy $S_2(A)$ we need to compute $\Tr[(\rho_A)^2]$ where $\rho_A$ is the reduced density matrix of the region $A$. In the recent Page curve computation one makes a replacement of the type 
\be
\Tr\big[(\rho_A)^2\big]~\r~ \Big ( \Tr\big[(\rho_A)^2\big]+ C\big(\Tr[\rho_A]\big)^2 \Big ) \ ,
\label{mfive}
\ee
for some constant $C$. Thus we will be starting with a quantity that appears to be \emph{different} from the entanglement entropy that we wished to compute.  

\item \label{(b)}
It has been argued that the prescription (\ref{mfive}) is justified because it takes into account the fact that there can be topology change in gravity. We consider the role of topology changing processes in section \ref{secchange} (using (1+1)-dimensional gravity as an example) and find that topology change does not allow for the prescription (\ref{mfive}). The second R\'{e}nyi entropy $S_2(A)$ is still given by $\Tr[(\rho_A)^2]$ on the full Hilbert space, which now includes disconnected line segments arising from the possibility of topology change.

\item \label{(c)}
So what can give the prescription (\ref{mfive})? It is crucial to note that we cannot make arbitrary `prescriptions' for the behavior of the effective semiclassical gravity theory and then use these to compute quantities like entanglement entropy. Let the variables of the exact gravity theory be denoted by $g_{exact}$ and of the approximate effective theory by $g_{ef\!f}$. The effective variables $g_{ef\!f}$ need to be some functionals of the exact variables $g_{exact}$; we write this symbolically as 
\be
g_{ef\!f}=F[g_{exact}] \ .
\label{dthreeqq}
\ee
 The Lagrangian of the exact theory then determines, through (\ref{dthreeqq}), the dynamics of the effective theory
 \be
{\mathcal L}_{exact}[g_{exact}] ~\r~ {\mathcal L}_{ef\!f}[g_{ef\!f}] \ .
\label{dfourqq}
\ee
Similarly, any quantity $Q_{exact}[g_{exact}]$ which is of interest in the exact theory will map, through (\ref{dthreeqq}), to a quantity $Q_{ef\!f}[g_{ef\!f}]$ in the effective theory
\be
Q_{exact}[g_{exact}]~\r~Q_{ef\!f}[g_{ef\!f}] \ .
\label{dfivepreqq}
\ee
Thus any prescription like (\ref{mfive}) for the semiclassical dynamics must have its origins in the dynamics of the exact theory. To summarize, in our investigations we have not been able to find any way that the effective theory emerging from the exact theory will give rise to a wormhole that will connect different replica copies. 
\end{enumerate}

\subsubsection{Investigating nonlocalities}

If the effective variables describing the black hole are made from the exact variables in the region of the black hole $r<10\,r_h$ then we have seen that the postulates \ref{EFF1}-\ref{EFF4} imply \ref{EFF5}; i.e. the Page curve does not come down. So to look for what feature in the exact theory can give a prescription like (\ref{mfive}) we consider nonlocal effects. We consider three kinds of these nonlocalities:

\begin{enumerate}[start=1,
    labelindent=\parindent,
    leftmargin =2\parindent,
    label=(\roman*)]

\item \label{(i2)}
\emph{Nonlocal effects in the exact theory that connect the interior of one hole to the interior of another hole.} The effective small correction theorem extends in a straightforward way in this case to say that the Page curve cannot come down. We look at a model where the interior regions of the holes disconnect to give rise to baby universes, and this correlates the different interiors. We find that in this case the evolution in the black hole interior violates unitarity.

\item \label{(ii2)}
\emph{Nonlocal effects in the exact theory that connect the interior of the hole to the radiation region.} In this case we can have models like that in section~\ref{nonlocal} where the Page curve comes down to zero, but the hole does not look like a piece of burning coal as seen from outside: \ref{WII-1} differs from \ref{C1}. It is also not clear how such nonlocal interactions lead to the prescription (\ref{mfive}) used in the Euclidean path integral.

\item \label{(iii2)}
\emph{Nonlocal effects between the radiation near infinity from one hole and the radiation near infinity from another hole.} It has been argued that such effects can change the way we measure the entanglement of the radiation $R$. This is because one needs to measure many identical copies of the radiation $R_1, R_2, \dots$ produced from identically prepared holes in order to judge the state of this radiation. If these different measurements interfered with each other, then one would have a novel effect with radiation from a black hole; i.e. we would have an effect that is not present when we check the entanglement of radiation from normal quantum objects. Note that since we can separate the different copies of the hole by an arbitrary distance, this interaction between the radiation regions $R_i$ must not fall off with distance. We do not believe that there are such nonlocal effects in string theory; it is also not clear how exactly such effects would lead to the prescription (\ref{mfive}).
\end{enumerate}

\subsection{Summary}

Let us return to our original issue: why has the wormhole paradigm been so confusing? One reason is that the wormhole paradigm is not addressing the information paradox itself, but a somewhat different question. The information paradox arises from a combination of two observations:

\begin{enumerate}[start=1,
    labelindent=\parindent,
    leftmargin =2\parindent,
    label=(\roman*)]

\item \label{(i)'}
The no-hair results suggest that all matter in a black hole rushes to the central singularity, leaving the vacuum state around the horizon.

\item \label{(ii)'}
Hawking's computation shows that entangled pairs are created from such a vacuum region, leading to a monotonically rising Page curve. The small corrections theorem (\ref{three}) makes this difficulty precise, since no small corrections to Hawking's computation can bring the Page curve down.

\end{enumerate}
The fuzzball paradigm resolves the paradox by showing that in string theory the no-hair theorems are violated: all microstates that have been constructed are horizon-sized quantum fuzzballs with no horizon or singularity. But these constructions need the full structure of string theory; there are no fuzzballs in (1+1)-dimensional gravity, we simply get a monotonically rising Page curve \cite{Callan:1992rs,Russo:1992ht,Keski-Vakkuri:1993ybv,Fiola:1994ir}.

The recent wormhole paradigm arguments do not seek to address the information paradox as summarized in points \ref{(i)'} and \ref{(ii)'} above. Instead, these arguments typically start with an \emph{assumption} that some hitherto unknown effects in the quantum gravity theory makes the black hole behave like a piece of coal as seen from outside; i.e. the Page curve comes down to zero at the end of evaporation. The question that is then asked is: \emph{given} this behavior of the Page curve, how can we recover some approximation to semiclassical dynamics  around the horizon? Note that this question is \emph{different} from the information paradox.

There is an immediate difficulty in answering the above question about semiclassical behavior at the horizon. As noted in section~\ref{seceff}, one cannot get this semiclassical behavior through \emph{any} combination of the degrees of freedom in the black hole region $r<10\,r_h$. In the fuzzball paradigm, we note this fact and observe that there will be no low-energy semiclassical dynamics at the horizon (property \ref{F4} in section~\ref{secfuzzball}). There is a possibility of getting some effective classical dynamics for infalling objects with \emph{high} energies $E\gg T$, where $T$ is the temperature of the hole; this possibility is called the conjecture of fuzzball complementarity \cite{Mathur:2017fnw,Mathur:2017wxv}. However, this conjecture has no bearing on the discussions of the information paradox and the Page curve since these discussions only involve the Hawking quanta which have energy $E\sim T$. For such $E\sim T$ quanta, the fuzzball paradigm says that there is no effective dynamics that yields (\ref{twopp}).

The wormhole paradigm seeks to get the effective dynamics (\ref{twopp}) through a variety of postulates that involve \emph{nonlocal} effects connecting the hole to its far away radiation. Since the question being asked is about effective low-energy dynamics, the computations with the wormhole paradigm typically involve simple (1+1)-dimensional theories like JT gravity, not the full structure of string theory. But (1+1)-dimensional gravity has been well studied and here one finds no resolution of the information puzzle: the Page curve keeps rising monotonically. So what can we hope to learn by using such simple theories? What one does in the wormhole paradigm is to add `prescriptions' to the behavior of the (1+1)-dimensional theory. These prescriptions can, for example, be in the form of a modification (\ref{mfive}) of how the R\'{e}nyi entropy should be written in terms of path integrals. With these new prescriptions for the low-energy dynamics, it is then argued that one has found a Page curve that comes down to zero.

Here we come to a crucial issue. One cannot make an arbitrary prescription for low-energy effective dynamics. Instead, these low-energy effective variables have to emerge from some map of the form (\ref{dthreeqq}) This map then  determines the low-energy effective dynamics and the rules for computing low-energy quantities as in (\ref{dfourqq}) and (\ref{dfivepreqq}). The wormhole paradigm does not seek to give us the map (\ref{dthreeqq}). But in that case, how do we know that the low-energy prescriptions are correct? We have tried to list various prescriptions that have been considered in the wormhole paradigm, and then ask what effect in the exact theory these would emerge from. These effects in the exact theory must take the form of  nonlocal effects over long distances, since constructions of effective variables that use only the degrees of freedom in the region of the hole $r<10\,r_h$ cannot bring the Page curve down. We do not believe such nonlocal effects are actually present in string theory. But in the present article we will not discuss the existence of nonlocality in string theory; we will take up this issue in a following article. Instead, we will elaborate on the observations made in the sections above in an effort to concretize the nonlocalities that are explicitly or implicitly part of  the wormhole paradigm.

\subsection{The plan of the paper}
 
The plan of this paper is as follows.\\
\\
In section~\ref{sec small} we derive the `effective small corrections theorem'. This theorem extends the small corrections theorem of \cite{Mathur:2009hf} to the case where we have only  approximate semiclassical behavior at the horizon in terms of effective variables made out of all the degrees of freedom in the region of the hole. Thus, we do not assume that the actual spacetime around the horizon is close to the classical one. This theorem shows that we cannot have both, (i) the black hole radiating like a piece of coal as seen from outside (conditions \ref{C1}-\ref{C3} of section~(\ref{seccoal}) and, (ii) some effective degrees of freedom in the region $r<10\,r_h$ giving rise to a `code subspace' where we have approximate semiclassical behavior satisfying the weak requirements of \ref{EFF4}. This theorem thus implies that the wormhole paradigm must have some kind of nonlocal effects as an essential ingredient in getting the Page curve to come down\footnote{To be precise, getting around the effective small corrections theorem requires the violation of one of its assumptions, the least radical of which is the introduction of non-locality. Otherwise, something like non-unitarity would be required; something that is not very appealing.}.

This is followed by section~\ref{secdef}, where some notation and background for black holes is recalled: the Penrose diagram, good slices, baby universes etc.

In sections~\ref{secchange}-\ref{pageiii} we examine recent suggestions that the Page curve for a black hole can be computed by adding certain prescriptions to how we use semiclassical gravity. In these computations one finds the entanglement entropy by taking a suitable limit of R\'{e}nyi entropies, and argues that the resulting Page curve will come down like the Page curve of a normal body. We find that in these computations the R\'{e}nyi entropies are replaced by a new quantity that is \emph{not} the R\'{e}nyi entropy, so the curve that is argued to come down is not the Page curve, in the sense of entanglement entropy. It has sometimes been argued that the replacement of the R\'{e}nyi entropies by the new quantities is dictated by the possibility of topology change in gravity. We examine the role of topology change in the computation of entanglement entropies via path integrals and find that, at least in the (1+1)-dimensional quantum gravity example studied, topology change does \emph{not} imply a replica wormhole connecting different copies in the R\'{e}nyi entropy computation. We note that one cannot make an arbitrary prescription for how semiclassical geometries should behave in an effective theory, since this effective theory must descend from the exact theory through the relations (\ref{dthreeqq}), (\ref{dfourqq}) and (\ref{dfivepreqq}). We therefore argue that the recent Page curve computations differ from the Gibbons-Hawking computation of black hole entropy in a fundamental way: while the Gibbons-Hawking computation starts with a path integral that would yield the entropy for \emph{any} physical system, the Page curve computation modifies the starting path integral in a way that yields a quantity different from the entanglement entropy.

We could not identify, in section~\ref{nonlocalchoices}, any clear nonlocality postulate for the exact theory that could yield the prescriptions used for the effective theory in the recent Page curve computations. We therefore proceed by examining various kinds of nonlocalities that have been postulated and find that the postulate that baby universes connect the interiors of different black holes leads to a nonunitarity of evolution. Alternatively, nonlocalities that connect the hole to infinity lead to an asymptotic observer finding different behaviors for quanta radiated from coal and from a black hole. 

In section~\ref{requirement} we give an explicit set of conditions that must be met by any bit model for the wormhole paradigm. These conditions impose the requirement of an effective low-energy semiclassical dynamics at the horizon and the requirement of a Page curve that comes down.

We then conclude in section~\ref{discussion} by a summary. The appendix~\ref{appA} contains a review of some aspects of the fuzzball paradigm and appendix~\ref{appb} details a bit model for the process of Hawking pair creation for a classical black hole horizon.\\
\\
In this paper we have just tried to isolate the nonlocality postulates that are implied by the wormhole paradigm.\footnote{We do not discuss approaches like that of \cite{Penington:2019kki} which require the exact theory to be an ensemble averaged theory; this is because we have in mind string theory as our exact theory and we believe that string theory is not an ensemble averaged theory.} The literature of the wormhole paradigm spans a large number of papers, but we have included very few references. This is because we are not seeking to analyze in detail any particular paper in this paradigm but to instead explore the different categories of models that have been proposed. Thus where we do include references, they just point to the kind of model that we are analyzing. As we already noted above, it is possible that the proponents of some wormhole models have ideas in mind different from those we cover and it is the hope of this paper that these authors would explain their work in the bit model language used in the present paper and thus clarify the physics that leads to the Page curve coming down in their model.

In a follow-up paper we plan to present a discussion of why we believe that such nonlocalities do not exist in string theory. In particular some confusion has been caused by suggestions that AdS/CFT duality implies a nonlocality in gravity. But such is not the case; the CFT and the gravity theory are both completely local, and the nonlocality of the map between the two cannot be used to argue for a nonlocality in gravity itself.

\section{The effective small corrections theorem}\label{sec small}

The small corrections theorem, proved in \cite{Mathur:2009hf}, shows that any small corrections to semiclassical horizon dynamics will not change Hawking's conclusion that the Page curve monotonically rises. We are now interested in a situation where the actual state of the hole is not necessarily close to the semiclassical geometry; in fact this state can be a very complicated mess of the quantum gravitational degrees of freedom in a region which for concreteness we take to be $r<10\,r_h$. We then ask that in some effective variables made out of these complicated degrees of freedom, we get the low-energy semiclassical dynamics described by the conditions \ref{EFF4} listed in section~\ref{seceff}. In this situation we can immediately extend the small corrections theorem to an `effective small corrections theorem'; here the term effective denotes the fact that instead of the exact bits $\{b,c\}$ we now have effective bits $\{b_{ef\!f}, c_{ef\!f}\}$. 

The effective small corrections theorem provides a strong constraint which relates the exact theory to any effective theory.  In short, the result of the theorem is the following. Suppose that in the effective theory the dynamics of low-energy modes around the horizon is the traditional semiclassical dynamics, then in this effective description, we will find the production of entangled pairs of the form (\ref{twopp}). Suppose further that far from the hole (say for $r>100\,r_h$) the physics decouples from the physics of the hole, by which we mean: (i) the effective degrees of freedom $g_{ef\!f}$ describing the hole do not involve the degrees of freedom $g_{exact}$ at $r>100\,r_h$ and, (ii) quanta in the exact theory that reach $r>100\,r_h$ are no longer influenced significantly by the hole. \emph{Then the entanglement entropy $S_{ent}$ will keep rising monotonically in the exact theory, i.e. the Page curve of the exact theory will not come down.} We will outline the derivation of this theorem below, setting it up in the context of our present discussion. As we proceed with this outline we will make clear the assumptions that go into the proof.

\subsection{The proof of the effective small corrections theorem}\label{proof}

We proceed in the following steps.
\begin{enumerate}[start=1,
    labelindent=\parindent,
    leftmargin =1.3\parindent,
    label=(\Alph*)]
\item \label{(AA)}{\bf The exact theory:} First consider the exact theory. The black hole has a mass $M$ as seen from infinity; let the classical black hole for this mass have Schwarzschild radius $r_h$. We assume that far from the hole the exact theory is just given by standard string theory around gently curved space (`normal physics'). For pedagogical convenience, let us say that this far-away region is $r>100\,r_h$. We assume that the degrees of freedom in this far region are independent of the degrees of freedom in the hole, just as is the case in normal quantum field theory.
\begin{figure}[h]
\begin{center}
\includegraphics[scale=0.45]{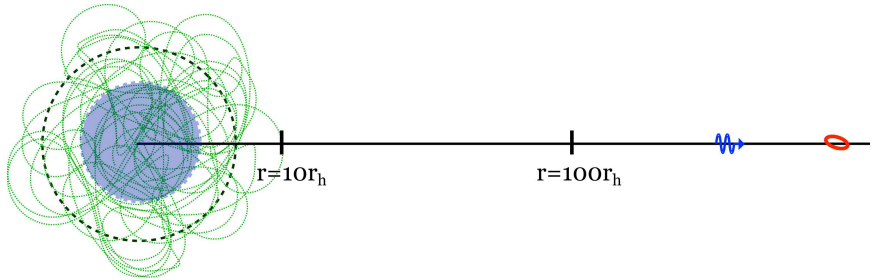}\hskip20pt
\end{center}
\caption{The gravitational mess. The blue region depicts the classical black hole (within $r<r_h$) and the green region depicts `the region of the hole' ($r<10\,r_h$) where complicated quantum gravity effects may occur. Also shown is the region far away ($r>100\,r_h$) where `normal physics' occurs.}
\label{fig1a}
\end{figure}
In and around the black hole the classical metric has a low curvature; this may suggest that we should take the traditional picture of the hole where the metric is given by (\ref{mfour}) with $\bar g_{\mu\nu}$ the classical black hole metric. However, we will not limit ourselves to such a semiclassical picture, instead, allowing for the possibility that due to some unknown quantum gravity effects, the entire region of the hole is a complicated quantum gravitational mess; we depict this symbolically in fig.~\ref{fig1a}. For pedagogical concreteness we let the `region of the hole' be the region $r<10\,r_h$.
\item \label{(BB)} {\bf Requiring an effective semiclassical description:} Now consider the effective theory. In the distant region $r>100\,r_h$ we will not define any effective theory, since we have already assumed that the low-energy physics in the far region is just low-energy string theory in gently curved space (`normal physics'). So the exact theory in the far region already has the behavior we would want for any effective theory. 
In the region of the hole ($r<10\,r_h$) the exact theory is very complicated. We assume that from the large number of degrees of freedom describing the exact theory in this region, a small subset can be used to describe dynamics that approximate the dynamics expected from the semiclassical black hole. This subset is described by effective variables $g_{ef\!f}$ which are some complicated functionals of the exact degrees of freedom $g_{exact}$
\be
g_{ef\!f}=F[g_{exact}] \ .
\label{dsix}
\ee
This small subset $g_{ef\!f}$ is sometimes called the `code subspace' which captures semiclassical dynamics from all the complicated degrees of freedom in the region. For pedagogical concreteness, we assume that the degrees of freedom $g_{ef\!f}$ describe the metric and a scalar field $\phi$ satisfying
\be
\square \phi\approx 0 \ .
\label{dfive}
\ee
The approximation sign here indicates that the effective semiclassical description is only required to be approximate; we will be more explicit about the accuracy of this approximation below. We will be quite generous in the freedom which we allow for this effective theory. We require the effective behavior (\ref{dfive}) only for low-energy modes. Thus for a hole with $r_h=3\, \rm{km}$, we can ask that (\ref{dfive}) need only hold for wavelengths between, say, $1\, \rm{cm}$ and $20\, \rm{km}$, the range where we need to follow vacuum modes to see the emergence of particle pairs in the Hawking computation. Further, we require that the effective variables $g_{ef\!f}$ given by (\ref{dsix}) describe the semiclassical dynamics (\ref{dfive}) only for the duration of production of a few Hawking pairs; after which, one may need a different choice of effective variables $\t g_{ef\!f}=\t F[g_{exact}]$ to get the semiclassical dynamics. These are the minimum conditions that we need to describe the requirement that there be some kind of effective semiclassical behavior. Note that the only assumption we have made in these conditions is that the effective degrees of freedom describing the region of the hole emerge from the exact degrees of freedom in the region of the hole; thus in particular they do not involve the exact degrees of freedom in the far region $r>100\,r_h\,$.
\item \label{(CC)} {\bf Pair production in the effective description:} Given (\ref{dfive}), we will have the production of entangled pairs in the effective theory (fig.~\ref{fig1c}). We lose no generality of the argument by taking a simple form for the state of the pair 
\be
|\psi_{ef\!f}\rangle_{pair}={1\over \sqrt{2}}\Big (|0\rangle_{b,ef\!f}|0\rangle_{c,ef\!f}+|1\rangle_{b,ef\!f}|1\rangle_{c,ef\!f}\Big ) +O(\epsilon) \ .
\label{dseven}
\ee
Here the $O(\epsilon)$ corrections encode the fact that the evolution (\ref{dfive}) was only approximate; we will specify the magnitude of these corrections more precisely below. In the region immediately around the hole (say the region $r< 2\,r_h$) spacetime is curved and the definition of particles is somewhat ambiguous. Once the quantum $b_{ef\!f}$ gets far from the hole, the definition of particles becomes well defined. We will use this latter fact to remove part of the ambiguity from $b_{ef\!f}$ in the step below. (There will be no need to remove any ambiguity in the definition of particles from $c_{ef\!f}$.)
\begin{figure}[H]
\begin{center}
\includegraphics[scale=0.5]{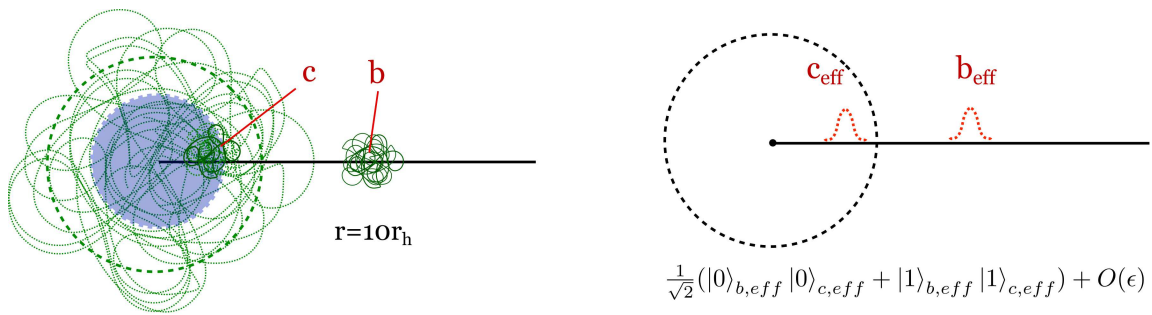}
\end{center}
\caption{The left-hand figure depicts an entangled pair $\{c,b\}$ in the exact theory. The right-hand side depicts the entangled degrees in the effective theory, where the entangled pair emerges in the state  $|\psi_{ef\!f}\rangle_{pair}$ as shown above (also mentioned in (\ref{dseven})).}
\label{fig1c}
\end{figure}
\item \label{(DD)} {\bf The movement of $b_{ef\!f}$ away from the hole:} The Hawking pair (\ref{dseven}) is created in the vicinity of the hole, say in the region $r\lesssim 10\,r_h$, with the degrees of freedom in $b_{ef\!f}$ then moving from this region towards infinity. When these degrees of freedom reach the region $r>100\,r_h$, they must be described as excitations of standard string theory around gently curved space by the assumptions in \ref{(AA)} above (this is depicted in fig.~\ref{fig1d}). That is, the effective degrees of freedom become some set of degrees of freedom of the \emph{exact} theory. Let these exact degrees of freedom resulting from $b_{ef\!f}$ be denoted by $b$. 

This step in the argument is very important, since it connects the effective theory to the exact theory. If there were no such connection, then the effective theory would likely be an irrelevant figment of our imagination. Note that in step \ref{(AA)} we have assumed that the degrees of freedom in the far region $r>100\,r_h$ are independent of the degrees of freedom in the region of the hole $r<10\,r_h$. Thus the degrees of freedom making up $b$ will be independent of the degrees of freedom in $r<10\,r_h$. We use this fact to partially fix the ambiguity in the definition of the quantum $b_{ef\!f}$, which was noted in \ref{(CC)} above. We do this by choosing the definition of $b_{ef\!f}$ such that the \emph{exact} degrees of freedom giving rise to $b_{ef\!f}$ are independent of the \emph{exact} degrees of freedom making up $c_{ef\!f}$. This can always be done, since $b_{ef\!f}$ turns into the excitations of the exact theory that reach the far region, while $c_{ef\!f}$ remains in the hole. We will denote by $c$ the exact bit that $b$ is entangled with.
\begin{figure}[h]
\begin{center}
\includegraphics[scale=0.43]{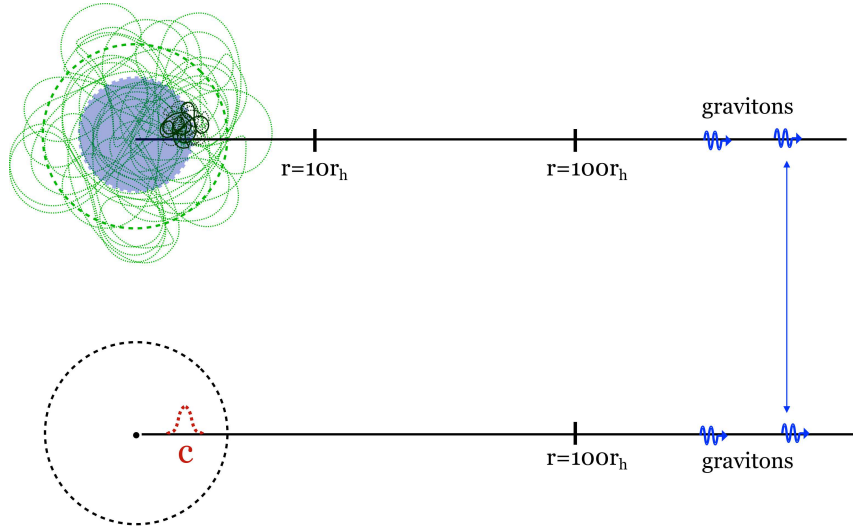}
\end{center}
\caption{ Movement of $b_{eff}$ away from the hole. When these degrees of freedom reach the region $r>100\,r_h$, they must be described as excitations of standard string theory around gently curved space by the assumptions in \ref{(AA)}. This is shown in the figure by there being no difference between the exact and effective theory for $r>100\,r_h$. }
\label{fig1d}
\end{figure}
\item \label{(EE)} {\bf Entanglements:} We label the successive emissions from the hole by steps $1, 2, 3, \dots$ and denote the quanta emitted at emission steps $1, 2, \dots, N$ as $\{ b_1, b_2, \dots, b_N\} \equiv \{ b\}$ (an `emitted' quantum here is one that has reached the region $r>100\,r_h$). Consider any subset $A$ of the degrees of freedom for the \emph{exact} theory. We write $S(A)$ for the entanglement entropy of the degrees of freedom in $A$ with the rest of the degrees of freedom of the exact theory. Then the entanglement entropy $S_N$ of the radiation with the remaining hole in the exact theory after $N$ emissions is
\be
S_N=S(\{b\}) \ .
\ee
We are now interested in the next step of the emission (the $(N+1)$th step). It is convenient to break this emission process into two steps: \ref{(aa)} the process in \ref{(CC)} where a pair is created in the effective theory in the region of the hole $r<10\,r_h$; \ref{(bb)} the process in \ref{(DD)} where the degrees of freedom in $b_{ef\!f}$ move to the far region $r>100\,r_h$. Let us consider the entanglement entropy in these steps.

\begin{enumerate}[start=1,
    labelindent=\parindent,
    leftmargin =1.12\parindent,
    label=(\alph*)]
    
\item \label{(aa)} Here the pair (\ref{dseven}) is created in the region of the hole $r<10\,r_h$ and it is assumed that the far region decouples from the region of the hole. Thus this process of pair creation must be given by a unitary transformation of the exact degrees of freedom in the region $r<10\,r_h$. The entanglement of the far region with the region of the hole cannot change in this process.\footnote{If two parts of a system are entangled, and we make a unitary action on one part, the entanglement between the two parts does not change.} Thus after this $(N+1)$th step of pair creation in the region of the hole the entanglement of the exact degrees of freedom in the far region with exact degrees of freedom in the region of the hole is still $S_N$. The entanglement of the newly created quanta $\{b_{N+1},c_{N+1}\}$ are such that the leading order part of the state (\ref{dseven}) of the effective quanta gives 
\be
S(b_{N+1, eff} + c_{N+1, eff})=0\ , \quad S(c_{N+1, eff})=\ln 2 \ .
\ee
The entanglement of the exact degrees of freedom corresponding to these excitations must then satisfy
\be
S(b_{N+1} + c_{N+1})<\epsilon_1 \ , \quad S(c_{N+1})>\ln 2-\epsilon_2 \ ,
\label{dten}
\ee
for some $\epsilon_1,\epsilon_2\ll1$. The relation (\ref{dten}) finally specifies the magnitude of the small corrections that we have mentioned in the above steps.\footnote{The steps relating the corrections to the state of the pair to $\epsilon_1$ and $\epsilon_2$ are discussed in \cite{Mathur:2009hf}.}

\item \label{(bb)} The degrees of freedom that give rise to the new radiation quantum $b_{N+1}$ move out to the far region. The value of the entanglement entropy at this emission step is then, by definition, given by
\be
S_{N+1}=S(\{ b\}+b_{N+1}) \ ,
\label{dnine}
\ee
since now $b_{N+1}$ has joined the earlier quanta $\{ b\}$ in the outer region $r>100\,r_h$. 
\end{enumerate}

\item \label{(FF)} {\bf The inequality:} We now recall the strong subadditivity relation
\be
S(A+B)+S(B+C)\ge S(A)+S(C) \ .
\label{del}
\ee
Here $A,B,C$ are three subspaces made from degrees of freedom that are independent of each other. We set
\be
A=\{ b\}\ , \quad B=b_{N+1}\ , \quad C=c_{N+1} \ ,
\ee
and from (\ref{del}) we get 
\be
S(\{ b\}+b_{N+1})+S(b_{N+1}+c_{N+1}) \ge S(\{ b\})+S(c_{N+1}) \ .
\label{dtw}
\ee
From (\ref{dten}) and \eqref{dnine} this can be written as
\be
S_{N+1}> S_N+\ln 2 -(\epsilon_1+\epsilon_2) \ .
\label{dsixt}
\ee
Thus for $\epsilon_1, \epsilon_2\ll 1$, the entanglement entropy keeps growing monotonically with the number of emission steps; it does not behave like the entanglement for a normal body which rises till the halfway point of evaporation and then falls back to zero. 
\end{enumerate}
In brief, as a result of the steps \ref{(AA)}-\ref{(FF)}, the effective small corrections theorem says the following. Suppose that the far region decouples from the region of the hole (as is the case for the burning away of a piece of coal), then if semiclassical dynamics (\ref{dfive}) emerges in any effective description, it will force the Page curve of the \emph{exact} theory to keep rising monotonically.

\section{Some definitions}\label{secdef}

In this section we summarize the meaning of some terms that will be used in the discussions of later sections.

\subsection{The classical black hole}

\begin{figure}[ht]
\centering
\begin{subfigure}{0.47\textwidth}
\centering
\includegraphics[width=0.5\linewidth]{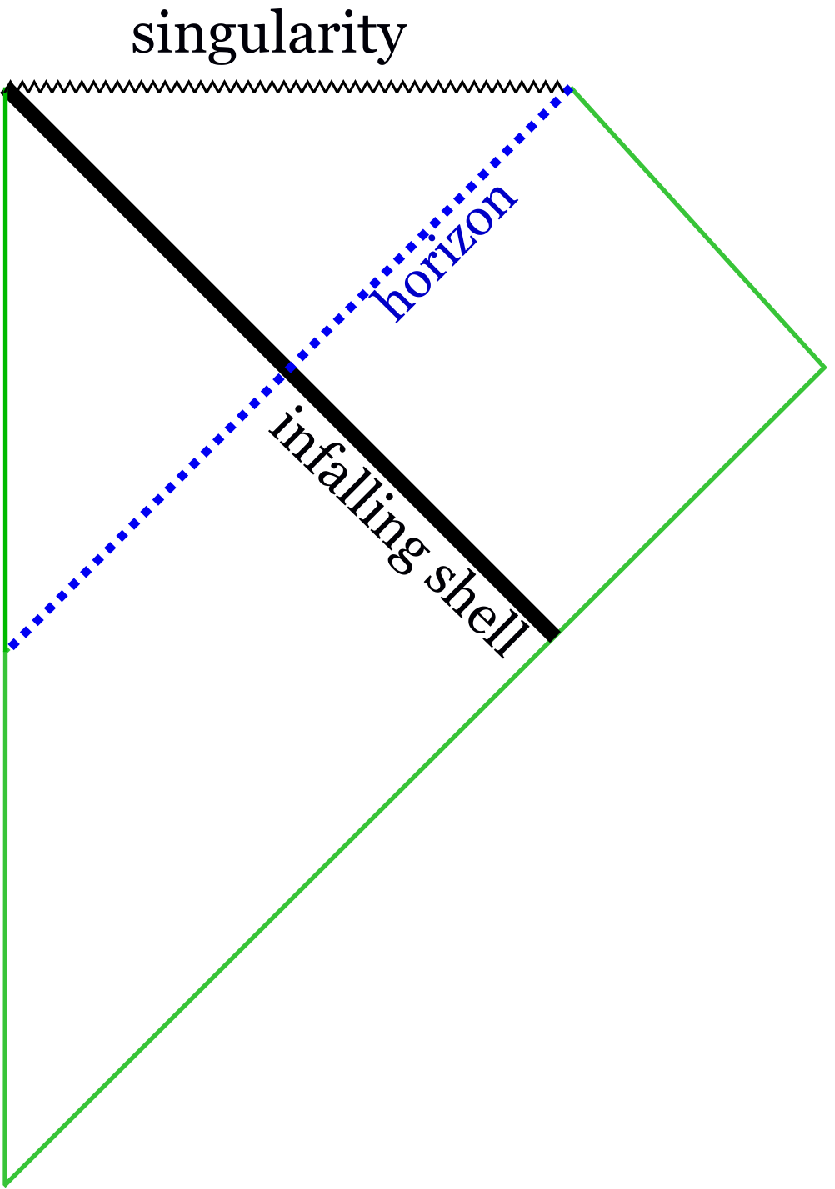}
\end{subfigure}
\begin{subfigure}{0.47\textwidth}
\centering
\includegraphics[width=1\linewidth]{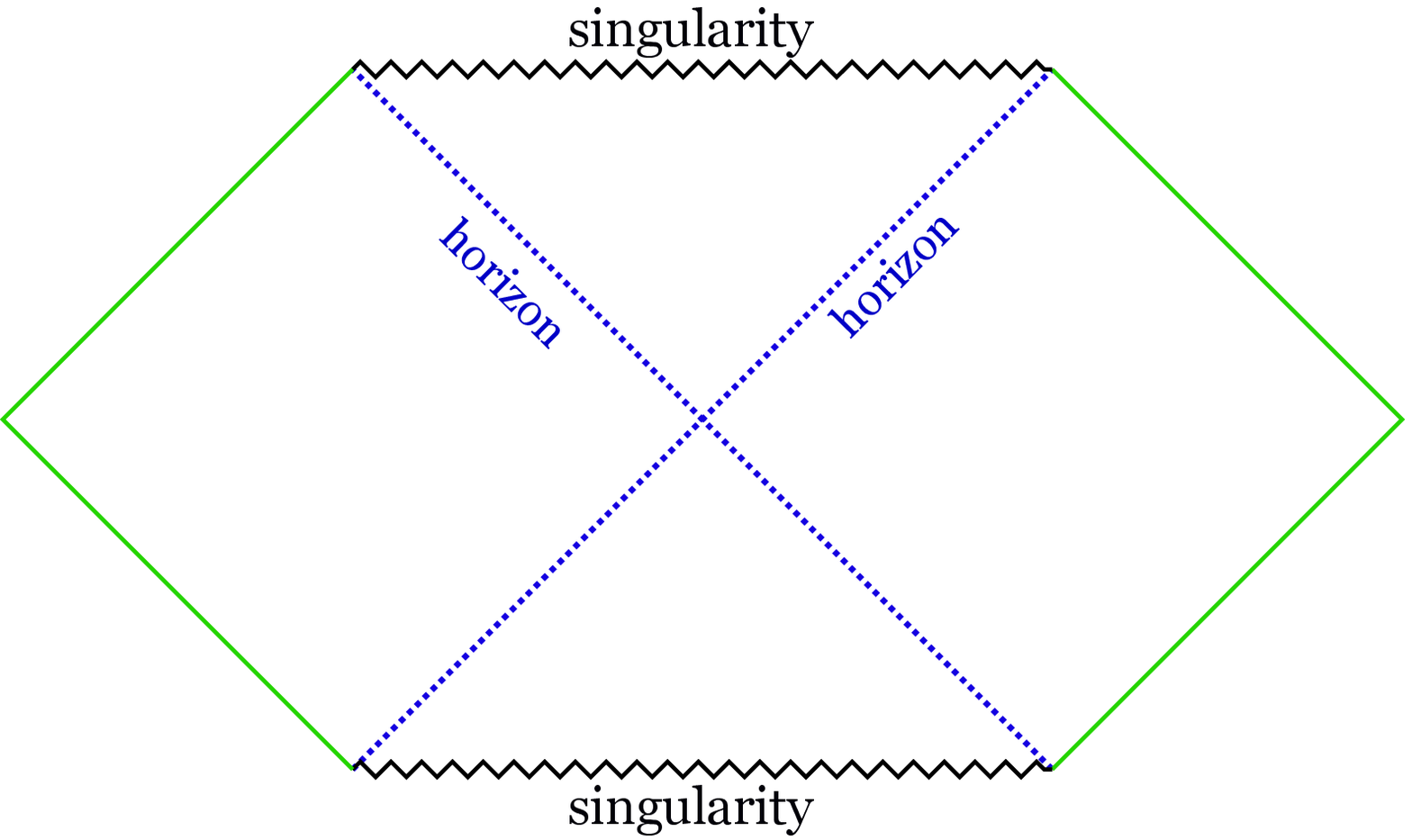}
\end{subfigure}
\caption{The Penrose diagrams for a classical black hole forming from an infalling null shell and for an eternal black hole. The blue dotted line represents the classical horizon. Each asymptotic region of the eternal black hole sees a future and past horizon.}\label{fig2}
\end{figure}

The classical Schwarzschild hole in 3+1 dimensions is given by the metric
\be
ds^2= -\Big(1-{r_h\over r}\Big)dt^2  + {dr^2\over 1-{r_h\over r}} + r^2 \big(d\theta^2+\sin^2\!\theta\, d\phi^2\big) \ ,
\label{dmetric}
\ee
with $r_h=2GM$. These Schwarzschild coordinates describe only the exterior of the horizon $r>r_h$. To see both the outside and inside of the horizon in a common coordinate patch, we can use the Eddington-Finkelstein coordinate 
\be \label{EddFink}
u=t+r^*=t+r+r_h\log\Big({r\over r_h}-1\Big) \ ,
\ee
in which the metric (\ref{dmetric}) becomes
\be
ds^2=-\Big(1-{r_h\over r}\Big)du^2+2 dudr+r^2 \big(d\theta^2+\sin^2\!\theta\, d\phi^2\big) \ .
\ee
We can start from flat space and form the black hole by sending in a shell of energy $M$ composed of radially infalling massless particles. The Penrose diagram for the corresponding classical hole is given in fig.~\ref{fig2}. If we just analytically continue the metric (\ref{dmetric}) as far as it can be continued, we find the eternal hole whose Penrose diagram is depicted in fig.~\ref{fig2}. This `eternal' hole has a singularity in the past (bottom) quadrant and a second asymptotic infinity in the left quadrant. Thus the eternal hole does not correspond to a physical situation that we can create in the lab.

\subsection{The semiclassical black hole}

\begin{figure}[H]
\begin{center}
\includegraphics[scale=1.2]{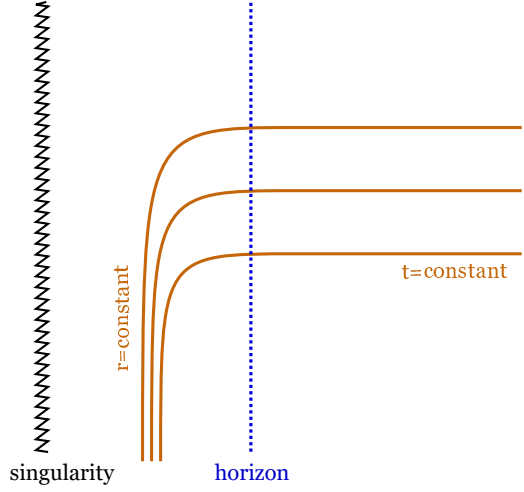}\hskip20pt
\end{center}
\caption{Good slices of a classical black hole. Asymptotically these are constant-time slices, while inside the horizon time and space switch roles, so they become constant-$r$ slices. These segments are then linked smoothly. The good slices do not pass close to the singularity.}
\label{fig3}
\end{figure}

We can study the process of black hole formation and evaporation using `good slices'; i.e. slices that are smooth and pass only through regions where the curvature is low (i.e. the Ricci scalar ${\mathcal R}\ll l_p^{-2}$). This is important because if we were forced to have slices that passed through a singularity, then we could not be sure of how quantum fields evolve past this singularity. The good slices in Eddington-Finkelstein coordinates are depicted in fig.~\ref{fig3}. 

Outside the horizon $r=r_h=2GM$ a spacelike slice can be taken as $t=\bar t$ for some constant $\bar t$. Inside the horizon space and time interchange roles and a spacelike slice can be taken as $r=\bar r$ for some constant $\bar r$. As a concrete example we may take $\bar r={r_h\over 2}=GM$, so that this part of the slice is neither near the horizon nor near the singularity. The inside and outside parts of this spacelike slice can be joined by a smooth `connector' segment. To move to a later slice we can advance the $t=\bar t$ part of the slice to $t=\bar t +\Delta \bar t$. We keep the shape of the connector part the same. We then join up with the $r=\bar r$ part of the slice by making this part \emph{longer}. This shows the stretching of hypersurfaces that leads to the creation of particle pairs in the region around the horizon (see appendix~\ref{appb} for a bit model description of this. Each time we advance the slice by $\Delta \bar t\sim r_h$, we create $\sim 1$ particle pairs whose entangled state can be schematically represented by (\ref{pair}).

The essential feature of a classical black hole is the existence of a horizon. It is this horizon that allows a constant $r$ segment $r=\bar r$ to be \emph{spacelike} rather than timelike. There are two aspects of this segment that are important. Firstly, the part of the slice given by $r=\bar r$ can be made arbitrarily long, while still staying within the black hole radius $r_h$. Secondly, excitations on this part can have either sign of the energy $E$, as measured from infinity. Given the above two aspects of the $r=\bar r$ segment of the slice, we can make states inside the black hole as follows. We place $n_{quanta}$ quanta of wavelength $\lambda \sim r_h$ along this segment, each having a spin that can be $\uparrow$ or $\downarrow$. Let the proper distance between quanta be $\sim r_h$. Furthermore, let alternating quanta have energies that are of opposite signs (i.e. $E, -E, E,-E,\dots$). By choosing different values for the spins, the number of states ($N_{states}$) on this segment is given by
\be
N_{states}=2^{n_{quanta}} \ .
\label{dtone}
\ee
By choosing the $r=\bar r$ part of the slice to be sufficiently long, we can place an arbitrarily large number of such quanta and thus get an arbitrarily large value for $N_{states}$. Thus the entropy
\be
S\equiv \log N_{states} \ ,
\ee
can be made arbitrarily large and in particular we can make
\be
S>S_{bek} \ ,
\ee
where $S_{bek}={A\over 4G}$, where $A$ is the area of the black hole horizon. This is called the `bags of gold' problem or the problem of `unbounded entropy': we can store an entropy in the hole which is arbitrarily larger than the Bekenstein entropy $S_{bek}$.

The bags of gold problem is closely related to the Hawking puzzle. The evaporation of the hole generates negative energy quanta on the $r=\bar r$ part of the slice, of the kind used in the above construction of states. If we keep feeding the black hole with quanta of wavelength $\lambda\sim r_h$, then these quanta end up on our $r=\bar r$ slice as the positive energy quanta used in the above construction. The entanglement entropy $S_{ent}$ of the black hole with its radiation can be made arbitrarily large and in particular we can have
\be
S_{ent}>S_{bek} \ .
\ee

\subsection{Some definitions}

\begin{figure}[hb]
\centering
\begin{subfigure}{0.5\textwidth}
\centering
\includegraphics[scale=0.45]{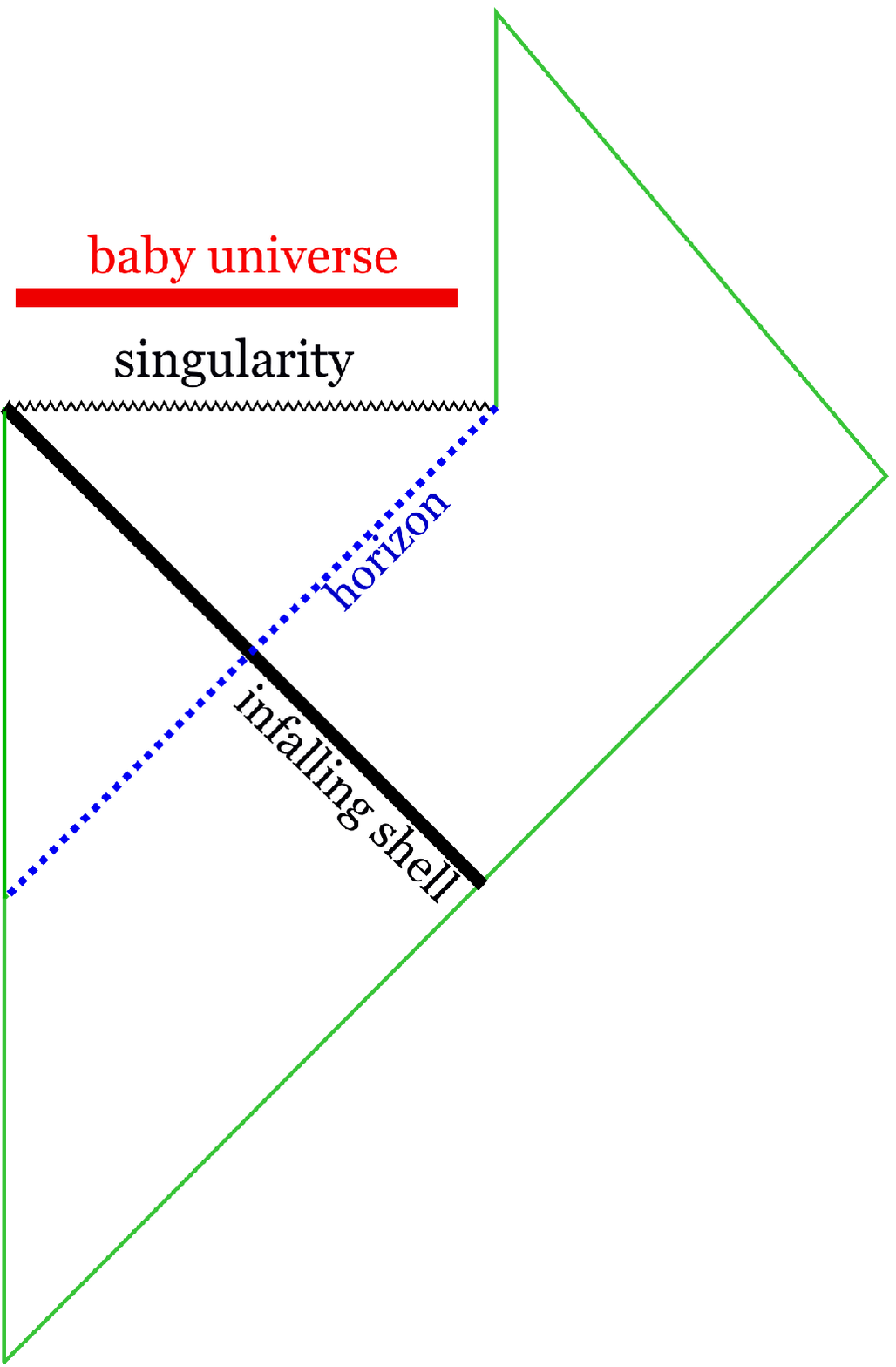}
\end{subfigure}%
\begin{subfigure}{0.5\textwidth}
\centering
\includegraphics[scale=0.55]{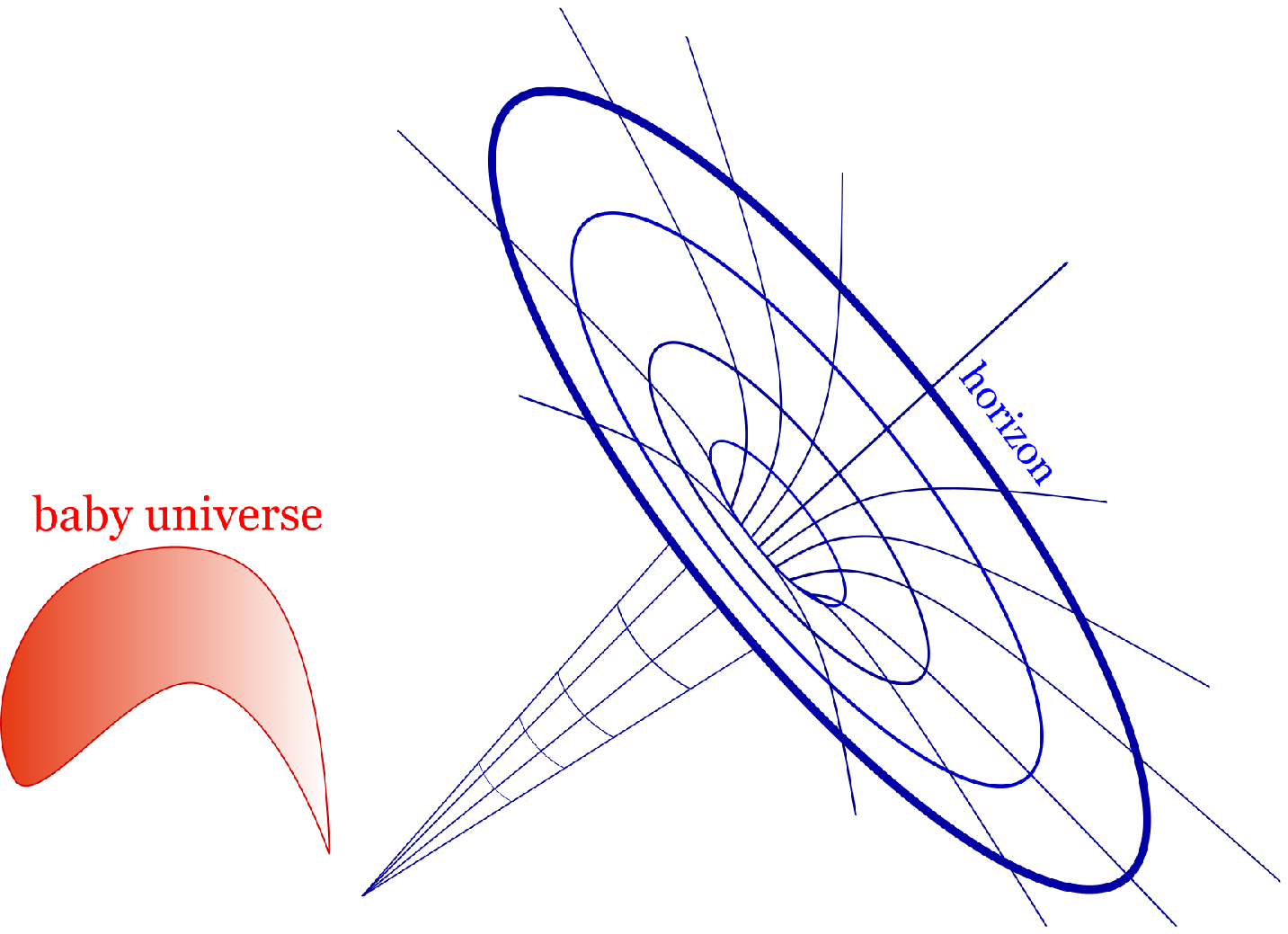}
\end{subfigure}
\caption{The disconnected baby universe. Before the endpoint of evaporation, the black hole interior contains the matter which made the hole, as well as the negative energy Hawking quanta from the evaporation process. One may imagine that this interior region pinches off into a `baby universe' that is disconnected from the rest of the spacetime.}
\label{fig4}
\end{figure}

If we include the backreaction of the negative energy quanta that fall inside  the hole, then the mass $M$  of the hole slowly decreases, and we reach the endpoint of evaporation as $M\r 0$. One possibility at this stage is that the hole evaporates away completely as far as the usual part of spacetime $r\ge 0$ is concerned; i.e. the location $r=0$ returns to being a normal part of spacetime with the vacuum state in its vicinity. But  before the endpoint of evaporation, the interior of the hole contained the matter which made the hole as well as the negative energy quanta which fell into the hole in the evaporation process. We can imagine that this interior region pinches off into a `baby universe' that is disconnected from the usual $r\ge 0$ part of spacetime. This situation is depicted in fig.~\ref{fig4}.
In some situations we will call some part of a spacelike slice an `island'. With the good slices we have chosen, this island will be, roughly speaking, the $r=\bar r$ segment of the spacelike slice. The exact location of the upper endpoint of this island will be determined by an optimization process, but this exact location will not be of importance for the physical argument. The relevant aspect of the island will be that it contains the negative energy members of the Hawking pairs (except perhaps for the final few, depending on where the exact upper endpoint of the island is).

\subsection{The Euclidean hole}

The analytic continuation $t\r -i\tau$ converts the Schwarzschild metric (\ref{dmetric}) to the metric of the Euclidean hole
\be
ds^2=  \Big(1-{r_h\over r}\Big)d\tau^2 + {dr^2\over 1-{r_h\over r}} + r^2 \big(d\theta^2+\sin^2\!\theta\, d\phi^2\big) \ .
\label{dttwo}
\ee
The radial coordinate ranges over $r_h\le r<\infty$ and the `Euclidean time' direction $\tau$ is taken to be compact, with $0\le \tau < 4\pi r_h$; this period corresponds to the inverse temperature of the hole
\be
T^{-1}\equiv \beta=4\pi r_h=8\pi GM \ .
\ee
The $r,\, \tau$ directions form a cigar whose tip lies at $r=r_h$; the metric is smooth at this tip with the chosen periodicity of $\tau$. Note that the Euclidean hole has no horizon or any region interior to a horizon. There is no place where quanta can have negative energy as seen from infinity, and there is no analogue of pair creation. Thus the Euclidean metric (\ref{dttwo}) does not exhibit the `bags of gold' problem or the problem of growing entanglement entropy ($S_{ent}$). The metric (\ref{dttwo}) can however be thought of as a saddle point for some path integral in the gravity theory.

\subsection{(1+1)-dimensional gravity}

Arguments in the wormhole paradigm have often been made with the help of two dimensional gravity theories.  
In two dimensions, the Einstein action is topological. For Euclidean signature, we have
\be
{1\over 2\pi}\left ( \int d^2 x \sqrt{g\,}\, R + 2 \int_\partial dy \sqrt{h\,}\, K\right ) =\chi \ ,
\label{dtsix}
\ee
where $R$ is the 2-d bulk curvature, $K$ is the extrinsic curvature at boundaries and $\chi$ is the Euler number. Varying such an action does not determine the metric. We can get an action which does have stationary points by including a scalar field $X$; we can also write this field as 
\be
X=e^{-2\Phi} \ ,
\ee
where $\Phi$ is the dilaton. The most general form of the action is then
\be
S=C\int d^2 x \sqrt{-g\,} \left ( \h R X -\h U(X) (\nabla X)^2+V(X) \right ) \ .
\label{dtfive}
\ee
In terms of $\Phi$ this has the form
\be
S=\h C \int d^2 x \sqrt{-g\,} \ e^{-2\Phi}\left ( R-\t U(\Phi) (\nabla \Phi)^2+2 \t V(\Phi)\right ) \ .
\ee
Two dimensional theories of this form can be obtained by dimensionally reducing a $D>2$ dimensional Einstein gravity theory on the angular sphere; in that case $X\sim r^{D-2}$. 

In the CGHS model~\cite{Callan:1992rs}, Hawking radiation was computed with a particular choice of the two-dimensional theory. They considered the action suggested by the worldsheet action in string theory, and coupled this to $N_f$ free scalar fields $f_i$, giving
\be
S_{CGHS}={1\over 2\pi} \int d^2 x \sqrt{-g\,} \left( e^{-2\Phi} \big( R+4 (\nabla \Phi)^2 +4\lambda^2\big) - \h\sum_i  (\nabla f_i)^2\right) \ .
\label{dcghs}
\ee
Hawking radiation was computed for the scalar fields $f_i$ and the Page curve was found to be monotonically rising, just as in Hawking's original computation. The important difference from Hawking's computation is the following. In the CGHS model, the gravity theory is fully quantum; we do not make the assumption that the matter fields travel on a fixed curved space. The reason that we anyway find a definite gravity background $\{g,\Phi\}$ is that in two dimensions the gravity theory has no propagating degrees of freedom, so the path integral over gravity variables can be gauge fixed to a particular configuration of $g$ and $\Phi$. Thus the CGHS model tells us that treating gravity quantum mechanically in 1+1 dimensions does not change Hawking's conclusion that the Page curve will keep monotonically rising as the evaporation proceeds. In the CGHS model we do not take into account the backreaction created by the negative energy quanta falling into the hole. 

In the RST model \cite{Russo:1992ht} the action (\ref{dcghs}) was slightly modified to a form where the backreaction could be easily computed. The hole can then be seen to evaporate away as the radiation proceeds. The Page curve again keeps rising till the endpoint of evaporation. The entanglement entropy of this radiation was computed as a function of time in \cite{Fiola:1994ir,Keski-Vakkuri:1993ybv}.

From these computations in (1+1)-dimensional theories, one can see that the monotonically rising nature of the Page curve should not depend on precisely which action of the type (\ref{dtfive}) we take. One theory which has been considered in recent computations is  
Jackiw-Teitelboim gravity (JT gravity). The action of JT gravity plus a matter CFT is
\be
{S_0\over 4\pi} \left ( \int d^2 x \sqrt{-g\,}\, R + 2\int_\partial  dy \sqrt{h}\,K + {1\over 4\pi}\int d^2 x \sqrt{-g\,}\, X (R+2) + {1\over 2\pi} \int_\partial dy \sqrt{h}\, X_b K \right ) +S_{CFT} \ .
\ee
Here the first two terms are the topological terms (\ref{dtsix}); they have been included since one may need to sum over different topologies in some Euclidean computations. The term $S_{CFT}$ denotes matter fields that we add to the gravity action; we take these matter fields to define a conformal field theory (CFT). A particular example of this CFT could be one given by a set of free scalar fields. 

Note that all these two dimensional theories are `incomplete' theories of gravity in the following sense. Black holes in these theories have a Bekenstein entropy $S_{bek}>0$ given in terms of the value of the field $X$ (or equivalently $\Phi$) at the horizon. However, the theories do not have the necessary degrees of freedom to manifest $\exp[S_{bek}]$ orthogonal quantum states to account for this entropy. String theory on the other hand is a `complete' theory, since we expect that  there are in fact $\exp[S_{bek}]$ states to account for the entropy. Consider, for example, the black hole in 4+1 noncompact dimensions studied in \cite{Strominger:1996sh}. The Bekenstein entropy is reproduced by counting the number of states of branes carrying the given mass and charges. Note however that these states differ from each other in the configurations of branes in the 5 \emph{compact} directions. If we dimensionally reduce on these compact directions, then we cannot manifest the states required to account for the entropy. Similarly, fuzzball microstates differ from each other in the way the compact directions fiber over the noncompact directions. If we consider the dimensionally reduced theory, then there will be no fuzzballs; there will be a unique state for the black hole as indicated by the no-hair theorems.

\section{The Page curve - I: What topology change can and cannot do}\label{secchange}

It has been argued that using a simple theory of gravity like JT gravity, one can deduce that the Page curve of a black hole must come down like the Page curve of a normal body. But how can this be, when we have already noted in the above section that the Page curve in simple (1+1)-dimensional theories of gravity keeps rising montonically? As we will see below, the important step in the recent computations will be that a new `prescription' will be added to the (1+1)-dimensional theory. This prescription will, in turn, be equivalent to requiring a certain nonlocality in the \emph{exact} theory. Our task is to clarify this nonlocality requirement.\\
\\
Our discussion of the Page curve will span three  sections, so we start by summarize the points that we will make in these three sections:

\begin{enumerate}[start=1,
    labelindent=\parindent,
    leftmargin =2\parindent,
    label = (\arabic*)]

\item \label{(1)} 
It has been argued that the recent Page curve computations are similar to the Gibbons-Hawking computation of entropy in the following sense. In the Gibbons-Hawking computation,  a semiclassical computation is able to reproduce the entropy, while the actual significance of this entropy as a count of states will only emerge when we know the exact quantum gravitational structure of the hole. But we will argue that the recent semiclassical Page curve computations are \emph{not} similar to the Gibbons-Hawking computation in this way. In the Gibbons-Hawking computation we start with a path integral of the exact quantum gravity theory that should yield the entropy; this path integral is then argued to have a semiclassical saddle point which we use. But in the Page curve computations, we start with a quantity that is \emph{not} the entanglement entropy that we wanted to compute. Instead, we modify the path integral for the R\'{e}nyi entropies by a `prescription'. It is this prescription that contains the nonlocal effects that the wormhole paradigm must invoke in order to avoid a monotonically rising Page curve.

\item \label{(2)}
It has been argued that the prescription arises automatically when we take into account the fact that in gravity one can have topology change. We will argue that such is \emph{not} the case. We will see explicitly how topology change affects the structure of the Hilbert space and the definition of the inner product. We will then note that these effects of topology change do \emph{not} generate a link between different replica copies, either in the Euclidean setting or in the Lorentzian setting. 

\item \label{(3)}
In the wormhole paradigm, one seeks to nevertheless introduce such links between copies, arguing that they are a feature of the semiclassical Gibbons-Hawking type of path integral that emerges as an approximation of the exact theory. But here we must remember that the exact variables are related to the effective semiclassical variables through the relation $g_{ef\!f}=F[g_{exact}]$ (eq.(\ref{dthreeqq})), which then forces the behavior of all other effective quantities through (\ref{dfourqq}) and (\ref{dfivepreqq}). Thus we cannot postulate that the effective semiclassical theory will have links between copies if there is no corresponding dynamics in the \emph{exact} theory that corresponds to the prescription of introducing these links. We will not be able to identify any clear postulate in the exact theory which can give the replica wormhole prescription in the effective theory. In a later section we will look at some models of nonlocality that have been proposed in the exact theory. We will see that for models where the nonlocalities stay within the black hole interiors, there is a loss of unitarity, while with models that have nonlocalities involving also the region far from the hole, there is a violation of normal low-energy physics far from the hole.
\end{enumerate}
In the remainder of this section we will make a first pass at the role of different topologies in Lorentzian and in Euclidean signature.

\subsection{Topology change in (1+1)-dimensional gravity: the Lorentzian theory} \label{sectopology}

It is sometimes said that the new `prescription' in the (1+1)-dimensional theory just takes into account the fact that we must allow topology change in a gravity theory. We will see that such is not the case. It is true that in the CGHS or RST computations the topology of the (1+1)-dimensional spacetime was taken to be the trivial one, similar to the topology that Hawking assumed in his (3+1)-dimensional computation. But we will also see that allowing topology change in the black hole region will \emph{not} change Hawking's conclusion that the Page curve keeps monotonically rising. The reason for this is that the effective small corrections theorem does not care about which topologies contribute to the  dynamics of the black hole interior. 

One might argue that since we do not really know how quantum gravity behaves, it is possible that there are  new rules for amplitudes in quantum gravity, which need not hold in non-gravitational quantum theories. But actually we do know a lot about quantizing gravity, especially in the (1+1)-dimensional case. The string worldsheet theory is a (1+1)-dimensional quantum gravity theory, since we need to sum over both the quantum fields $X^\mu(\tau,\sigma)$ on the worldsheet as well as the metric $g_{ab}(\tau,\sigma)$ on the worldsheet. The worldsheet theories with central charge $c<1$ yield quantum gravity theories that have been solved in multiple ways: through dynamical triangulations \cite{Agishtein:1990fs,Jain:1992bs}, in light cone gauge (KPZ) \cite{Knizhnik:1988ak} and in conformal gauge (DDK) \cite{David:1988hj,Distler:1988jt}. All these ways of studying 2-d quantum gravity include the possibility of topology change. Let us therefore review from first principles what topology change means in 1+1 dimensions and what constraints we have from unitarity on such a theory. In this way we will understand what aspects of the dynamics we can change and what we cannot.

Let us start with a very simple model; this model will have all the features that we wish to highlight. In 1+1 dimensions, the spatial sections are 1-dimensional, thus we first consider a single line segment. Let this segment have $N$ lattice points along it and on each lattice point we have a single bit whose value can be $0$ or $1$. Thus there are $2^N$ states on this segment;  we label these states as 
\be
|\psi_i^{(N)}\rangle \ , \quad i=1, \dots, 2^N \ .
\ee
We can assume that the spacing between lattice points is the Planck length.

Now suppose this line segment described a (1+1)-dimensional cosmology at some fixed time. The cosmology can expand, so that at some later time we have a longer line segment. Since we have kept the spacing between lattice sites as the Planck length, we will have $N'>N$ lattice sites on this new segment and thus a larger number of states allowed for the quantum bits on the segment. At first this looks confusing, since the dimension of the Hilbert space should not change during time evolution. But the answer is simple. Since this is a theory of quantum gravity, the spacelike slice is described not only by the quantum fields on it, but also by the metric on it. Our Hilbert space consists of the union of the Hilbert spaces for segments of all different lengths $N\ge 0$, with $2^N$ states $|\psi_i^{(N)}\rangle$ on each such segment. The evolution can then take us from one quantum state on a segment of one length $N$ to some quantum state of a segment of a different length $N'$. The transition Hamiltonian $H$ must satisfy 
\be
\langle \psi_j^{(N')}|H|\psi_i^{(N)}\rangle=\left (\langle \psi_i^{N}|H|\psi_j^{(N')}\rangle \right )^* \ .
\label{dthfive}
\ee
That is, the amplitude for any given transition must be the complex conjugate of the amplitude of the reverse transition. So, even though the metric on the segment can change, we still have a well defined notion of unitarity.

This setup is sufficient to understand the quantum gravitational setting of (1+1)-dimensional models like the CGHS model or the RST model, where spacetime could stretch but not change topology. Now suppose we do wish to allow topology change. What should we do? In our (1+1)-dimensional situation the answer is simple: all we can do is to allow the possibility that a line segment can break into two segments or two such segments can join to form one. A basis of our Hilbert space now consists of the following configurations. Each configuration has some number $k\ge 0$ of line segments with the $i$th segment having length $N_i$ and $2^{N_i}$ possible states of the quantum bits on the segment. Note that if two segments in a configuration have the same length and the same configuration of quantum bits, then they are indistinguishable, i.e. they are like two bosons of the same species. This aspect will be important when we talk about baby universes later. The Hamiltonian then gives transition amplitudes between these basis states, which can be nonzero between states having a different number of line segments. The amplitudes must still satisfy the analogue of (\ref{dthfive})
\be
\langle \{ \psi_j^{(N')}\} |H|\{\psi_i^{(N)}\}\rangle=\left (\langle \{\psi_i^{(N)}\}|H|\{\psi_j^{(N')}\}\rangle \right )^* \ ,
\label{dthfivep}
\ee
where we have used the symbol $\{\psi_i^{(N)}\}$ to denote a collection of line segments. We have not specified what the transition amplitudes are and choosing different values for these amplitudes will define different (1+1)-dimensional quantum gravity theories. With our choice of matter field (a single bit per lattice site) there are no other freedoms in defining the overall \emph{structure} of the theory. In particular, we must satisfy (\ref{dthfivep}) if we wish to preserve unitarity. This is important to note since we will find that in some of the recent models of black hole evaporation using baby universes, the evolution chosen turns out to be \emph{not} unitary.

We have gone through these simple points in detail since there has been much confusion about what postulates we can or cannot add to a quantum gravity theory like JT gravity. The above discussion was in the Lorentzian theory, where the black hole problem is actually defined. We will now turn to the Euclidean theory, which has also been used to make indirect computations of entanglement entropy. Again, our goal will be to understand the significance of different postulates that we may or may not add to a theory like JT gravity.

\subsection{The small corrections theorem and topology change}\label{topsmall}

Having seen the nature of topology change in gravity, we remark here on the fact that the possibility of topology change in the black hole interior has no effect on the derivation of the effective small corrections theorem. To see this, consider the (1+1)-dimensional case in which we studied topology change in section~\ref{sectopology} above. Topology change could lead to a 1-dimensional spatial segment to break into two segments, or two such segments could join to form one. There are two possibilities to consider:
 
\begin{enumerate}[start=1,
    labelindent=\parindent,
    leftmargin =2\parindent,
    label=(\alph*)]
 
\item \label{(aaa)}
Suppose this breaking of segments happens at the horizon and in such a way that it invalidates the effective semiclassical dynamics at the horizon. Furthermore, this breaking happens often enough that this effective semiclassical description is invalidated for a significant fraction of emitted quanta (i.e. for a fraction that is order unity). In that case the central assumption of the wormhole paradigm is invalidated, since we do not have semiclassical dynamics at the horizon in some effective variables. So we will not consider this possibility further.
 
\item \label{(bbb)}
Suppose the line segment breaks in two very occasionally, so that such a break typically happens when the segment becomes very long. As a example, we may say that such a break happens when the segment holds $\sim S_{bek}$ negative energy quanta $\{c\}$. But we see that such a process of breaking segments has no effect on the effective small corrections theorem, since its derivation does not need to know what happens to the $c$ quanta that fall into the hole, only that the evolution of the black hole region is unitary. 
\end{enumerate}

\subsection{Using Euclidean path integrals for (1+1)-dimensional gravity}\label{eucl}

The physical gravity theory is the Lorentzian one, however, its Euclidean continuation can be a useful tool to compute certain quantities; for instance, the Gibbons-Hawking computation of $S_{bek}$. We will now consider another use of the Euclidean path integral: as a means of generating entangled states between two disconnected regions. We will then make the observation that this kind of Euclidean path integral does \emph{not} imply that there is an interaction between the two disconnected spaces. If we do argue for such an interaction, then this interaction would be a \emph{new} postulate; i.e. not something that follows from a Euclidean path integral in the gravity theory.  This issue will be relevant in our understanding of how JT gravity has been used in the recent computations of the Page curve.

\begin{figure}[htbp]
\begin{center}
\includegraphics[scale=0.5]{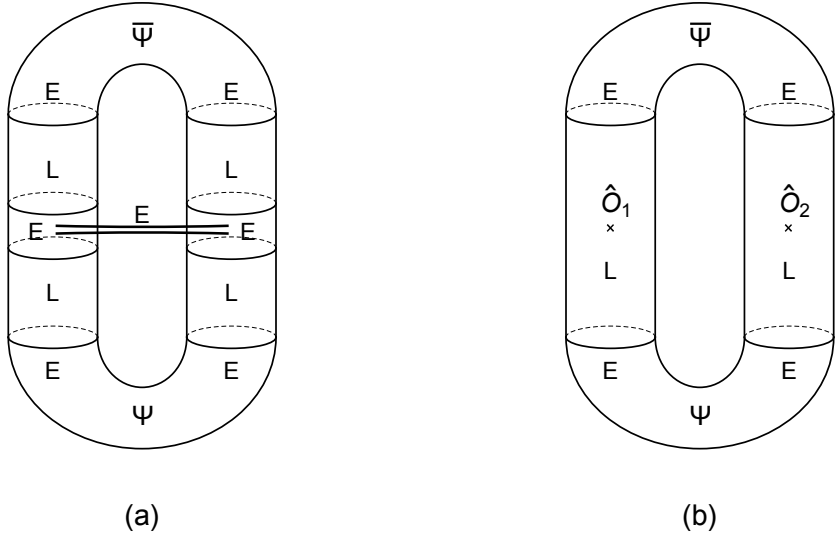}\hskip20pt
\end{center}
\caption{The bottom and top of figure (a) depicts the generation of the states $|\Psi\rangle$ and $\langle\Psi|$ respectively on the union of two circles by Euclidean evolution (E). Under Lorentzian time evolution (L), these states of their respective circles evolve independently; there is no interaction. An interaction, shown by a horizontal Euclidean evolution, can equally be written as operator insertions; one per circle, as in (b).}
\label{fqes8}
\end{figure}

In the previous subsection we took our line segments to be open. We can equally well consider closed loops; let us do that here since then the corresponding Euclidean manifolds will be simpler in that they will not have a boundary. Consider the space made of two disconnected circles; let the spatial coordinate on these circles be $\sigma_1$ and $\sigma_2$ respectively. On each circle let there be a free scalar field $X$. There is no interaction Hamiltonian connecting these two circles, but the overall state $|\Psi\rangle$ of the system can have entanglement between the two circles, as in the case of the generically nonfactorizable state
\be
|\Psi\rangle =\sum_n c_n |\psi_{1,n}\rangle|\psi_{2,n}\rangle \ ,
\label{zten}
\ee
where $|\psi_{1,n}\rangle$ and $|\psi_{2,n}\rangle$ are states on the circles $1$ and $2$ respectively. Despite there being no interaction between the two circles, we may introduce an interaction as an artificial technique to generate the entangled state. Thus consider the scalar field on the Euclidean cylinder of length $\tau$; the two ends of this cylinder are the two circles $1$ and $2$ that we had started with. The path integral over $X$ on this cylinder generates the entangled state on the two circles
\be
|\Psi\rangle_{thermal}=\sum_n e^{-\tau E_n} |E_{1,n}\rangle |E_{2,n}\rangle \ ,
\label{zel}
\ee
where $|E_{1,n}\rangle$ and $|E_{2,n}\rangle$ are states of energy $E_n$ on circles $1$ and $2$. While this generates a particular entangled state, more general entangled states of the form (\ref{zten}) can be made by including higher genus surfaces in place of the cylinder and/or  adding operator insertions $\hat O$ on this Euclidean cylinder; such possibilities are depicted in fig.~\ref{fqes8}(b).

The important point to note is that the Euclidean manifold generating the entangled state in the above way is an artificial construction whose sole purpose is to obtain the entangled state; this manifold joining the two circles does not imply that there is an interaction Hamiltonian between the two circles. Nevertheless, we will now modify the dynamics by adding a `prescription' using the above kinds of Euclidean surfaces linking the two circles to generate an interaction between the two circles. 

Suppose we construct the state (\ref{zel}) using a path integral on the Euclidean cylinder as described above. We can now evolve the state on each circle in Lorentzian time, using the Hamiltonian of the free fields on the respective circles. There is of course no interaction between the two circles in this evolution.

Now suppose we want to introduce an interaction between the circles. We wish to introduce a cylinder that stretches from one circle to the other; one may call this a `wormhole'. We cannot do this while staying in Lorentzian signature, since it is not possible to put a continuous light cone structure on a geometry of the kind in fig.~\ref{fqes8}(a) where the wormhole is a horizontal cylinder connecting the two circles $1,2$. Thus we assume that the prescription for the new interaction is as follows: (i) the Lorentzian evolution changes to Euclidean on each circle, (ii) the Euclidean wormhole joins these two Euclidean sections as in fig.~\ref{fqes8}(a) and, (iii) we return to Lorentzian evolution on the two circles. If we wish to compute a complete amplitude, we can take the inner product with the state (\ref{zel}) again; the geometry giving this full amplitude is depicted in fig.~\ref{fqes8}(a).

We have now added a prescription that gives an interaction between the field theories on the two circles $1,2$. Let us see what the nature of this interaction is. In the physical problem of the black hole, one circle (say circle $1$) will correspond to the space inside a black hole, while the other circle (circle $2$) will describe the degrees of freedom far from the hole. Since these two regions are far from each other, we should think of the  wormhole connecting them as `long'. 

\begin{figure}[htbp]
\begin{center}
\includegraphics[scale=0.6]{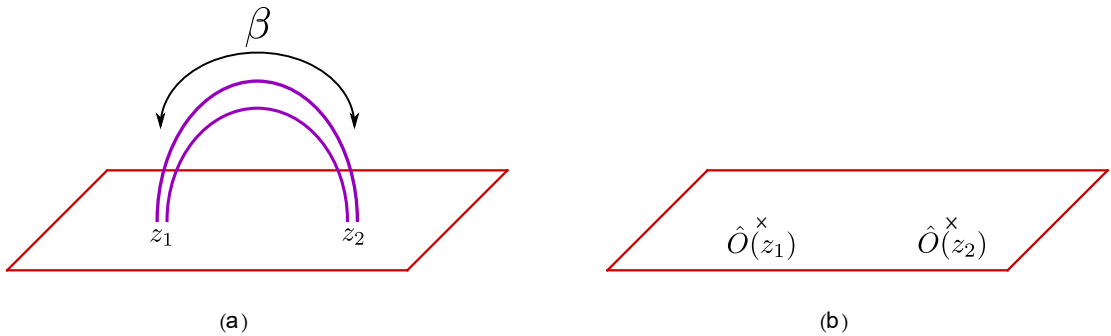}\hskip20pt
\end{center}
\caption{Two equivalent pictures where a wormhole linking two points $z_1,\,z_2$ on the plane can be thought of as operators inserted at these two points in the computation of, for instance, correlation functions.}
\label{fqes7}
\end{figure}

The wormhole interaction can instead be written as an operator that acts on two Hilbert spaces: one on the worldsheet of the theory on circle $1$ and one that acts of the worldsheet of the theory on circle $2$, as in fig.~\ref{fqes8}. We take local coordinate patches $z_1, z_2$ on the two worldsheets and let $\hat O_{h_i}$ be a basis of operators in each patch. Then we can write the effect of the wormhole interaction as an effective operator  
\be
\hat W=|0\rangle_1~{}_2\langle 0|+\sum_i e^{-\beta' h_i}|\hat O_{h_i}\rangle_1~{}_2\langle\hat O_{h_i}|~+~{\rm Hermitian ~ conjugate} \ ,
\ee
where $\beta'$ governs the length of the wormhole and we have separated the identity term and the contributions of higher dimension operators (see fig.~\ref{fqes7}). The identity term gives no interaction between the two circles. In the limit of a long wormhole, only operators $\hat O_{h_i}$ with low dimensions $h_i$ contribute significantly. An example of such a low dimension operator is $\hat O=\p X$, for which the effect of the wormhole becomes
\be
\hat W \r e^{-\beta'}\p X(z_1)\p X(z_2) \ .
\label{ztw}
\ee
Let us look at the effect of this interaction on any of the components making up the state (\ref{zel}). The state $|E_{1,n}\rangle$ contains excitations of the form
\be
\hat a^\dagger_{1, k_1}\!\cdots \hat a^\dagger_{1, k_p}|0\rangle_1 \ ,
\label{zthir}
\ee
where the $\hat a^\dagger_{1,k_i}$ are creation operators on circle $1$ and $|0\rangle_1$ is the vacuum state on this circle. We have a similar structure for the state $|E_{2,n}\rangle$  on circle $2$. In the operator (\ref{ztw}) we can expand each $\p X$ in creation and annihilation operators on the respective circles and by doing so we see that its action on \eqref{zel} yields terms of the following kind. One of the oscillator excitations $\hat a ^\dagger_{1,k}$ in (\ref{zthir}) is annihilated by $\p X(z_1)$, and an oscillator excitation $\hat a^\dagger_{2,k'}$ is created on circle $2$. Thus we can say that a particle vanishes from circle $1$ and another particle appears on circle $2$. This is the nonlocal transport of quanta that results from the prescription that we have added to the theory of free scalar fields on circles $1$ and $2$.

\subsection{Summary}

Let us summarize what we have seen in this section. Firstly, considering the Lorentzian theory. Whilst it is true that there can be topology change in gravity, if we take a (1+1)-dimensional theory for instance then the role of this topology change is well understood: all that can happen is that the spacelike slice -- which is a 1-dimensional manifold -- can split into two segments, or two such segments can join to form one. The rules for this splitting and joining should ensure that the the evolution is unitary. We understand these rules in many formalisms where (1+1)-dimensional gravity has been studied and do not have the freedom to add arbitrary rules to the quantum gravity theory on the grounds that we do not know what role topology change should play.

Now consider Euclidean computations. Here we have to be careful about a new issue: there is a `technical tool' that we can use to generate an entangled state between two noninteracting regions. This tool is a path integral over a cylinder that connects two noninteracting circles. We must be careful to not confuse this technical tool with a real interaction between the two noninteracting circles. If we nevertheless postulate such Euclidean cylinders between two circles imply actual interactions in the theory, then roughly speaking such interactions are of the form where we take a quantum from one circle and place it on the other circle (eq.(\ref{zthir})); one could call this a wormhole-type interaction between otherwise noninteracting regions.

\section{The Page curve - II: The prescription of replacing R\'{e}nyi entropies by new quantities}

In section~\ref{secentropyreview} we review the definition of entanglement entropy and then how R\'{e}nyi entropies are given by appropriate traces when the entanglement is between one part of a spacelike slice and its complement in section~\ref{secrenyi}. In section~\ref{secprescription} we see how a `prescription' is used in the wormhole paradigm to replace the R\'{e}nyi entropy by a new quantity.\footnote{For other critiques on path integral justifications of the island formula, see~\cite{Karlsson:2020uga,Karlsson:2021vlh}.} In section~\ref{sectop} we recall our discussion of section~\ref{secchange} about what topology change can do, and we note that the above prescription does \emph{not} follow from considerations of topology change.

\subsection{Entanglement entropies: review of notation}\label{secentropyreview}

Consider a quantum system in a pure state $|\Psi\rangle$.  Suppose there is some way to separate the degrees of freedom of this system into two sets, which we call subsystem $A$ and subsystem $B$. The Hilbert space ${\mathcal H}$ is assumed then to decompose as
\be
{\mathcal H}={\mathcal H}_A \otimes {\mathcal H}_B \ .
\ee
If $|\psi_i\rangle$ and $|\chi_j\rangle$ are orthonormal bases of states on subsystems $A$ and $B$ respectively, then we can write the full pure state on $\mathcal{H}$ as
\be
|\Psi\rangle = \sum_{i,j} C_{ij} |\psi_i\rangle|\chi_j\rangle\ , \quad \sum_{i,j} |C_{ij}|^2=1 \ .
\label{dfone}
\ee
The inner product of the full system $A\cup B$ factorizes into inner products on the subsystems. For instance, using the product states $|\Psi_1\rangle=|\psi_1\rangle|\chi_1\rangle$ and $|\Psi_2\rangle=|\psi_2\rangle|\chi_2\rangle$ we get
\be
\langle \Psi_1|\Psi_2\rangle= \langle \psi_1|\psi_2\rangle \langle \chi_1|\chi_2\rangle \ .
\ee
For more general $|\Psi_1\rangle$ and $|\Psi_2\rangle$ the inner product is obtained from the above relation using linearity.

Consider the state $|\Psi\rangle$ in (\ref{dfone}) and suppose we wish to trace out subsystem $B$ to get a density matrix $\rho_A$ describing system $A$ (a reduced density matrix). We first write the bra corresponding to the ket $|\Psi\rangle$
\be
\langle \Psi|=\sum_{i',j'} C^*_{i',j'} \langle \psi_{i'}|\langle \chi_{j'}| \ ,
\ee
and then obtain the density matrix of the full system as
\be
 |\Psi\rangle \langle \Psi|= \Bigg(\sum_{i,j} C_{ij} |\psi_i\rangle|\chi_j\rangle\Bigg) \Bigg( \sum_{i',j'} C^*_{i',j'} \langle \psi_{i'}|\langle \chi_{j'}|\Bigg) \ .
\ee
Finally we trace out subsystem $B$, getting the reduced density matrix for subsystem $A$ to be
\be
\rho_A= \Bigg( \sum_{i,j} C_{ij} |\psi_i\rangle\Bigg) \Bigg( \sum_{i',j'} C^*_{i',j'} \langle \psi_{i'}|\Bigg) \delta_{jj'} ~=~\sum_{i,i'}\Bigg(\sum_j C_{ij}C^*_{i'j}\Bigg) |\psi_i\rangle\langle\psi_{i'}| \ .
\label{dftwo}
\ee
Note that the partial trace in (\ref{dftwo}) is done using the inner product on the Hilbert space ${\mathcal H_B}$
\be
\langle \chi_{j'} |\chi_j\rangle=\delta_{j'j} \ .
\label{dfthree}
\ee
If we use some other matrix in place of the identity matrix in (\ref{dfthree}) to perform the partial trace, then we are \emph{not} computing the desired reduced density matrix but some other quantity. It is important to note this fact, since we will see that in the wormhole paradigm the prescriptions that are added are equivalent to replacing (\ref{dfthree}) by a different matrix.

The entanglement entropy $S_{ent}(A)$ (also called the von Neumann entropy or the fine grained entropy of subsystem $A$) is given by
\be
S_{ent}(A)=-\Tr\big[ \rho_A \log \rho_A\big] = -\sum \lambda_i \log \lambda_i \ ,
\ee
where $\lambda_i$ are the eigenvalues of $\rho_A$. The R\'{e}nyi entropies $S_n(A)$ are defined by
\be
S_n(A)=-{1\over n-1} \log \big[\Tr\rho_{\!A}^{\,n}\big]= -{1\over n-1} \log \Big[\sum_i \lambda_i^n\Big] \ .
\label{dsfive}
\ee
This family of quantities provides a useful way of obtaining the (generally difficult to calculate) $S_{ent}$ via the analytic continuation to $n\in \mathbb{R}$, followed by taking the limit $n\r 1^+$. If this can be well defined, then we get
\be
\lim_{n\r 1^+} S_n(A) = S_{ent}(A) \ .
\ee
To get a rough sense of what these measures of entanglement describe, consider the maximally entangled state between subsystems $A$ and $B$
\be
|\Psi\rangle={1\over \sqrt{N}} \sum_{i=1}^N |\psi_i\rangle |\chi_i\rangle \ ,
\label{dffour}
\ee
from which the eigenvalues of the reduced density matrix $\rho_A$ are $\lambda_i={1\over {N}}$ and we get
\be
S_{ent}(A)=-\sum_{i=1}^N {1\over N} \log {1\over N} = \log N \ .
\ee
Therefore, $S_{ent}(A)$ measures the number of terms in the sum in (\ref{dffour}). For the R\'{e}nyi entropies, we have 
\be
\Tr[\rho_{\!A}^{\,n}]=\sum_{i=1}^N {1\over N^n}={1\over N^{n-1}} \ ,
\ee
and so
\be
S_n(A)=-{1\over n-1} \log\Big[{1\over N^{n-1}}\Big] = \log N \ .
\label{dffive}
\ee
Hence, in this maximally entangled case the R\'{e}nyi entropies are the same as $S_{ent}(A)$. The R\'{e}nyi entropies are less useful as a description of entanglement when the $\lambda_i$ are not all comparable to each other. If one eigenvalue $\lambda_1$ is large while the other $N-1$ are all equal, $S_{ent}$ reflects the largeness of $N$ while the $S_n$ saturate to a value set by $\lambda_1$ and do not reflect the large entanglement encoded by the other $\lambda_{i\neq1}$.

\subsection{Computing the R\'{e}nyi entropies}\label{secrenyi}

We now set up the computation of the second R\'{e}nyi entropy $S_2(A)$ for a 1-dimensional system and then see how this computation is modified by a prescription in the wormhole paradigm. 
\begin{figure}[h]
\begin{center}
\includegraphics[scale=0.6]{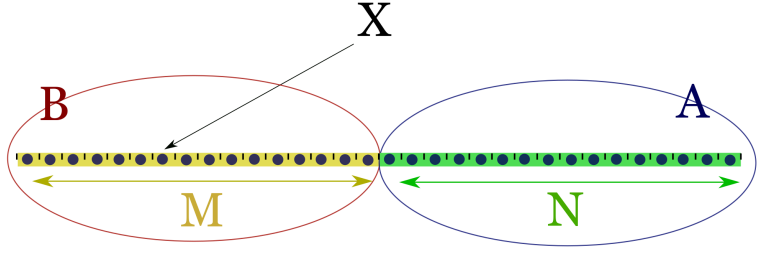}\hskip20pt
\end{center}
\caption{A 1-d discrete system where the first $M$ segments make up the subset $B$ and the last $N$ segments make up subset $A$. At the center of each segment shown is a scalar field degree of freedom $X$; this is the matter field on our 1-dimensional spacelike slice.}
\label{fig6}
\end{figure}
In fig.~\ref{fig6} we depict a 1-d system consisting of $M+N$ line segments. The first $M$ segments are labeled by an index $j=1, \dots, M$ and the last $N$ segments are labeled by $i=1, \dots, N$. At the center of each segment we place a scalar field degree of freedom $X$; this is the matter field on our 1-dimensional spacelike slice. When we will discuss the gravitational theory later, we will think of the first $M$ segments as the gravitational region containing the black hole, while the last $N$ segments will describe the spacetime away from the hole, including the region near infinity. 

Let us consider the last $N$ segments as defining a subsystem $A$ and the first $M$ segments as describing a subsystem $B$. Our goal is to trace out system $B$ to get a density matrix $\rho_A$ for system $A$. We start with a pure state $|\Psi\rangle$ for the full system $A\cup B$. Similarly to (\ref{dfone}), we write this state as
\be
|\Psi\rangle = \sum_{i_1,j_1} C_{i_1j_1} |\psi_{i_1}\rangle|\chi_{j_1}\rangle \ , \quad\sum_{i_1,j_1} |C_{i_1j_1}|^2=1 \ ,
\label{dfoneq}
\ee
where $|\psi_{i_1}\rangle$ and $|\chi_{j_1}\rangle$ are orthonormal bases for the states of subsystems $A$ and $B$ respectively. We now take a second set of $M+N$ line segments, on which we place the state dual to $|\Psi\rangle$
\be
\langle\Psi| = \sum_{i'_1,j'_1} C^*_{i'_1j'_1} \langle\psi_{i'_1}|\langle\chi_{j'_1}| \ .
\label{dfoneqq}
\ee
To trace over the subset $B$ we take the outer product $|\Psi\rangle \langle \Psi|$ using \eqref{dfoneq} and \eqref{dfoneqq}, and act with the delta function $\delta_{j_1, j'_1}$.
\begin{figure}[hb]
\begin{center}
\includegraphics[scale=0.3]{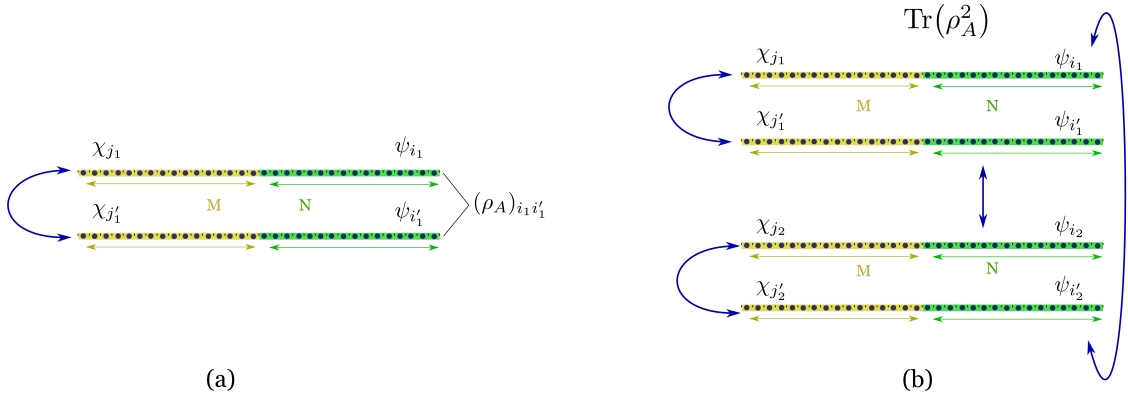}\hskip20pt
\end{center}
\caption{A schematic representation of a method for computing the quantity $\Tr(\rho^{\,2}_{\!A})$, displaying only the spatial slice on which a state is defined. This is in place of the full Euclidean manifold over which a path integral generates the given state on that slice. In (a), a partial trace (represented by the blue arrow) over the $B$ subsets, results in the reduced density matrix $(\rho_A)_{i_1i'_1}$. In (b) we multiply two copies of $(\rho_A)_{i_1i'_1}$ and perform a trace over the remaining indices to get $\Tr(\rho^{\,2}_{\!A})$.}
\label{fig7}
\end{figure}
In the 1-dimensional slices depicted in fig.~\ref{fig7}, this operation corresponds to identifying the values of the field variable $X$ in the first $M$ segments of the bra and the ket states and summing over all possible values, yielding the reduced density matrix $\rho_A$ on the $N$ segments describing set $A$. Expanding $\rho_A$ in a basis as $\sum_{i_1, i'_1}(\rho_A)_{i_1 i'_1}|\psi_{i_1}\rangle \langle \psi_{i'_1}|$ we see that the state on the last $N$ segments of (\ref{dfoneq}) gives the ket of the density matrix while the state on the last $N$ segments of (\ref{dfoneqq}) gives the bra of the density matrix.

Since our goal is to compute $S_2(A)$, we need a second copy of the density matrix $|\Psi\rangle \langle \Psi|$ on $A\cup B$. To get this second copy of $\rho_A$, we take one more set of $M+N$ line segments, with the states
\begin{equation} \label{dfoneqp}
|\Psi\rangle = \sum_{i_2,j_2} C_{i_2j_2} |\psi_{i_2}\rangle|\chi_{j_2}\rangle \ , \quad \langle\Psi| = \sum_{i'_2,j'_2} C^*_{i'_2j'_2} \langle\psi_{i'_2}|\langle\chi_{j'_2}| \ ,
\end{equation}
such that $\sum_{i_2,j_2} |C_{i_2j_2}|^2=1$. To trace over the set $B$ and get the second copy of $\rho_A$ we take $|\Psi\rangle \langle \Psi|$ and act with the delta function $\delta_{j_2, j'_2}$. Now we have two copies of $\rho_A$, one with components $(\rho_A)_{i_1 i'_1}$ and the other with $(\rho_A)_{i_2 i'_2}$. To compute $\rho_{\!A}^{\,2}$ we must act with $\delta_{i'_1,i_2}$ and then to get $\Tr[(\rho_A)^2]$ we must further act with $\delta_{i_1, i'_2}$. Collecting together all these steps gives\footnote{We write this as explicitly as possible in order to make very clear the difference once a prescription is introduced in the following subsection.}
\begin{align} \label{dsone}
\Tr\big[(\rho_A)^2\big] = \Bigg(\sum_{i_1,j_1} C_{i_1j_1} |\psi_{i_1}\rangle |\chi_{j_1}\rangle\sum_{i'_1,j'_1} C^*_{i'_1j'_1} \langle\psi_{i'_1}|\langle\chi_{j'_1}|\Bigg)& \Bigg( \sum_{i_2,j_2} C_{i_2j_2} |\psi_{i_2}\rangle |\chi_{j_2}\rangle\sum_{i'_2,j'_2} C^*_{i'_2j'_2} \langle\psi_{i'_2}|\langle\chi_{j'_2}|\Bigg)\nonumber \\
&\qquad\times \Big(\delta_{j_1, j'_1} \delta_{j_2, j'_2}\Big) \Big(\delta_{i_1, i'_2} \delta_{i'_1i_2} \Big) \ .
\end{align}

\subsection{The `prescription'} \label{secprescription}

We have gone through these elementary steps in detail so that we can now state the `prescription' that will be used in the wormhole paradigm to modify the above computation. This prescription makes the following replacement of the first bracket in the second line of (\ref{dsone})
\be \label{dstwo}
    \delta_{j_1, j'_1}\delta_{j_2, j'_2} ~\r~ \delta_{j_1, j'_1} \delta_{j_2, j'_2} + C \delta_{j_1, j'_2} \delta_{j_2, j'_1} \ ,
\ee
where $C$ is a constant that will be specified below\footnote{See for example an overview in \cite{Almheiri:2020cfm}.}.
\\
\\
\begin{figure}[H]
\begin{center}
\includegraphics[scale=0.35]{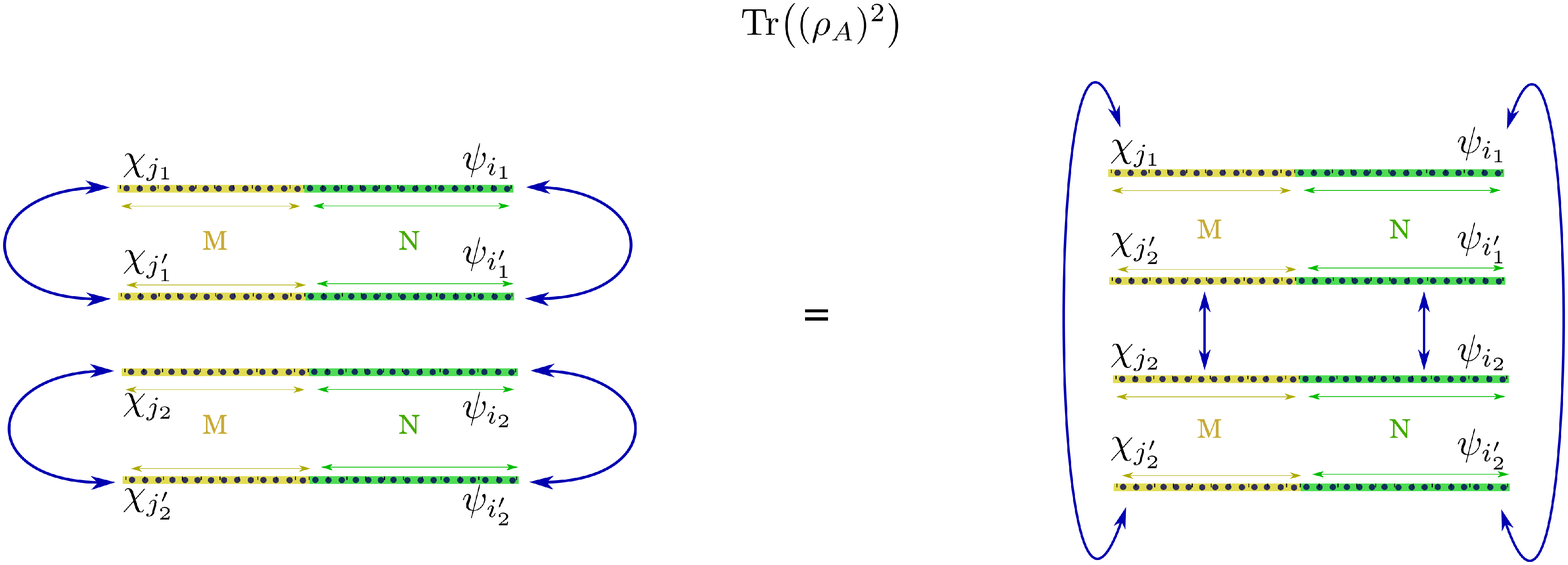}\hskip20pt
\end{center}
\caption{In the wormhole prescription, the computation of the second R\'{e}nyi entropy depicted in fig.~\ref{fig7} is modified by a term generated by the partial traces show in blue on two copies of the total density matrix. This corresponds to $(\Tr[\rho_A])^2$. A simple redefinition of indices gives the right-hand side.}
\label{fig8}
\end{figure}
The indices of type $j$ run over the subsystem $B$ that we trace over, which in the black hole context describes the gravitational region containing the black hole. With a little relabelling of indices we can see that the effect of the prescription (\ref{dstwo}) in the computation (\ref{dsone}) can be written in terms of the reduced density matrix on $A$ as
\be
\Tr\big[(\rho_A)^2\big]~\r~ \Big ( \Tr\big[(\rho_A)^2\big]+ C\big(\Tr[\rho_A]\big)^2 \Big) \ .
\label{mfivep}
\ee
Let us note the effect of this prescription on entanglement entropies. Suppose the state $|\Psi\rangle$ has the form
\be
|\Psi\rangle={1\over \sqrt{k}} \sum_{i=1}^k |\psi_i\rangle |\chi_i\rangle \ ,
\ee
then one finds
\be
\Tr[\rho_A]=1\ , \quad \Tr\big[(\rho_A)^2\big]={1\over k} \ .
\ee
The second R\'{e}nyi entropy $S_2(A)$ for this state is given by (cf. (\ref{dsfive}))
\be
S_2(A)=- \log \Big[\Tr\big[\rho_{\!A}^{\,2}\big]\Big]= \log k \ .
\ee
The entanglement between $A$ and $B$ rises with increasing $k$ and so here $S_2(A)$ correctly reflects this. However, with the above prescription we get
\be
S_2^{prescription}(A)=- \log \Big[ \Tr\big[\rho_{\!A}^{\,2}\big]+ C \, \big (\Tr[\rho_A]\big)^2 \Big]= - \log \bigg[ \left( {1\over k}+ C \right ) \bigg] \ .
\label{dsseven}
\ee
Suppose we take the constant $C$ to be
\be
C=e^{-S_{bek}} \ ,
\ee
then we see from (\ref{dsseven}) that for $k\ll S_{bek}$ we get
\be
S_2^{prescription}(A)\approx \log k \ ,
\ee
while for $k\gtrsim S_{bek}$ we get
\be
S_2^{prescription}(A)\approx S_{bek} \ .
\ee
Thus $S_2^{prescription}(A)$ is a quantity that behaves like the usual R\'{e}nyi entropy for low amounts of entanglement, but saturates to the value $S_{bek}$ for large values of the entanglement. It is important to note that $S_2^{prescription}$ is \emph{not} the original quantity $S_2$ that we were supposed to compute. So what is the reason that we are computing $S_2^{prescription}$? It has sometimes been said that the modification $S_2\r S_2^{prescription}$ arises because we must take topology change into account in a theory of gravity. We now argue that this is not the case; topology change can indeed occur in gravity, but it does not imply the replacement $S_2\r S_2^{prescription}$.

\subsection{Topology change}\label{sectop}

Topology change may appear to be something mysterious, however, we have already seen in section \ref{sectopology} how to handle topology change in (1+1)-dimensions. The fundamental structure of the quantum theory is \emph{not} altered, as far as the notion of Hilbert space, inner products and unitarity are concerned. In the (1+1)-dimensional case we considered, the spatial sections simply did not have to consist of only a single line segment but instead could be made of multiple line segments, with the full Hilbert space being the union of these different possibilities. There has to be an inner product on this Hilbert space and the evolution has to be unitary with respect to this inner product. 
\begin{figure}[H]
\begin{center}
\includegraphics[scale=0.6]{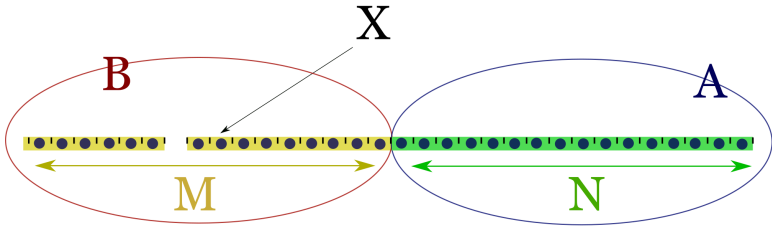}\hskip20pt
\end{center}
\caption{Introducing topology change in the 1-d system, the first $M$ segments which up the subset $B$ can be broken into sub-segments. The last $N$ segments still make up the subset $A$. Again, $X$ represents the matter field on our 1-dimensional spacelike slice.}
\label{fig9}
\end{figure}
In fact this structure of having multiple line segments is not peculiar to gravity, We could consider the quantum dynamics of a 1-dimensional polymer, which can break into multiple line segments. A similar Hilbert space and evolution will emerge, though the details of the Hamiltonian will depend on the physics of the system.

With this in mind, let us turn to the computation of the R\'{e}nyi entropies. In the discussion of section \ref{secrenyi}, the subregion $B$ represents the region with gravity. Allowing topology change then only changes the fact that the states $|\chi_j\rangle$ describing this subset $B$ consist not just of one segment but a linear combination of states with different number of segments. Recall that we had taken a line segment to be made up of a certain number of links and at the center of each link we had placed a scalar degree of freedom $X$. In a connected segment, the Hamiltonian will typically have an interaction linking the scalar field values at neighboring links $\alpha, \alpha+1$ of the form
\be
H^{int}_{\alpha,\alpha+1} = {1\over 2}{(X_\alpha-X_{\alpha+1})^2\over \delta^2} \ .
\ee
There will be no such term between links that are on different line segments; this fact will tell us when we have topologically disconnected segments. Crucially, any entanglement entropy is simply a function of the state $|\Psi\rangle$ on a spacelike slice with the inner product being needed to trace out the subsystem $B$. The Hamiltonian is, however, \emph{not} involved in the computation of entanglement. In a constrained system the allowed states may be subject to a Hamiltonian constraint and we will discuss this issue below, but once we have a state $|\Psi\rangle$ that is a physical state for the system $A\cup B$, then to compute a R\'{e}nyi entropies $S_n(A)$ all we need is the inner product on the system $B$. 

But now we see immediately that topology change does not lead to any prescription like (\ref{dstwo}). All that happens is that the states $|\chi_j\rangle$ now span both single segment and multi-segment possibilities. The states labeled by indices $i_1, j_1, i'_1, j'_1$ in (\ref{dsone}) pertain to the first copy of $\rho_A$ and the states labeled by $i_1, j_2, i'_2, j'_2$ pertain to the second copy of $\rho_A$. We do not end up mixing these two copies of $\rho_A$ in a new way a suggested by the prescription (\ref{dstwo}). It is true that gravity can allow for topology change, but this just changes the structure of the Hilbert space, without changing how a quantity like $S_2(A)$ is to be computed.

So let us ask: why was a prescription like (\ref{dstwo}) suggested? To understand the answer to this question, we will now cast the above computation of $S_n(A)$ in the language of path integrals.

\section{The Page curve - III: Path integrals and the difference between R\'{e}nyi and Gibbons-Hawking type computations} \label{pageiii}

In section~\ref{secpathintegral} we see how to recast states in terms of path integrals and observe that allowing topology change in the gravity theory does \emph{not} give a wormhole that should connect different replica copies. In section~\ref{seceuclideancomparison} we recall the Gibbons-Hawking computation of entropy and observe that the R\'{e}nyi entropy-inspired computations using added prescriptions are fundamentally different from the Gibbons-Hawking computation: with the R\'{e}nyi entropy computations we start with a path integral prescription that is not the correct one for the R\'{e}nyi entropy for a general quantum system, while in the Gibbons-Hawking computation the starting point is the correct path integral that should count all microstates.

Section~\ref{secworment} discusses further, in relation to the sewing procedure in 2-d CFTs, how having `wormholes' in a Euclidean picture has no relation to interactions in the real, Lorentzian theory (and thus cannot answer a Lorentzian problem such as that of black hole information loss).

In section~\ref{liu} we note that there are computations which show that the Page curve comes down, but these computations hold for \emph{normal} systems where the Page curve would automatically come down by the standard computation of Page \cite{Page:1993df}. In these systems there is no analogue of the effective pair production (\ref{twopp}), so these computations do not address the goal of the wormhole paradigm.

\subsection{Expressing states through path integrals}\label{secpathintegral}

As we have already noted, the computation of entanglement entropies is related to the state $|\Psi\rangle$ on a spacelike slice, it does not involve the dynamics of the system. Thus there is no natural connection between the computation of a quantity like $S_n(A)$ and a path integral in the theory. So why should we try to use path integrals? 

In 2-d CFTs, the path integral has been useful in the computation of entanglement in the following way. We often wish to compute the entanglement of a subregion $A$ when the overall state $|\Psi\rangle$ on our spatial slice is the \emph{vacuum} $|0\rangle$. In this case we can generate the state $|\Psi\rangle=|0\rangle$ on our spacelike slice as follows. Working in Euclidean signature a path integral over the lower half plane generates the state $|0\rangle$ on the upper boundary of this half-plane. Let us call the 2-d manifold spanning the lower half-plane ${\mathcal M}_1$. We generate the dual state $\langle \Psi|=\langle 0|$ by a similar path integral on an upper half-plane, calling this manifold $\widetilde{\mathcal M}_1$. We then perform the trace over the subset $B$ by joining ${\mathcal M}_1$ and $\widetilde{\mathcal M}_1 $ along the region representing $B$ (the complement of $A$). The states on the edges of ${\mathcal M}_1$ and $\widetilde{\mathcal M}_1$ that are in the segment $A$ then give the density matrix $\rho_A$. If we wish to compute the entanglement for some state other than $|\Psi\rangle=|0\rangle$, we can insert appropriate operators in the manifolds ${\mathcal M}_1$ and $\widetilde{\mathcal M}_1 $ to alter the evolution.

But here we note that getting $\rho_A$ from a path integral in this way is just a trick that makes the computation easier; the original definition of the density matrix as a partial trace over the subset $B$ actually involves only the state $|\Psi\rangle$ on the 1-dimensional slice. Thus we should be careful to not modify the path integral computation in an arbitrary way through a `prescription', since then we will not be actually computing the density matrix $\rho_A$ or the entropies $S_n(A)$.

\begin{figure}[h]
\begin{center}
\includegraphics[scale=0.4]{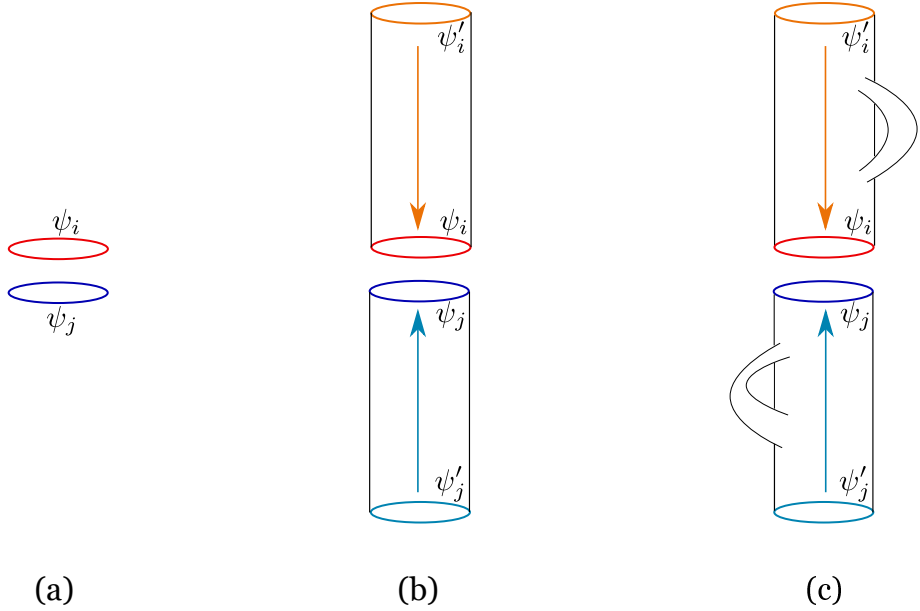}\hskip20pt
\end{center}
\caption{In (a) states $\psi_i$ and $\psi_j$ are shown on the circles that they are defined on. In (b) the states $\psi'_i$ and $\psi'_j$ evolve in Euclidean time to the states $\psi_i$ and $\psi_j$ respectively in a 2-dimensional gravity theory, depicted as cylinders. In (c) we allow the 2d gravity theory to have topology change, which is shown as the addition of handles to the cylinders.}
\label{innprod}
\end{figure}
As an example, let us start with an ordinary CFT on a circle. In fig.~\ref{innprod}(a) we depict the bra and ket states $|\Psi_i\rangle$ and $\langle \Psi_j|$ for such a CFT with the inner product between these states denoted by
\be
\langle \Psi_j|\Psi_i\rangle \ .
\label{dsfive2}
\ee
How should we get this inner product from a path integral? In fig.~\ref{innprod}(b), we depict an \emph{amplitude} between states $|\Psi'_i\rangle$ and $\langle \Psi'_j|$ which we write this
\be
\langle \Psi'_j|\Psi'_i\rangle_{amplitude} \ .
\label{dssix}
\ee
But why should this amplitude (\ref{dssix}) have anything to do with the inner product (\ref{dsfive2})? If we expand the state $|\Psi'_i\rangle$ into energy eigenstates, then these different eigenstates evolve with different factors
\be
\sum_n C_n |E_n\rangle \r \sum_n C_n e^{-E_n \tau}|E_n\rangle \ .
\ee
So the states $|\Psi'_i\rangle$ and $\langle \Psi'_j|$ in the amplitude (\ref{dssix}) will have to be different from the states $|\Psi_i\rangle$ and $\langle \Psi_j|$ in the inner product (\ref{dsfive2}). If we evolve the states in the amplitude through a large time $\tau\r \infty$, then we will end up with just the vacuum states at the middle of the cylinder, as the coefficients of all higher energy states are subleading.

But now consider not a simple CFT but a theory of gravity on our 2-d cylinder; the string world sheet theory is an example of such a gravity theory. The physical states then all have the same energy since they satisfy the momentum and Hamiltonian constraints (in terms of the stress tensor modes $L_0, \bar{L}_0$)
\be
(L_0-1)|\Psi\rangle=0\ , \quad(\bar L_0-1)|\Psi\rangle=0 \ .
\ee
Thus any state from the physical Hilbert space does not suffer a change in the relative weights of its parts as it evolves down the cylinder and after absorbing a suitable power of $e^\tau$, we can identify $|\Psi'_i\rangle$, $\langle \Psi'_j|$ with $|\Psi_i\rangle$, $\langle \Psi_j|$. Thus the inner product (\ref{dsfive2}) in this case can be written as an amplitude of the type (\ref{dssix}). 

Now suppose the 2-d gravity theory describing our cylinder is allowed to have topology change. This means that the 1-dimensional circle describing the spatial sections of the cylinder can break up into two such circles and vice versa. The evolution of the state on the cylinder can now have handles, as depicted in fig.~\ref{innprod}(c). Note that the physical quantity that we need for the definition of entanglement entropy is an inner product like (\ref{dsfive2}); recasting this as an amplitude (\ref{dssix}) is just something that we have done for our present purposes.

\begin{figure}[h]
\begin{center}
\includegraphics[scale=0.37]{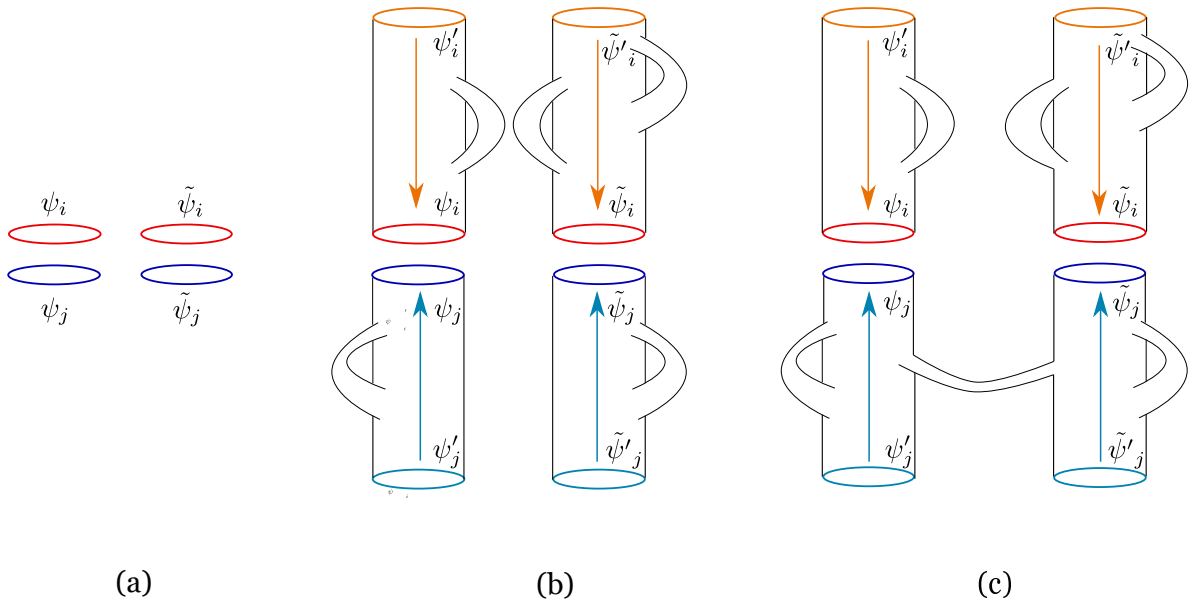}\hskip20pt
\end{center}
\caption{In (a) states are shown on their respective circles. In (b) the states evolve in Euclidean time in a 2-dimensional gravity theory, with all topology changes allowed. This is shown as cylinders with handles. In (c) we show handles which surpass the cylinders and join two different cylinders. This is forbidden, as we argue in text.}
\label{trhand}
\end{figure}

Now we come to the crucial step. Suppose we wish to compute the second R\'{e}nyi entropy $S_2(A)$. For this purpose we create two copies of the bra/ket pair, as in fig.~\ref{trhand}(a), and identify the segments along the subset $B$ of each bra/ket pair. The further traces in the second line of \eqref{dsone} then give $\Tr[(\rho_A)^2]$, which we then use in (\ref{dsfive}) to compute $S_2(A)$. But now suppose someone wishes to rewrite this computation using path integrals. The bra and ket states for each copy are created by a time evolution, similar to the evolution we did for the string world sheet to write (\ref{dsfive2}) as an amplitude (\ref{dssix}).
This time evolution can have handles on the 2-d worldsheet if the Hamiltonian allows for topology change, \emph{but there will be no handles that connect the 2-d manifolds of one bra/ket pair to the manifolds for the other bra/ket pair.} This is because the time evolution over these 2-d manifolds was just a trick to reproduce the inner products needed for the computation of $\Tr[(\rho_A)^2]$ and in this trace the only delta functions are the ones given in (\ref{dsone}). If we use a path integral to reproduce this result for $\Tr[(\rho_A)^2]$, then we can only have those handles that arise in the evolution that gives rise to the states in fig.~\ref{trhand}; we cannot include any other kinds of handles at will.

To summarize, we have seen that entanglement entropies are a property of a given state; the dynamics of the theory is not involved. There is no natural appearance of a path integral in the computation of entanglement entropies, since the path integral describes the dynamical evolution of states. We can however use the path integral as a trick to recast the computation of a quantity such as R\'{e}nyi entropy. But while this path integral may manifest handles within the computation of each copy of the density matrix, there are no handles between different copies; i.e. we find no analogue of a `replica wormhole'.

\subsection{The Gibbons-Hawking computation}\label{seceuclideancomparison}

Let us consider the Gibbons-Hawking computation of entropy, starting from first principles; this will allow us  to see the differences between the Gibbons-Hawking computation and the Euclidean arguments for the Page curve. The Gibbons-Hawking argument proceeds in the following steps:

\begin{enumerate}[start=1,
    labelindent=\parindent,
    leftmargin =2\parindent,
    label=(\Alph*)]
\item \label{(AAAA)}
First consider \emph{any} quantum system, not necessarily one with gravity. We assume that the system is described by a Hilbert space ${\mathcal H}$ and a Hamiltonian $H$. The eigenstates of this Hamiltonian satisfy
\be
H|\psi_i\rangle= E_i |\psi_i\rangle \ ,
\ee
and their time evolution is governed by
\be
|\psi_i(t)\rangle=e^{-iHt}|\psi_i(0)\rangle \ .
\label{dthone}
\ee
Note that this is Lorentzian time evolution. Now we analytically continue to Euclidean time using $t\r-i\tau$ and consider the quantity
\be
Z(\beta)={\Tr} [ e^{-\beta H}]=\sum_i e^{-\beta E_i} \ .
\label{dthtwo}
\ee
From this quantity we can extract the entropy using the standard expressions of statistical mechanics. We write the temperature $T$, free energy $F$ and average energy $\langle E \rangle$ as
\be
T={1\over\beta} \ , \quad F=-T\log Z\ , \quad\langle E \rangle = -{\p\over \p\beta} \log Z \ ,
\ee
from which the relation $F=\langle E \rangle -TS$ can be used to find the entropy $S$.
 
\item \label{(BBBB)}
We now note that the analytic continuation $t\r -i\tau$ applied to the evolution (\ref{dthone}) gives the evolution 
\be
|\psi_i(\tau)\rangle=e^{-E_i\tau}|\psi_i(0)\rangle \ .
\label{dthonep}
\ee
The trace we need in (\ref{dthtwo}) then implies that the quantity $Z$ is a one loop path integral with period $\beta$ for the loop. Note that so far the steps we have outlined hold for \emph{any} quantum theory.

\item \label{(CCCC)}
Now we specify to the gravity theory. Following the steps above, we make a periodic identification of Euclidean time with period $\beta$. The full path integral for the exact theory with this identification should give the entropy $S$ that we seek. Since quantum gravity is complicated, we find that we do not know how to carry out this full path integral. However, we observe that there is a saddle point of the classical action for the Euclidean geometry (\ref{dttwo}), with $M$ related to $\beta$ via $\beta=8\pi GM$. We make an \emph{assumption} that this saddle point will give a good leading order approximation to the full path integral of the exact quantum gravity theory. With this, following the steps above, we find
\be
S=4\pi G M^2= {A\over 4G}\equiv S_{bek} \ ,
\label{dththree}
\ee
where $A$ is the area of the black hole horizon.
\end{enumerate}

We have recalled this well known Gibbons-Hawking computation to emphasize the fact that in steps \ref{(AAAA)} and \ref{(BBBB)} we were setting up a computation that makes sense for \emph{any} quantum theory. Then we come to the gravity theory and do a saddle point evaluation of the quantity that we had \emph{already} defined; i.e. the one loop path integral $Z$. This yields the entropy $S_{bek}$. Our belief in the assumption in \ref{(CCCC)} is bolstered by the fact that the result (\ref{dththree}) agrees with the Lorentzian computation of entropy that was already known.\footnote{Hawking's Lorentzian calculation showed that the hole emits a temperature $T=1/(8\pi GM)$; putting this in the standard thermodynamics relation $TdS=dE$ gives $S=S_{bek}=A/(4G)$.} 

The assumptions in the recent Euclidean computations of the Page curve appear to be fundamentally different in the following way. The prescription (\ref{mfivep}) changes what we call a R\'{e}nyi entropy, replacing this entropy by a different quantity. Thus we will not be starting with a definition for the R\'{e}nyi entropy that is standard for \emph{all} systems, whether gravitating or not. We have looked at the role of topology change for (1+1)-dimensional quantum gravity and, at least for this case, we have found that topology change does not imply the prescription (\ref{mfivep}). Thus our arguments indicate  that the recent computations of the Page curve seem to be addressing a quantity that is not the entanglement entropy that is described by the usual Page curve.

\subsection{Wormholes that represent entanglement} \label{secworment}

In the above discussion we have looked at the exact gravity theory and asked if the replacement (\ref{mfivep}) can arise from the possibility of topology change in this exact theory. We did not find any motivation for (\ref{mfivep}) from the possibility of topology change in the exact theory in our discussion using (1+1)-dimensional quantum gravity. We now ask a different question: is it possible that when we are computing the correct R\'{e}nyi entropy ($\sim\Tr(\rho^2)$ for the second R\'{e}nyi entropy) in the exact theory, there is an emergence of an approximate saddle point of the effective theory that does yield something resembling (\ref{mfivep}). The result of our analysis below will be that we will not find the emergence of a prescription like (\ref{mfivep}). 
\begin{figure}[h]
\begin{center}
\includegraphics[scale=0.37]{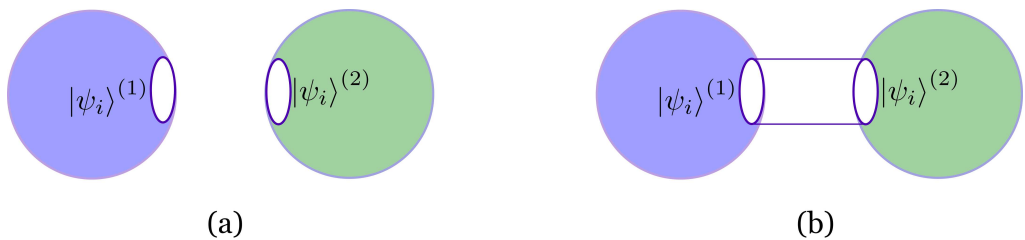}\hskip20pt
\end{center}
\caption{The sewing procedure. In (a) two states, $|\psi_i\rangle^{(1)}$ and $|\psi_i\rangle^{(2)}$, are defined on the circle forming the boundary of the respective sphere. In (b) the sum \ref{none} generates a connection (`wormhole') between the two states, shown as the tube connecting the two spheres.}
\label{sewing}
\end{figure}
To see the kind of effective descriptions that we will investigate, consider the `sewing' procedure in 2-d CFTs. In fig.~\ref{sewing}(a) we depict two spheres. In each sphere we cut a hole and at the boundary of each hole we place the same state $|\psi_i\rangle$. Then we consider the combination
\be
|\Psi\rangle=\sum_i C_i |\psi_i\rangle^{(1)} |\psi_i\rangle^{(2)} \ ,
\label{none}
\ee
where the superscripts $(1), (2)$ denote the two different spheres. The state $|\Psi\rangle$ is entangled between the two spheres, but we have no Hamiltonian connection between the two spheres. With an appropriate choice of the $C_i$ and $|\psi_i\rangle$, the sum in (\ref{none}) generates a `sewing' of the two spheres, where the spheres are now joined by a `wormhole' as in fig.~\ref{sewing}(b). The length of the wormhole can be altered by changing the coefficients $C_i$. 

What can we use the manifold in fig.~\ref{sewing}(b) for? The entanglement in the state (\ref{none}) generates correlations between the two spheres, so if we compute a correlator between the two spheres the result will be generically nonzero
\be
\langle\phi(z_1)\phi(z_2)\rangle \ne 0 \ ,
\label{ntwo}
\ee
where $z_1$ and $z_2$ are patches on the first and second sphere respectively. If $\phi$ represents a high-dimension field, then the correlator may be well approximated by the action of a geodesic joining $z_1, z_2$ along a path that goes through the wormhole. More generally, a path integral over the field $\phi$ on the sewn manifold will yield the correlator (\ref{ntwo}). 

Note that the correlator (\ref{ntwo}) just measures the correlations that we established between the two spheres by taking an entangled state (\ref{none}). There was  no Hamiltonian connection between the two spheres to start with, so if we had switched to a Lorentzian theory on the two spaces (with an entangled state between them) then we could not send a signal from one space to the other through such a wormhole. This is the issue we discussed in section~\ref{eucl}, where it was noted that Euclidean connections between manifolds did not help with the problem of resolving the information paradox -- a problem arising from dynamical evolution in the Lorentzian section. 

At the present time our interest is in looking for a Euclidean path integral prescription that may yield the R\'{e}nyi entropy. So in line with the above example of sewing, we ask the following. Suppose we see that there are 1-dimensional segments in our (1+1)-dimensional gravity theory on which the states are entangled in the manner (\ref{none}). Then we may try to add a wormhole connection between these segments to represent this entanglement, as in the above example of sewing. Can such a connection give rise to a prescription like (\ref{mfivep})?

We refer again to the depiction in fig.~\ref{fig7}(b) of the states in the computation of the second R\'{e}nyi entropy $S_2(A)=\Tr(\rho_{\!A}^{\,2})$. Let the two entangled subspaces be $A$ and $B$, and let the overall entangled state be of the diagonal form
\be
|\Psi\rangle={1\over \sqrt{N}}\sum_{I=1}^N |\psi_i\rangle_A |\chi_i\rangle_B \ ,
\label{nthree}
\ee
We see then the following identifications in fig.~\ref{fig7}(b):
\begin{itemize}
    \item The state labeled $\psi_{i_2}$ on one copy of the subset $A$ is equal to the state $\psi_{i'_1}$ on the next copy because of the matrix multiplication in $\Tr(\rho_{\!A}^{\,2})$.
    \item The index on the state $\psi_{i_2}$ becomes equal that of the state $\chi_{j_2}$ on that subset $B$ if we have taken the entangled state (\ref{nthree}).
    \item The index of the state $\chi_{j'_1}$ is equal to that of the state $\psi_{i'_1}$ again because we have taken the entangled state (\ref{nthree}).
\end{itemize}
From the above, we conclude that the state $\chi_{j'_1}$ will equal the state $\chi_{j_2}$. Following the rough idea of sewing that we saw above, we can be tempted to then draw a manifold $M_1$ connecting the line segments containing the states $\chi_{j'_1}$ and $\chi_{j_2}$. Similarly, we can repeat similar arguments and also draw a connecting manifold $M_2$ between the segments containing the states $\chi_{j'_2}$ and $\chi_{j_1}$, since the states on these segments are again the same. We note that these connections $M_1, M_2$ do arise in the computation of $(\Tr(\rho_{A}))^2$, for which we have fig.~\ref{fig8} (in particular the right-hand side). So do we have a suggestion that `sewing' entangled segments will give something like $(\Tr(\rho_A))^2$?

The answer is no, for the following reason. In our computation where we start with the R\'{e}nyi entropy $S_2(A)=\Tr(\rho_{\!A}^{\,2})$, we have a trace that identifies the state $\chi_{j'_2}$ with the state $\chi_{j_2}$. This identification arises from the definition of one copy of the density matrix $\rho_A$. Similarly, we have an identification between $\chi_{j'_1}$ and $\chi_{j_1}$ from the other copy of $\rho_A$. If we were to \emph{remove} these identifications then, yes, we would get $(\Tr(\rho_A)^2)$. However, we cannot arbitrarily remove the identifications and so do not get the prescription (\ref{mfivep}).

More generally, the issue we face is the following. The idea behind the replica wormhole is that we should fix the way we trace over the different states \emph{outside} the gravity region, but allow all possible ways of joining manifolds \emph{inside} the gravity region. How then do we know that the quantity we end up computing has anything to do with the (second) R\'{e}nyi entropy $S_2(A)=\Tr(\rho_{\!A}^{\,2})$? Any entanglement entropy is a property of the entangled \emph{state} that we start with. The state we are interested in is the entangled state of radiation and the remaining hole. In a Euclidean formulation, we do not know how to ensure that this state is the one whose entanglement we are computing. In fact if we do a sum over all manifolds without boundary in the gravity region, then we have no place to input which state in the gravity region we are interested in considering. We have tried various ways to start with a situation that does compute the R\'{e}nyi entropy $S_2(A)=\Tr(\rho_{\!A}^{\,2})$, but have not been able to map this to a computation where a replica wormhole appears and gives the modification (\ref{mfivep}).

\subsection{Modeling the evaporation of coal}\label{liu}

We now describe a nice computation of R\'{e}nyi entropies performed in \cite{Liu:2020jsv} which gives a set of diagrams that seem to resemble replica wormholes. But as we will argue below, this resemblance is superficial and the computation of \cite{Liu:2020jsv} cannot be used to support the idea of replica wormholes in gravity.

It is possible to get a result that superficially resembles the replica wormhole computation in the following way. Starting with a normal radiating body like a piece of coal, write a path-integral expression for the normal R\'{e}nyi entropy (i.e. without any prescription modifying the definition of the R\'{e}nyi entropy). In this path integral, one may find that certain paths dominate. For one such dominating path, the \emph{value} of the action ($S_0$) for the segment of the path joining a pair of replica copies may cancel the value of the action for a different segment of the overall path. We can represent this cancellation by a schematic diagram that links the different copies involved. Such a linkage diagram will have a superficial visual similarity to the different `replica wormholes' that are assumed to be saddle points of the gravity path integral in the wormhole paradigm. The following shows that these linkage diagrams and the replica wormholes are different things:

\begin{enumerate}[start=1,
    labelindent=\parindent,
    leftmargin =2\parindent,
    label=(\alph*)]
\item \label{(aaaa)}
The computation of \cite{Liu:2020jsv} starts with assuming that one has a normal body like a piece of coal. \emph{But such a normal body does not have the semiclassical low-energy dynamics of a black hole horizon where we get entangled pairs (\ref{twopp}).} Thus the computation of \cite{Liu:2020jsv} can be taken as an explanation for why the Page curve comes down for a normal body; it tells us nothing about any theory of the black hole where we require the production of (\ref{twopp}).

\item \label{(bbbb)}
Now suppose we \emph{do} assume that the black hole horizon has an effective semiclassical description where we get (\ref{twopp}). The computations of \cite{Liu:2020jsv} do not assume any nonlocal interactions between the hole and its radiation. Thus the effective small corrections theorem of section~\ref{sec small} tells us that in this situation the Page curve \emph{cannot} come down. This is counter to the goal of the computation of \cite{Liu:2020jsv}.

\item \label{(cccc)}
Thus the crucial issue in all such computations is the following. If we perform any computation for the Page curve in the wormhole paradigm, then we have to also check if the effective semiclassical horizon behavior (\ref{twopp}) holds for the system being analyzed. If we do not demonstrate (\ref{twopp}) then we may just as well be describing the Page curve of a normal body, which is known to come down at the end of the evaporation process by the computation of Page \cite{Page:1993df}. 
\end{enumerate}

\subsection{Summary}

Let us summarize what we have seen in this and the previous two sections on the Page curve. Quantum gravity should certainly allow the possibility of topology change: such topology change leads, for example, to handles in the evolution of a string world sheet. However, we have not found handles between \emph{different} replica copies. The reason is that each replica is an independent copy of the theory, and while the dynamics can generate handles within any one theory, this same dynamics cannot generate handles between different copies of the theory. 

In this context, an example that is often cited is that of the Gibbons-Hawking computation. Here one starts with a path integral where the Euclidean time circle has a period $\beta$ and then that the saddle point is a `cigar' with a novel topology where the Euclidean time circle shrinks to zero size. This situation might suggest that saddle points can somehow appear with topologies that were not built into the original path integral. Let us see the origin of this new topology in some detail and then we will see that the case of a wormhole between replicas is different.
\begin{figure}[h]
\begin{center}
\includegraphics[scale=0.45]{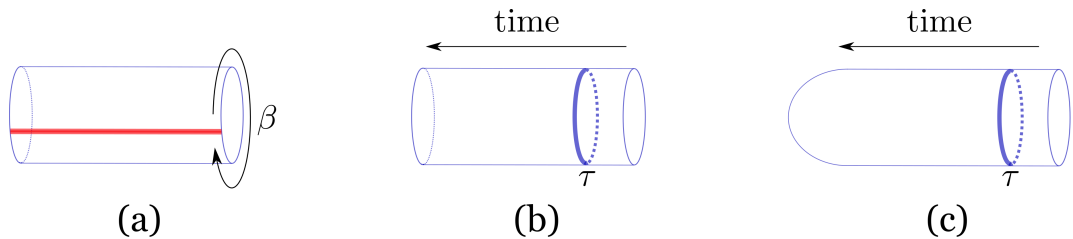}\hskip20pt
\end{center}
\caption{The Gibbons-Hawking-fuzzball understanding. In (a) all the states along the horizontal red slice are taken and weighted by $e^{-\beta E}$ to compute the path integral. In (b) the same path integral is calculated in a different way; states along the circle $\tau$ are evolved along the horizontal direction. In (c) the cylinder is shown closed with a cap, assuming only vacuum state survives after evolution by a long enough Euclidean time.}
\label{vpath}
\end{figure}
To understand the Gibbons-Hawking case, let us look at a toy example that resembles the Gibbons-hawking computation. In fig. \ref{vpath} we show a 2-dimensional CFT on a cylinder. In fig.~\ref{vpath}(a), we compute the path integral by taking all the states along the horizontal slice and weighting them with $e^{-\beta E}$. Here one can think of the complicated states defined along the horizontal line as black hole microstates, so that we are computing the path integral for the black hole. In fig.~\ref{vpath}(b) we note that we can compute this path integral in a different way. Now we use the other channel: we define states on the $\tau$ circle and evolve them along a `time' in the horizontal direction. If the horizontal direction of the cylinder is long, then only the vacuum state survives in this channel. Thus we can `cap-off' the cylinder as shown in fig.~\ref{vpath}(c), where this cap generates the vacuum state. We therefore see that the path integral can be obtained by a `cigar' geometry.\footnote{We still have to answer why in the black hole case, the analogue of the horizontal direction in fig.~\ref{vpath}(a), is long. The reason is that we have $g_{tt}\r 0$ as $r\r r_h$; this effectively generates the disc geometry of fig.~\ref{vpath}(c). Using the usual map between disc and cylinder coordinates, this disc maps to a cylinder that is effectively infinite on the left side.} This way of understanding the Gibbons-Hawking computation in terms of fuzzballs was discussed in \cite{Mathur:2014nja} and follows the discussion by Hawking in \cite{Hawking:1979ig}.

In all of the above, we started with a path integral in fig.~\ref{vpath}(a) which was the correct path integral for counting the states of the black hole. A similar situation occurs for the Hawking-Page transition \cite{Witten:1998zw}. For this transition we again have just one copy of the gravity theory, and are required to do a path integral over all manifolds that end on the boundary of AdS. For different values of the Euclidean time compactification $\Delta \tau=\beta$, we can have different topologies for the saddle point: one where the $\tau$ circle remains nonzero everywhere, and one where it shrinks to zero and generates a cigar geometry. But if we start with two \emph{different} replica copies of AdS, then the starting path integral has no paths that go from one copy of AdS to another. In this case we do not see how one could get a saddle that connects the replica copies. 

The reason this discussion is crucial is for the following. Suppose one argues that topology change should generate a wormhole between replicas and that this means that we should replace the R\'{e}nyi entropies by new quantities, as per (\ref{mfive}). Since these new quantities are not the original R\'{e}nyi entropies, how do we know that we are computing the Page curve and not some other quantity? Ultimately, the effective theory must emerge from the exact theory by some map $g_{ef\!f}=F[g_{exact}]$ and the rules for whether replica copies should be connected or not emerges from this map. The exact theory does not have a connection between replicas, since these are independent copies of the theory. But why should the effective variables $g_{ef\!f}$ have connections between replicas? Usually effective variables are just a low-energy subset of the exact variables and such a choice will not give a connection between replicas.

The reason why we need to be vary careful to find the source of an ansatz like (\ref{mfive}) is the following. With the black hole we always have the cigar geometry that computes the Bekenstein entropy. If we replace the R\'{e}nyi entropy by a new quantity which brings in the cigar geometry, then this new quantity might just be computing the Bekenstein entropy, at least for some domain of parameters.\footnote{E.g. when the replica saddle is argued to dominate.} In that case the replacement (\ref{mfive}) would amount to using a prescription which replaces the
entanglement entropy with the Bekenstein entropy. As the hole evaporates the Bekenstein entropy goes to zero and so it might appear that the entanglement has come down. But then the whole question hinges on why we could make the replacement (\ref{mfive}). After all, if we were allowed to say that one cannot entangle with the black hole by more than the Bekenstein entropy, then there would be no puzzle with the Page curve; as the hole evaporates, the entanglement would automatically go to zero. In fact the whole information paradox can be restated as a mismatch between the Euclidean computation (which implies a maximal entanglement $S_{bek}$) and the Lorentzian computation where we can get an entanglement that is arbitrarily larger than $S_{bek}$.\footnote{We can get the entanglement of the traditional Lorentzian hole to be arbitrarily larger than $S_{bek}$ by feeding the hole at the same rate that it evaporates, or by looking near the endpoint of evaporation where the entanglement is large but $S_{bek}$ goes to zero.} Thus we should be careful that we are not doing the following: (i) making an ansatz that somehow replaces the entanglement entropy by $S_{bek}$ and then (ii) arguing that since $S_{bek}$ goes to zero as the hole evaporates, the Page curve goes down to zero.

\section{Postulating nonlocalities}\label{nonlocalchoices}

We have seen above that abstract arguments using semiclassical gravity do not allow us to show that the Page curve for a black hole will come down like that of a normal body. Furthermore, the effective small corrections theorem makes it impossible to get semiclassical dynamics (\ref{twopp}) around the horizon if we do not use any kind of nonlocality between the hole and its radiation. In section~\ref{secintro} we looked at several ways in which one might postulate some kind of nonlocality for the gravity theory. We now look in more detail at these possibilities, illustrating our understanding of various proposals by making very simple bit models to illustrate the essential idea of the proposal.

Recall that if we are dealing with an effective theory, then we cannot postulate an arbitrary set of rules for this effective theory. We have seen in section~\ref{sec small} that the effective variables $g_{ef\!f}$ must descend from the exact variables $g_{exact}$ by a map $g_{ef\!f}=F[g_{exact}]$ (eq.(\ref{dthreeqq})). This map then forces the dynamics of the effective theory to descend from the dynamics of the exact theory as in (\ref{dfourqq}) and also any quantity in the exact theory maps to a definite quantity in the effective theory through a map like (\ref{dfivepreqq}). Thus any nonlocality in the effective theory must arise either from a nonlocality in the exact theory or from a nonlocal definition of variables for the effective theory. We proceed as follows:

\begin{enumerate}[start=1,
    labelindent=\parindent,
    leftmargin =2\parindent,
    label=(\roman*)]

\item \label{(i3)}
In section~\ref{secnonlocaldef} we will consider the postulate that the exact theory does not have any nonlocal interactions between the radiation $R$ and the remaining hole, but that the effective variables $g_{ef\!f}$ are made by combining degrees of freedom of the exact theory from both the region $r<10\,r_h$ and the radiation region. In this case we find that these effective variables have the property that acting on the exact bits in the radiation region changes the observations that would be made by an experimenter at $r=5\,r_h$. Note that this is not the behavior that we expect from radiation from a piece of coal.

\item \label{(ii3)}
We then turn to the case that there are nonlocal effects between the radiation and the hole in the \emph{exact} theory. In section~\ref{experiment} we describe explicitly an experiment that will check whether the radiation from the hole is in a pure state or in a state that is entangled with the hole. This experiment will allow us to be precise about what we mean by the exact degrees of freedom at infinity: these are just the bits that are measured by an experimental apparatus far from the hole. We consider three types of nonlocal effects:

\begin{enumerate}[start=1,
    labelindent=\parindent,
    leftmargin =2\parindent,
    label=(\Alph*)]

\item \label{(A2)}
Nonlocal interactions between the black hole interior of one copy and the black hole interior of another copy. A simple model for such effects is of the following form. The interior of the hole in one black hole disconnects as a `baby universe' and joins to the interior of another black hole. We will see that trying to bring the Page curve down using such effects leads to a violation of  unitarity in the black hole interior. This possibility \ref{(A2)} is discussed in section~\ref{seca}.

\item \label{(B2)}
Nonlocal interactions between one region near spatial infinity and another, well separated region near spatial infinity. Such nonlocal effects violate the conventional notion of locality in physics. This possibility \ref{(B2)} is discussed in section~\ref{secb}.

\item \label{(C2)}
Nonlocal interactions between the inside of the hole and the region spatial infinity. Such `wormhole' effects will also violate the conventional notion of locality in physics. This possibility \ref{(C2)} is discussed in section~\ref{secc}.  
\end{enumerate}
\end{enumerate}

\subsection{Nonlocal definition of effective variables}\label{secnonlocaldef}

One of the common ideas in the wormhole paradigm is the following:

\begin{enumerate}[start=1,
    labelindent=\parindent,
    leftmargin =2.5\parindent,
    label=(NL\arabic*)]

\item \label{(NL1)}
First assume that in the exact theory, the black hole radiates like a piece of coal. This means that in this exact theory the black hole satisfies the conditions \ref{C1}-\ref{C3} in section~\ref{seccoal}. Thus, there are no significant interactions of the radiated quanta with the remaining coal once these quanta leave the region $r>10\,r_h$ and the degrees of freedom defining the radiation at infinity are distinct from the degrees of freedom defining the hole in the region $r<10\,r_h$. The Page curve comes down to zero at the end of the evaporation process like the Page curve of a normal body.

\item \label{(NL2)}
It is possible to take some combination of the exact bits making up the radiation and the exact bits of the remaining hole to define a set of low-energy effective degrees of freedom around the horizon radius $r_h$. These effective degrees of freedom should reproduce semiclassical dynamics around the horizon, i.e. (\ref{twopp}) and (\ref{mtwo}). Very little need be demanded from these effective variables, only the conditions of \ref{EFF4} listed in section~\ref{seceff}. This effective semiclassical dynamics is then argued to be what was somehow seen in Hawking's original computation of entangled pairs, while the exact dynamics of the theory is similar to the burning of coal. Thus, the argument goes, the exact quantum gravity theory can resolve the information paradox even though Hawking's computation showed a problem with monotonically growing entanglement.
\end{enumerate}
We will see that the above scenario with \ref{(NL1)} and \ref{(NL2)} is actually not possible. We will see that trying to achieve \ref{(NL2)} forces an interaction linking the radiation to the hole, so the situation is \emph{not} like that of burning coal: condition \ref{C1} of section~\ref{seccoal} is violated. We are forced to a picture where we have to:

\begin{enumerate}[start=1,
    labelindent=\parindent,
    leftmargin =2.5\parindent,
    label=(NL\alph*)]

\item \label{(NLa)}
Collapse the radiation to a dense form, perhaps making a second black hole out of this radiation.

\item \label{(NLb)}
Argue that this black hole is connected to the original hole by a wormhole that provides an alternate path of interaction between the hole and the radiation. 
\end{enumerate}
Such a picture, comprised of \ref{(NLa)} and \ref{(NLb)}, has been suggested by Maldacena \cite{MaldacenaSaclayComment1}. We do not believe that \ref{(NLa)} and \ref{(NLb)} are actually the case in string theory, but we will not discuss this issue here; our goal will be to argue that \ref{(NL1)} and \ref{(NL2)} are not possible as a picture of what can happen and that any such attempt must end up in something like \ref{(NLa)} and \ref{(NLb)}.

\subsubsection{How can we differentiate such a black hole from coal?}

Let us begin with a very basic question. Suppose the black hole satisfies property \ref{(NL1)}. Then in its exact description it behaves just like a piece of coal. But for a piece of coal we have no analogue of \ref{(NL2)}; i.e. we do not expect that by using some combination of radiation and coal degrees of freedom we can see a smooth horizon. Thus we see an immediate problem with the proposal of \ref{(NL1)} and \ref{(NL2)}: how can we get \ref{(NL2)} for the degrees of freedom describing the hole and its radiation, but not for those describing the coal and its radiation? One might try to say the following: perhaps the bits that come out of a black hole are entangled with the remaining hole in some special way, which is different from the entanglement between the bits emitted from a piece of coal and the remaining coal. But we will now see that this is not possible since all possible entangled states of the black hole radiation can be obtained by emission from a normal body like a piece of coal.

Consider a box containing $N$ atoms of a gas, with each atom having two spin states $\pm$. Let this box sit for a long time $t$, so that the atoms collide multiple times and reach a state where their spin states are entangled in a generic way. Now open a small hole in this box such that the atoms escape one by one to infinity. After $n$ spins have emerged, the overall state of the radiation and the spins remaining in the box has the form
\be
|\Psi\rangle=\sum_{i=1}^{2^{N-n}} \sum_{j=1}^{2^n} C_{ij} |\chi_i\rangle|\psi_j\rangle \ ,
\label{bone}
\ee
where the $\{|\psi_j\rangle\}$ are a basis of $2^{n}$ spin states of the form $|\!\pm\pm\dots \pm\rangle$ for the $n$ spins in the radiation and $\{|\chi_i\rangle\}$ are a basis of $2^{N-n}$ spin states of the form $|\!\pm\pm\dots \pm\rangle$ for the $N-n$ spins remaining in the box. We now explore different values of the time $t$ for which the atoms are allowed to interact before the hole is opened. By the ergodic hypothesis, the state $|\Psi\rangle$ evolves through a dense subset of all allowed spin states for the $N$ atoms, with the same uniform measure that is used in the analysis by Page of the Page curve for a normal body \cite{Page:1993df}. Therefore, as we explore different values for $t$, we get a dense subset of all the allowed values of the coefficients $C_{ij}$. 

Now consider the black hole. Suppose the total number of quanta that will be emitted by the hole is $N$. Consider the point in the evaporation process where $n<N$ quanta have been emitted. The entangled state of this radiation must be of the form
\be
|\widetilde{\Psi}\rangle=\sum_{i=1}^M\sum_{j=1}^{2^n} \widetilde{C}_{ij} |\widetilde{\chi}_i\rangle|\widetilde{\psi}_j\rangle \ ,
\label{btwo}
\ee
where $|\widetilde{\psi}_j\rangle$ describe the $2^n$ states $|\!\pm\pm\dots \pm\rangle$ of the radiation and $|\widetilde{\chi}_i\rangle$ are some states describing the black hole. We do not know the number of states $M$ describing the hole, but for any $M$ and any $\widetilde{C}_{ij}$ we can get a state of the box of gas (\ref{bone}) that is arbitrarily close to (\ref{btwo}). We do this by taking $N$ such that
\be
2^{N-n}>M \ .
\label{bthree}
\ee
Then since the space of $C_{ij}$ in (\ref{bone}) is ergodically explored, we can get the entanglement structure of (\ref{btwo}) by taking an $N$ satisfying (\ref{bthree}) and waiting for an appropriate time $T$ before opening the hole in the box. Thus we see that there can be nothing special about the entangled state of the radiation originating from a black hole; any form of the entanglement of this radiation can also be produced by radiation from an ordinary body, such as a box of gas. So we are back to our original question: if some combination of bits in the radiation and the hole can give rise to an effective semiclassical description at the horizon yielding the pair production (\ref{twopp}), then why should we not get a similar semiclassical horizon behavior for a radiating piece of coal or box of gas?

The answer is, of course, that we \emph{cannot} both, (i) take a model for the black hole in which it radiates like a piece of coal and, (ii) take some combination of bits from the radiation and the hole and make effective bits that describe a semiclassical approximation to the traditional black hole. Let us explore this issue in more detail, since it has been a source of confusion in the field.

\subsubsection{The kinematics of effective bits}

Let us, therefore, take a look at how we might make effective degrees of freedom/bits. First consider a piece of coal. Suppose this coal emits a photon which has two spin states denoted by $|0\rangle_b$ and $|1\rangle_b$.\footnote{We have labeled the photon states with a subscript $b$ in line with our earlier notation that the radiation quanta are called $b$.} In a typical emission (relatively early in the evaporation), this photon will be close to maximally entangled with the remaining coal. Thus there will be two orthogonal states $|\chi_1\rangle$ and $|\chi_2\rangle$ of the remaining coal such that the overall state of the radiation and coal is
\be
|\Psi\rangle\approx {1\over \sqrt{2}} \Big ( |\chi_1\rangle |0\rangle_b+|\chi_2\rangle |1\rangle_b \Big ) \ .
\ee
Suppose someone were now to suggest the following. If we define
\be
 |\chi_1\rangle\equiv |0\rangle_c\ , \quad |\chi_2\rangle\equiv |1\rangle_c \ ,
 \label{bfour}
 \ee
then surely we have the total state
 \be
|\Psi\rangle\approx {1\over \sqrt{2}} \Big ( |0\rangle_c |0\rangle_b+|1\rangle_c |1\rangle_b \Big ) \ .
\label{bfive}
\ee
This looks like the entangled pair (\ref{twopp}). So have we not shown that the surface of coal can actually be described by a semiclassical horizon through which objects will fall smoothly? The answer is no, of course not; if we shine photons on the coal they will scatter back off the surface of the coal, they will not pass smoothly through a horizon. So what is wrong with the map (\ref{bfour})? The answer is that we can always make a map like (\ref{bfour}) between \emph{states}, but what is important is the \emph{dynamics} of these states. Just relabelling the states in the manner (\ref{bfour}) to get (\ref{bfive}) does not mean that the coal has a smooth horizon, since the bit $c$ does not have the correct Hamiltonian interaction with the bit $b$ to generate a vacuum state at a horizon.

The same situation holds if we consider a photon which has been emitted past the halfway point of evaporation. Let the states of the photon again be denoted $|0\rangle_b$ and $|1\rangle_b$. This time the photon is close to maximally entangled with the early radiation. Thus there are two states $|\widetilde{\chi}_1\rangle$ and $|\widetilde{\chi}_2\rangle$ such that the entangled state of the photon is described as
\be
|\widetilde{\Psi}\rangle\approx {1\over \sqrt{2}} \Big ( |\widetilde{\chi}_1\rangle |0\rangle_b+|\widetilde{\chi}_2\rangle |1\rangle_b \Big ) \ .
\ee
Making the map
\be
 |\widetilde{\chi}_1\rangle\equiv |0\rangle_c \ , \quad |\widetilde{\chi}_2\rangle\equiv |1\rangle_c \ ,
 \label{bfourp}
 \ee
we get something of the form (\ref{bfive}). But again this does not mean that the piece of coal has a semiclassical horizon that photons will fall through; if we shine photons on the coal, they will scatter back.

The above observations are of course very obvious, but they are key to understanding why no effective semiclassical variables at the horizon can be made just by taking combinations of the radiation and black hole degrees of freedom.

\subsubsection{Using the dynamics of the bits at $r<10\,r_h$}\label{dynamics}

We have seen above that, (i) for a normal body like a piece of coal we cannot get the semiclassical dynamics of the horizon just by using appropriate combination of bits from the radiation and the remaining coal and that, (ii) any entangled state of the radiation bits can be reproduced to an arbitrarily good approximation by radiation from an ordinary body. We call the above attempts \emph{kinematical} in nature, since we have played with the states of the radiation and the coal, but have not used any dynamical information about how these bits interact. This part of the discussion has already been useful, since it tells us that purely kinematical attempts at getting semiclassical horizon behavior using the radiation and the remaining hole will not work.

\begin{figure}[h]
    \centering
    \includegraphics[scale=0.4]{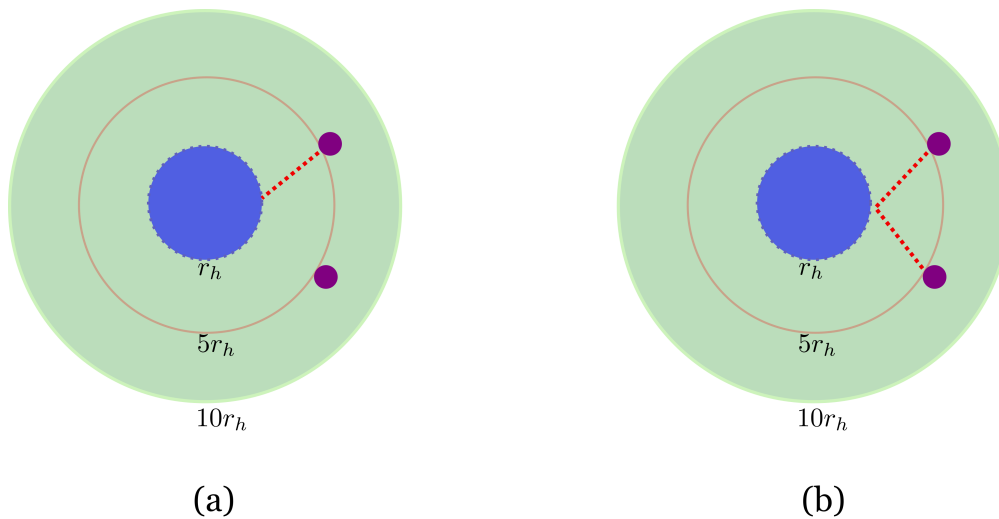}
    \caption{Experiment inside $10\,r_h$. In (a) we depict the semiclassical horizon, where a beam of photons (red) sent by an observer (purple) from $r=5\,r_h$ is absorbed at the horizon. In (b) we depict that if the horizon has structure, the beam sent can be collected by another observer at a different angular location of $r=5\,r_h$.}
    \label{exprh}
\end{figure}

Let us now consider the dynamics of the various degrees of freedom in the above discussion. First consider those of the radiation. Suppose the black hole emits $N\gg 1$ bits in its entire evaporation process. Then the energy of the typical quanta is small, much below the Planck scale. We also have an arbitrary amount of space to manipulate these radiation quanta; in particular we can increase their wavelengths so that they become just like the quanta emitted from a piece of coal. Therefore, we do not a priori have any nontrivial dynamics of these radiation quanta that would be different from the dynamics of the radiation quanta originating from coal. Let us then start by focusing on the dynamics of the degrees of freedom in the region $r<10\,r_h$. Here someone can say: we do not know the dynamics of a black hole, so these bits do not have to behave like the bits in a piece of coal. Perhaps we can then have the following situation:

\begin{enumerate}[start=1,
    labelindent=\parindent,
    leftmargin =2\parindent,
    label=(\roman*)]

\item \label{(i4)}
We make effective fields from all of the quanta in the region $r<10\, r_h$, as well as the radiation bits $b_i$. These effective fields describe low-energy dynamics around a semiclassical horizon; i.e. we have (\ref{twopp}) and (\ref{mtwo}) in a region $r_h<r<10\, r_h$ of this effective semiclassical geometry.

\item \label{(ii4)}
We check the fact that we have an effective semiclassical description in the region $r_h<r<10\,r_h$ as follows. We send a beam of photons from a source at $r=5\, r_h$ and check for a scattered beam at a different angular location, again at $r=5\, r_h$ (we depict this in fig.~\ref{exprh}). With the photon beam being a coherent wave with wavelength $\lambda\sim r_h$, if the horizon is the semiclassical one then a large part of this beam will be absorbed, while if we have structure at the horizon then there will be a strong scattered beam.
\end{enumerate}
Let us now see what the consequences are of such a claim. For the effective semiclassical horizon, the effective bits near this horizon are entangled in a state of the form 
\be
|\Psi\rangle_1={1\over \sqrt{2}}\Big (|0\rangle_{b,ef\!f}|0\rangle_{c,ef\!f}+|1\rangle_{b,ef\!f}|1\rangle_{c,ef\!f}\Big ) \ .
\ee
Now we modify the spins of the quanta $b_i$ at infinity whilst keeping them in their low-energy state, where they have negligible interactions between each other. Since the effective bits at the horizon involve the $b_i$ at infinity, there has to be some modification of the state $|\Psi\rangle_1$ of the effective bits at the horizon. For the sake of argument, suppose this manipulation of the $b_i$ changes the state of the effective bits at the horizon to 
\be
|\Psi\rangle_2={1\over \sqrt{2}}\Big (|0\rangle_{b,ef\!f}|1\rangle_{c,ef\!f}+|1\rangle_{b,ef\!f}|0\rangle_{c,ef\!f}\Big ) \ .
\ee
The state $|\Psi\rangle_2$ does \emph{not } give the local semiclassical vacuum at the horizon and the incident beam we send towards the horizon will scatter and be picked up by the detector. It does not matter of course what $|\Psi\rangle_2$ is, all that matters is that there will have to be \emph{some} manipulation of the $b_i$ which will change the state $|\Psi\rangle_1$ to some other state. This is because only the state $|\Psi\rangle_1$ gives the local vacuum at the horizon through which objects all through smoothly. Thus we conclude the following:\\
\\
\emph{If we accept \ref{(i4)} and \ref{(ii4)} above, then by manipulating the radiation bits $b_i$ at infinity, while keeping them in a low-energy state, we will change the observations that an experimenter at, say, $r=5\,r_h$ will make about the hole.}\\
\\
We consider one more aspect of this argument. Suppose someone were to dispute the above conclusion in the following way. We have changed the state of the $b_i$ to change the state at the horizon, but the apparatus we used to detect the nature of the horizon was at $r=5\,r_h$ and this apparatus can also be considered to be made of the effective bits that are used for the low-energy semiclassical dynamics around the horizon. Could it be that when we change the state of the radiation quanta $b_i$, we change the state at the horizon \emph{and} the state of the measuring apparatus, so that overall the observations of the experimenter appear to be unchanged?

We can, however, easily see that such a line of argument cannot succeed. Consider the Hamiltonian $H$ describing the horizon region, including the apparatus at $r=5\,r_h$ and let $|E_n\rangle$ be the eigenstates of this Hamiltonian. Suppose we start with an eigenstate $|E_1\rangle$ describing the smooth horizon along with a particular state of the apparatus; this corresponds to some states of the (exact) bits in the region $r<10\,r_h$ and the radiation bits $b_i$. Now we change the state of the $b_i$. Since the $b_i$ are involved in the construction of the state around the horizon, there has to be some change of the $b_i$ which will change the state around the horizon. Suppose the new state is
\be
|\Psi\rangle=\alpha_1 |E_1\rangle +\alpha_2|E_2\rangle +\cdots\ne |E_1\rangle \ .
\ee
The change from $|E_1\rangle$ to $|\Psi\rangle$ has to be detectable by local observations in the region around the horizon. Thus we are back to our conclusion above, that manipulating the bits $b_i$ at infinity will change the observations of an experimenter near the horizon. Such a change does not happen for a piece of coal: manipulating the radiation quanta will \emph{not} change the observations of an experimenter near the coal who is scattering photons off the coal. Thus we would reach the following conclusion:\\
\\
\emph{There are two different kinds of photons $b_i$ at infinity, those emitted by normal objects and those emitted by black holes. Manipulating the state of the photons from coal will not affect any observations in the interior region containing the coal, but manipulating the state of the photons from a black hole will change the observations in the interior region containing the hole.}

\subsubsection{Using dynamics of the radiation bits at infinity}

In section~\ref{dynamics} we have assumed that the radiation bits $b_i$ were low-energy quanta; this is the case for the typical quantum that is radiated by a large black hole. But we can consider the situation where we squeeze these bits $b_i$ into a small region so that the new state is $|\Psi\rangle_b$. Now the $b_i$ may interact with each other and we cannot just treat their state kinematically as we did above. To get any dynamics different from the case of coal we have to squeeze the $b_i$ to a state which is not a state of normal matter. For the purposes of our discussion, we will call this state a `black hole state'. 

All the steps in our analysis from section~\ref{dynamics} remain unchanged. We again conclude that if we squeeze the $b_i$ into a different state $|\Psi'\rangle_b$, then this difference can be detected by observations of the hole performed by an experimenter sitting outside the hole at, say, $r=5\,r_h$. The only difference compared with before is that this time we can argue that the interactions between the $b_i$ that have been squeezed together can generate effects that are not present when the $b_i$ are low energy, well separated quanta. The quanta $b_i$ are far separated from the region $r<10\,r_h$, so how will any interactions between the squeezed $b_i$ manage to affect the dynamics around the hole? The only way this can happen is if we postulate that a shorter path opens up between the radiation and the region hole, i.e. a wormhole. Thus we have the following picture:\\
\\
\emph{Suppose we collapse the radiation degrees of freedom $b_i$ into a black hole state $|\Psi\rangle_b$. Suppose further that we can use the bits in $r<10\,r_h$ and the bits $b_i$ in this state $|\Psi\rangle_b$ to get effective variables that yield the semiclassical vacuum at the horizon. Then if we collapse the bits $b_i$ to a different state of the black hole $|\Psi'\rangle_b$, this difference will change the observations of an experimenter at $r=5\,r_h$ who is scattering a photon beam off the hole.}

\subsubsection{Summary}

There have been attempts to have the following picture of the black hole: (i) the hole radiating like a piece of coal as seen from outside and, (ii) using some combination of bits in the radiation and the bits in the region $r<10\,r_h$ to get `effective bits', in terms of which one sees semiclassical low-energy dynamics at the horizon. We have seen that this set of goals cannot be met:

\begin{enumerate}[start=1,
    labelindent=\parindent,
    leftmargin =2\parindent,
    label=(\alph*)]

\item \label{(a5)}
Suppose we just use the kinematical properties of the bits, i.e. we just use different combinations of bits in the radiation and in the hole to get our effective bits. Then these effective bits cannot reproduce semiclassical horizon dynamics. The reason is that every entangled state of the hole and the radiation can be reproduced to an arbitrary approximation by radiation from a box of gas, for which we should not find a smooth horizon. 
 
\item \label{(b5)}
We can try to bypass the above conclusion by saying that the bits in the hole have a dynamics that is special and that this dynamics can somehow enable a semiclassical horizon in terms of the effective bits. But then we find that there are two kinds of photons at infinity; those radiated by a piece of coal and those radiated by a black hole. Manipulating the former does not change the observations that an experimenter may make in the vicinity of the black hole, while manipulating the latter \emph{will} change these observations.

\item \label{(c5)}
We can try to bypass the conclusion in \ref{(b5)} by arguing that the effective semiclassical behavior at the horizon only arises when we squeeze the radiation quanta to a small scale, where novel physical effects start; we can consider this process as similar to collapsing the radiation quanta to a black hole. But collapsing the radiation to a different state of the hole will destroy the semiclassical behavior at the horizon of the original hole. We can picture such models as saying that a wormhole opens up between the original hole and the radiation that has been collapsed to form a second hole. Models of this type have been suggested by Maldacena \cite{MaldacenaSaclayComment1}. 

\item \label{(d5)}
We note that in the cases \ref{(b5)} and \ref{(c5)} the radiation does not behave like the radiation from coal; manipulating the radiation from coal has no effect on observations near the coal. Thus overall we find no model where the black hole radiates like a piece of coal as seen from outside and yet an effective semiclassical horizon dynamics can be obtained using some combination of bits in the radiation and in the hole. 
\end{enumerate}

\subsection{The experiment}\label{experiment}

In the above section we have seen some ways that nonlocal effects have been postulated for black holes. To understand the role of these nonlocal effects on the entanglement entropy, we set up the gedanken experiment that answers the basic question of the information paradox: is the state of the radiation pure or mixed (i.e. whether or not the total state was entangled or not)?

Consider a black hole $B$ that radiates away, so that we are left with a collection of radiation quanta $R$. We wish to know if $R$ is in a pure or mixed state. How will we check which of these is the case? This is the relevant question to ask at this point, since the following has been an argument relevant to the wormhole paradigm \cite{MaldacenaStringsComment1}: to check the purity of the radiation $R$ we need to take many identical instances of an experiment where the black hole is formed and allowed to evaporate. Normally we would assume that these different instances of the experiment can be well separated in space or time (or both), so we do not need to think of any interaction between them. However, in one approach to the wormhole paradigm it is argued that these different instances of the experiment \emph{will} interact with each other. It has also been suggested that such nonlocal effects should lead to the prescriptions like (\ref{mfive}). To understand what this means, we recall what measurements we need to do to check for entanglement. 

Before we come to the black hole, we start with a simpler case. Consider a system of just two spins and take these spins to be well separated; we can call one spin $B$ and the other $R$. The overall system $B\cup R$ is assumed to be in a pure state. We wish to know whether $R$ by itself is in a pure state or if it is in a mixed state (and thus entangled with $B$). An example of a pure state for $R$ is one where the overall state $|\Psi\rangle$ of $B\cup R$ is 
\be
|\Psi\rangle_{factorized}=\bigg({1\over \sqrt{2}} \Big(|\uparrow\rangle_B -|\downarrow\rangle_B\Big) \bigg) \bigg({1\over \sqrt{2}} \Big(|\uparrow\rangle_R +|\downarrow\rangle_R\Big) \bigg) \ ,
\label{deone}
\ee
and an example of an entangled state on $B\cup R$ is
\be
|\Psi\rangle_{entangled}={1\over \sqrt{2}}\Big( |\uparrow\rangle_B |\downarrow\rangle_R - |\downarrow\rangle_B |\uparrow\rangle_R  \Big) \ .
\label{detwo}
\ee
For our experiment, we have access to the spin $R$ but not to the spin $B$. We also assume that we have access to several identically prepared copies of the state $|\Psi\rangle$ of the system $B\cup R$. How should we check if the state of $R$ is pure or mixed?

We proceed as follows. We pass the spin $R$ through a Stern-Gerlach apparatus oriented in the $z$ direction and see if the the spin comes out (deflected by the magnetic field) in the upper or lower path. We repeat this process with several instances of the identically prepared state. For each of the two example possibilities (\ref{deone}) and (\ref{detwo}), we will find that a fraction $\h$ of the time the spin $R$ will be in $+z$ direction and $\h$ of the time it will be in the $-z$ direction. Therefore, so far we have not been able to learn if the state of $R$ is pure or mixed. Now we try other orientations of the Stern-Gerlach experiment. Suppose we orient the apparatus along the $x$ direction. Then for the entangled singlet state (\ref{detwo}) we will still find that $\h$ of the time the spin $R$ emerges along the $+x$ direction and $\h$ of the time in the $-x$ direction. For the pure state (\ref{deone}), however, we would find that the spin is along the $+x$ direction \emph{every} time. We would thus conclude that in (\ref{deone}) the spin $R$ is in a pure state (and we would have found this state as being along the $+x$ direction). Similarly, we would conclude that in (\ref{detwo}) the spin $R$ is maximally entangled with $B$, since all orientations of the Stern-Gerlach give the same probabilities $\pm \h$ of emerging in the two branches.

We thus see that given many identically prepared copies of a system in quantum mechanics, we can do experiments on one part ($R$) to check if this part is entangled or not with the remainder of the system ($B$). In a similar manner, we can check the purity of radiation from a piece of coal. In the case of the coal, $R$ will consist of $n\gg1$ spins which generate a ${\mathcal N}=2^n$-dimensional Hilbert space. To check the purity of $R$ we will need to measure a large number of identically prepared copies of $R$; this number will typically be some power of ${\mathcal N}$. Such a measurement may seem complicated, but we are talking here as a matter of principle so it does not matter how many times we need to repeat the experiment. To summarize, for the radiation from any piece of coal we are in a position to check the purity of $R$ (when the coal has full evaporated) by measurements performed on a suitably large number of identically prepared copies of the system.\footnote{If we look at only a fraction of the $n$ spins that make up the radiation $R$, then we will not be able to determine if $R$ is pure or not. Thus measurements that look at just a fraction of the spins do not have any bearing on the information paradox.}

Let us now set up the experiment with black holes that will check the purity of Hawking radiation. We collapse a star to create a black hole with mass $M\gg m_p$. We let this hole evaporate to radiation $R$. We detect the state of this radiation by a complicated set of Stern-Gerlach apparatii placed far from the hole. We can repeat this experiment in an identical way as many times as we wish, so that we have many identically prepared states of the system. The question now is: will the measurements show that the radiation $R$ is entangled or not? Let us contrast the situation between the fuzzball and wormhole paradigms:

\begin{enumerate}[start=1,
    labelindent=\parindent,
    leftmargin =2\parindent,
    label=(\alph*)]

\item \label{(a6)}
\textbf{The fuzzball paradigm:} Here the collapse of the star creates an object that is just like a piece of coal. There are no effective variables where we get any approximation to pair creation like (\ref{twopp}). The region far from the hole has `normal' physics; i.e.  

\begin{enumerate}[start=1,
    labelindent=\parindent,
    leftmargin =2\parindent,
    label=(\roman*)]
    
    \item \label{(i5)} 
    Degrees of freedom far from the hole are independent of degrees of freedom in the hole.

    \item \label{(ii5)}
    There are no long-distance nonlocal effects in the region far from the black hole.
\end{enumerate}
The Page curve has the form of that of a normal body and the state of the radiation $R$ at the end of the evaporation is pure. This purity will be manifested in the above experiment by the measurements done by the Stern-Gerlach apparatii. Different instances of the experiment can be taken to be well separated in space and/or time and there will be no interaction between the bits describing these different instances of the experiment.

\item \label{(b6)}
\textbf{The wormhole paradigm:} This time we require that there be effective variables which yield an approximation to semiclassical physics at least for the description of a few created pairs at a time:
\be
|\psi_{ef\!f}\rangle_{pair}={1\over \sqrt{2}}\Big (|0\rangle_{b,ef\!f}|0\rangle_{c,ef\!f}+|1\rangle_{b,ef\!f}|1\rangle_{c,ef\!f}\Big ) +O(\epsilon) \ .
\label{twoppqq}
\ee
Now the effective small corrections theorem severely limits one's choices. If we want the hole to radiate like a piece of coal; i.e. to satisfy the conditions \ref{(i5)} and \ref{(ii5)} listed in \ref{(a6)} above, then the Page curve will not come down. Suppose we \emph{do} want the Page curve to come down at the end of evaporation. Then we have to violate at least one of the conditions \ref{(i5)}, \ref{(ii5)}. At the end of section~\ref{nonlocalchoices} we listed three different kinds of nonlocality that can be proposed. We will now look at each of these proposal one by one.
\end{enumerate}

\subsection{Nonlocal effects between black hole interiors: baby universes}\label{seca}

Suppose we say that the region near asymptotic infinity has `normal' physics. This means that degrees of freedom at infinity are independent of the degrees of freedom in the black hole region and also that there are no long-distance nonlocal interactions in this region, far from the hole. We can still say that the interior of black holes is a novel kind of region and so new postulates can be made for the dynamics of this interior region. 
 
We will investigate postulates of the following kind. When the black hole evaporates, its interior detaches from the parent spacetime as a `baby universe'. Thus this baby universe contains the initial matter that fell in to make the hole, as well as any negative energy members $\{c_i\}$ of the Hawking pairs that fell into the hole later. 
 
If we stop at this point, then we just have one version of the well-known remnant scenario, where the remnant here takes the form of a baby universe. In such a remnant scenario, the radiation $R$ is not in a pure state; it is entangled with the matter in the baby universe and this mixed nature of $R$ will manifest itself in the experiment described in section~\ref{experiment}. But we now extend the argument further. We have seen that to determine whether $R$ is pure or not, we have to take many instances of the experiment. Let the number of these instances be $K$, then each of these instances produces a baby universe. We argue that once the baby universe has detached from the parent spacetime, it does not have any memory about which instance of the experiment it came from. Thus if two of the $K$ baby universes are in the same state, we should treat these systems as if they were `identical particles'. This fact introduces a relation between the different instances of the experiment. There was no such relation between different instances when we were looking at the evaporation of a piece of coal. It is then argued that this novel aspect of black hole dynamics with baby universes may help resolve the information paradox. In fact this relation between different instances of the experiment looks somewhat similar to the prescription (\ref{dstwo}) where one seeks to link different replica copies. Models using baby universes in this kind of manner have been considered, for instance, in \cite{Marolf:2020xie,Marolf:2020rpm}.
 
As we will see below, such dynamics using baby universes violates unitarity of evolution in the black hole interior. We will see this violation explicitly, but we first note why such a loss of unitarity is naturally expected by what we know from the effective small corrections theorem. Consider the $K$ different instances of the experiment as being carried out at widely separated locations. Define the total black hole region as the union of the regions $r\lesssim 10\,r_h$ around each hole and the far region to be the union of the remainder of the spacetimes. Let $S_{ent}^{total}$ be the total entanglement entropy between the total interior region and the far region. 
 
We can now immediately extend the effective small corrections theorem to this situation with $K$ holes. Suppose that around the horizon of each hole we have effective variables in which we see semiclassical dynamics at least for the creation of a few pairs in the state
\be
|\psi_{ef\!f}\rangle_{pair}={1\over \sqrt{2}}\Big (|0\rangle_{b,eff}|0\rangle_{c,eff}+|1\rangle_{b,eff}|1\rangle_{c,eff}\Big ) +O(\epsilon) \ .
\label{twoppqqq}
\ee
Note that the baby universe model described above has the assumption that physics is `normal' in the region far from the hole -- we are only changing the dynamics in the interior regions of the $K$ holes. We then have all the conditions needed for the effective small corrections theorem, now for the case where we have $K$ black holes. The theorem will tell us that the entanglement of the exterior region with the total interior region will grow as
\be
S^{total}_{N+1}> S^{total}_N+K\ln 2 -K(\epsilon_1+\epsilon_2) \ .
\label{dsixt2}
\ee
To see that (\ref{dsixt2}) holds, one just has to repeat the steps in the derivation of the effective small corrections theorem listed in section~\ref{proof}. In particular, note that all we need to prove (\ref{dsixt2}) is that the evolution in the \emph{union} of the $K$ black hole regions be unitary. If bits leave from the interior of one black hole and join the interior of another black hole, then this will not affect the derivation of (\ref{dsixt2}). 
 
Note that processes like topology change do not by themselves affect the derivation of the (effective) small corrections theorem, as long as these processes preserve the unitarity of evolution in the black hole interior. We had noted this fact in the discussion of section~\ref{topsmall}. The creation of a baby universe is a particular case of such a breaking of the slice, where the segment containing all the $\{c\}$ quanta breaks off as a baby universe at the endpoint of evaporation. If this creation of the baby universe was a unitary process, then the effective small corrections theorem would yield (\ref{dsixt2}) and the Page curve would not come down. However, we will now see that if baby universes with the same content behave like `identical particles', then there is a violation of unitarity in the evolution. To see this violation of unitarity, consider the following simple model that illustrates the essential issue.\footnote{To present the essential idea, we let the baby universe contain just one bit, but the issue remains the same when we let the baby universe contain $\sim S_{bek}$ bits as we expect it to.}   
\begin{figure}[H]
\begin{center}
\includegraphics[scale=0.4]{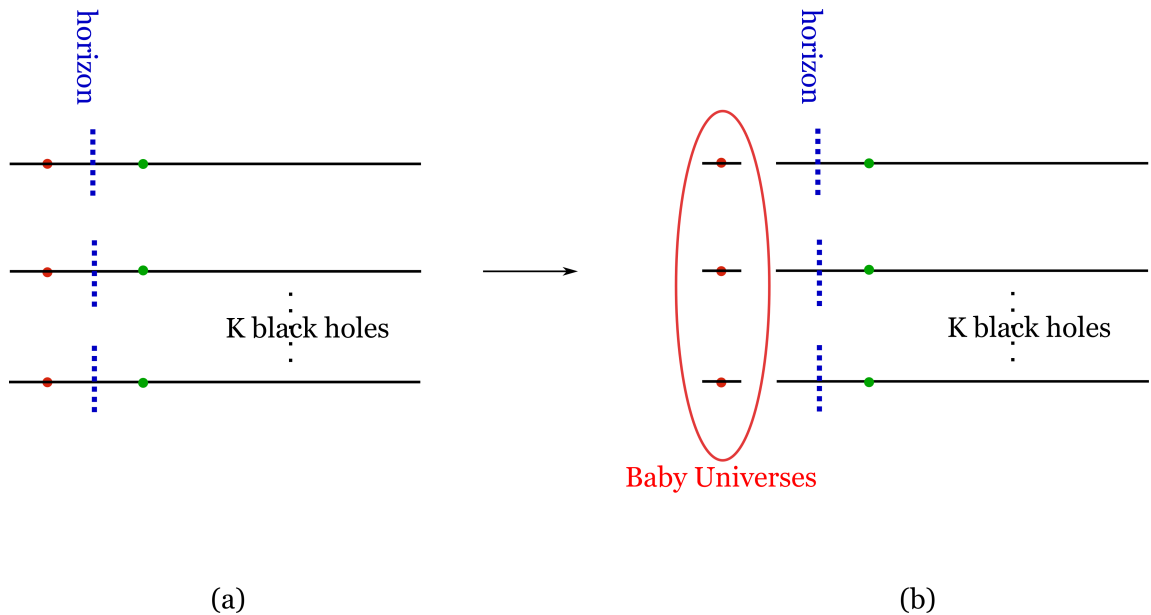}\hskip20pt
\end{center}
\caption{Non-unitarity in baby universes. In (a) we see spatial slices of $K$ different black holes with the region to the left of the horizon representing the black hole interior. In (b) a part of the interior breaks away and forms a baby universe.}
\label{fig10}
\end{figure}

In fig.~\ref{fig10}(a) we depict $K$ black holes, all well separated. Now we consider the following steps:

\begin{enumerate}[start=1,
    labelindent=\parindent,
    leftmargin =2.5\parindent]
\item \label{(11)}
An entangled pair is created at the horizon of each hole in the state
\be
|\psi\rangle_{pair}={1\over \sqrt{2}}\Big (|0\rangle_b|0\rangle_c+|1\rangle_b|1\rangle_c\Big ) \ .
\label{pairqqqq}
\ee
These quanta lead to an entanglement between the region outside the horizons and the region inside the horizons equal to
\be
K\log 2 \ .
\ee
We can understand this value as follows. The $K$ quanta of type $b$ outside the horizons have $2^K$ possible states and they are maximally entangled with the 
\be
N_1=2^K \ ,
\ee
states of the $K$ quanta of type $c$ inside the horizons.

\item \label{(22)}
In this step, we break off the interior regions of the slices. This generates $K$ different `baby universes', each of which contains a $c$ quantum. The state of this $c$ quantum can be $0$ or $1$. Two baby universes with the same state of the $c$ quantum will be in the same state and we treat them as identical bosonic particles. Then, the number of possible states of the $K$ baby universes is computed as follows. The number of baby universes where the $c$ quantum is $0$ can be $0, 1, \dots, K$ so there are $K+1$ possible choices. Thus the number of possible states of the $K$ baby universes is
\be
 N_2=K+1 \ .
 \ee
\end{enumerate}
Now we see the problem: the $N_1$ states at step \ref{(11)} have been mapped to $N_2$ states at step \ref{(22)}. But
\be
N_2<N_1 \ ,
\ee
for $K>1$. Thus the evolution `kills' some states, which means that it is not unitary. We took a very simple example above to show the essence of the problem, but the same problem arises when we use effective bits $|b\rangle_{ef\!f}, |c\rangle_{ef\!f}$ instead of $|b\rangle, |c\rangle$ and if we take more complicated rules for splitting off baby universes.
 
To summarize, we already knew from the effective small corrections theorem that the Page curve cannot come down in any model that manifests effective semiclassical dynamics (\ref{pairqqqq}) at the horizon, assumes infinity is `normal' and requires evolution in the black hole interior to be unitary. In the baby universe model described above one did want the Page curve to come down, so one must give up on one of the other conditions. From the analysis of the model we see that the model violates unitarity of evolution in the union of the black hole interiors.

\subsection{Nonlocal effects between different regions near spatial infinity}\label{secb}
 
\begin{figure}[h]
     \centering
     \includegraphics[scale=0.6]{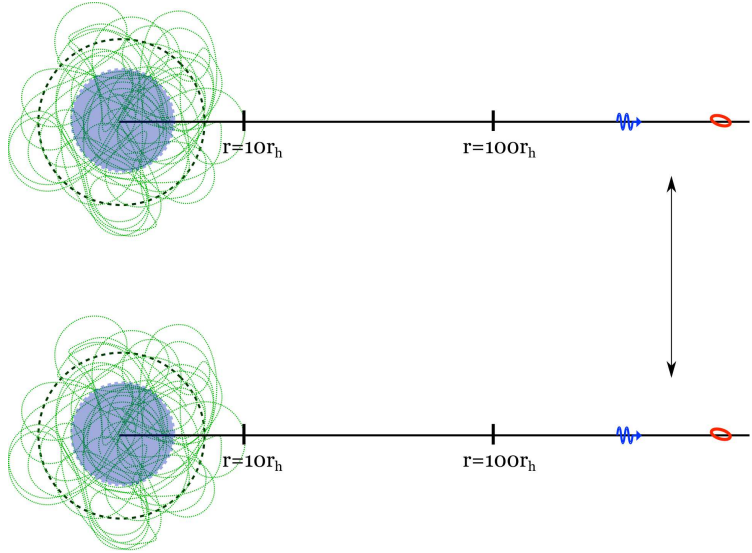}
     \caption{Radiation of different regions near spatial infinity separated by very long distances. The arrow shows the interfering such experiments.}
     \label{infexp}
 \end{figure}

Let us now consider the possibility that there are nonlocal interactions between the radiation sets produced by different black holes. The idea behind postulating such an interaction is the following. Our question is whether the radiation $R$ produced by a black hole is in a pure state or not. To check the purity of $R$, we need to take many instances of the experiment where we produce and evaporate identical black holes. If the radiation $R_a$ from these different experiments (labeled by $a$) interfere with each other in some way (fig. \ref{infexp}), then this fact would impact the standard method for checking entanglement. This in turn might offer some way out of the problem of the monotonically growing Page curve.
But we immediately notice an important feature that emerges from any such line of reasoning. It is true that the number of instances $N_{instances}$ of the black hole experiment we need will be  large, if the radiation $R$ contains a large number of quanta. But this number $N_{instances}$ is still a finite number. On the other hand, the \emph{separation} $D$ between the different instances  can be taken to be arbitrarily large. Thus for sufficiently large $D$, different instances of the experiment will not be near each other; rather, they will be separated by an arbitrarily large distance. Thus whatever interaction we postulate between different instances $R_a$ of the radiation, must be an interaction that does not fall off with the distance between the radiation sets.

If there were such an interaction in nature, would we not have detected it in some other way? To escape this difficulty, we may postulate that this nonlocal interaction arises only between quanta that have been radiated by black holes and not for example to quanta that have been radiated by a piece of coal. This possibility would in turn imply that there are two kinds of photons at infinity: those that have been radiated by black holes and those that have been radiated by coal. 
 
One may try to argue that the radiation from a black hole is very complicated, so that it would be hard to detect the nonlocal interactions that one has postulated. But black hole radiation is not more complicated than the radiation from a piece of coal. The only relevant microscopic scale for black hole physics is the Planck scale. A black hole with $M\sim 100 m_p$ should manifest the general physics expected from a large black hole, since $100\gg 1$. Such a hole will emit $\sim 10^4$ radiation quanta. On the other hand, a piece of coal radiates $\sim 10^{23}$ photons. In any case we are asking a question of principle here and it is immaterial how difficult it is to actually measure the nonlocal interaction that has been postulated.

\subsection{Nonlocal effects between the hole and its radiation}\label{secc}

We begin in section~\ref{holography} with a discussion of the idea that a holographic approach can help resolve the paradox. Here the exact theory is described by some holographic bits at infinity, while the spacetime bits are approximate effective bits, which need not be exactly independent between the hole and the radiation. One might then think that this non-factorization of degrees of freedom between the hole and infinity might help resolve the information puzzle. But we will see that there is a difficulty with this: the bit at infinity has to behave like a `normal' bit to accord with experiments and then we are back to the monotonically rising Page curve.
 
Then we look for possible models where nonlocal effects between the hole and the radiation could have an impact on the Page curve. As noted in section~\ref{secintro}, there are two kinds of models in this category. The first is where we have nonlocal Hamiltonian interactions between the radiation and the hole. We give a bit model of this type in section~\ref{secnonlocal}. The second approach is where we try to identify bits between infinity and the hole. We give a model that attempts to get such an identification in section~\ref{identify}.

The Hawking process produces entangled pairs. We have previously written these pairs using states $|0\rangle_b, |1\rangle_b$ and $|0\rangle_c, |1\rangle_c$. In this section it will be more convenient to map these states to states of an internal spin degree of freedom, which for convenience we call isospin. We let the map be as follows:
\be
|0\rangle_b ~\r~ |\uparrow\rangle_b \ , \quad |1\rangle_b ~\r~ |\downarrow \rangle_b \ , \quad |0\rangle_c ~\r~ |\downarrow\rangle_c\ , \quad |1\rangle_c ~\r~ -|\uparrow
\rangle_c \ ,
\ee
so that the entangled pair (\ref{pairqqqq}) transforms to a spin singlet
\be
{1\over \sqrt{2}}\Big( |0\rangle_b|0\rangle_c+|1\rangle_b|1\rangle_c\Big) ~\r ~ 
{1\over \sqrt{2}}\Big( |\uparrow\rangle_b|\downarrow\rangle_c-|\downarrow\rangle_b|\uparrow\rangle_c\Big) \ .
\ee

\subsubsection{The difficulty with invoking holography}\label{holography}

Suppose someone makes the following argument. The exact theory is described in a holographic way. Since we are in asymptotically flat space, we are assuming that some form of flat space holography exists; let us make this assumption. Then we say that spacetime itself arises as some approximate construct made out of these holographic bits. In this spacetime let us look at the bits $b, c$ that are involved in the semiclassical picture of the black hole. Since spacetime emerges only as an approximation, it may well be that the exact bits used to make $b$ are not fully independent of the exact bits that are used to make $c$. The small corrections theorem assumes that once the bit $b$ is far from the hole, then it is made of degrees of freedom that are independent of the degrees of freedom making up the bit $c$. One would then argue that if $b$ and $c$ are not made of strictly independent degrees of freedom when we write them out in terms of the exact holographic bits, then there may be a possibility that the small correction theorem is bypassed and the Page curve somehow comes down.

But there is a difficulty with any argument of this kind. We have to begin by asking: what is the meaning of saying that the spacetime bits $b$ and $c$ are not exactly independent of each other? In our spacetime analysis, we assumed that the bit $b$ has two states; let us call these $|\uparrow\rangle_b$ and $|\downarrow\rangle_b$. The bit $c$ also has two states $|\uparrow\rangle_c$ and $|\downarrow\rangle_c$. Let the exact holographic states at infinity be $|k\rangle_H$, where $H$ denotes the fact that these are the holographic bits. We must then write 
\be
|\uparrow\rangle_b=\sum_k C_{1k}|k\rangle_H\ , \quad|\downarrow\rangle_b=\sum_k C_{2k}|k\rangle_H\ , \quad |\uparrow\rangle_c=\sum_k C_{3k}|k\rangle_H\ , \quad |\downarrow\rangle_c=\sum_k C_{4k}|k\rangle_H \ .
\label{ccone}
\ee
Now let us ask again: what is the meaning of saying that $b$ and $c$ are not exactly independent? The \emph{dimension} of the space formed by $b$ and $c$ in our spacetime picture was $2\times 2=4$. Regardless of how we make the bits in a holographic picture, this dimension cannot become a slightly smaller number $3.9$; it must remain $4$ if we are to be close to the semiclassical picture. So (\ref{ccone}) must describe $4$ linearly independent combinations of the holographic bits $|k\rangle_H$.

But then what is the meaning of $b$ and $c$ not being independent? All we can say is that the linear combinations (\ref{ccone}) may not be exactly \emph{orthogonal}, though they are orthogonal in the usual spacetime picture of quanta. This idea of altering orthogonality runs into a problem with observations. The bit $b$ moves off to infinity, while the bit $c$ stays in the vicinity of the hole. We can now do experiments on the bit $b$ to see if the states in (\ref{ccone}) are orthogonal or not. Suppose we start with the state 
\be
|\uparrow\rangle_b|\uparrow\rangle_c \ ,
\ee
and apply a magnetic field for the appropriate amount of time to $b$ in order to rotate its spin to $|\downarrow\rangle_b$. Then we reach the state
\be
|\downarrow\rangle_b|\uparrow\rangle_c \ .
\ee
Now perform a measurement on $b$ to check if it in the $|\uparrow\rangle_b$ state. In normal quantum theory, the probability to find $b$ in this state is zero because
\be
\Big({}_c\langle \uparrow| {}_b\langle \uparrow|\Big)\Big(|\downarrow_b\rangle|\uparrow_c\rangle\Big)=0 \ ,
\ee
but if we change the dot products between the $4$ states spanned by $b$ and $c$ then we can have
\be
\Big({}_c\langle \uparrow| {}_b\langle \uparrow|\Big)\Big(|\downarrow_b\rangle|\uparrow_c\rangle\Big)=\epsilon\ne 0 \ .
\ee
This will then force one to the conclusion that photons radiated from a black hole behave differently at infinity than photons radiated from coal, as measured by experiments at infinity. The general problem with any picture which tries to say that the true nature of bits is holographic and that spacetime is only an approximate, emergent construct is outlined as follows:

\begin{enumerate}[start=1,
    labelindent=\parindent,
    leftmargin =2\parindent,
    label = (\alph*)]
\item \label{(a9)}
One might for instance imagine that spacetime is built of a tensor network. There is nothing wrong with a tensor network, just as there is nothing wrong with modeling spacetime as a cubical lattice of points with field variables on them.

\item \label{(b9)}
But for any such model, we must insist that the normal lab physics must emerge for low-energy variables at infinity. 

\item \label{(c9)}
Once we require this, it does not matter how we model our spacetime, the conclusion is the same:
the bits at infinity \emph{have} to behave as degrees of freedom that are independent of those in the hole. 

\item \label{(d9)}
It is crucial that the hole has finite mass and so emits a finite number of particles, while space itself is \emph{infinite}; thus we can take the finite number of emitted quanta as far away from the hole and from each other as we want, and then analyze them by standard lab apparatus at leisure. Since we have all the space and time to make measurements, we can require that these quanta behave as normal quanta to \emph{arbitrary} accuracy; in particular, they cannot be made of degrees of freedom that are dependent of degrees of freedom elsewhere in spacetime.
\end{enumerate}
To us it appears that this leaves no room for resolving the puzzle by saying that the exact bits are holographic while the spacetime bits are approximate and not really independent between the hole and infinity.

\subsubsection{Using small nonlocal interactions}\label{secnonlocal}

Suppose that at the horizon we create Hawking pairs in the usual way in the state
\be
|\psi\rangle_{pair}={1\over \sqrt{2}} \Big(|\uparrow\rangle_b |\downarrow\rangle_c- |\downarrow\rangle_b |\uparrow\rangle_c\Big) \ .
\label{dnone}
\ee
This is an entangled state, with an entanglement entropy of $\log 2$ between $b$ and $c$. An example of a nonentangled state would, for example, be simply
\be
|\widetilde{\psi}\rangle_{pair}= |\uparrow\rangle_b |\downarrow\rangle_c \ .
\ee
We make a model where the entangled state $|\psi\rangle_{pair}$ changes to the nonentangled state $|\widetilde{\psi}\rangle_{pair}$ \emph{gradually}, so that the effect is difficult to detect if we are looking at the $b$ quantum for a time that is short compared to the Hawking evaporation time. Let
\be
\sigma^+=\h\big(\sigma^1+i\sigma^2\big)\ , \quad\sigma^-=\h\big(\sigma^1-i\sigma^2\big) \ ,
\ee
and consider the operator
\be
\hat{O}=\sigma^-_b\sigma^+_c- \sigma^+_b\sigma^-_c \ .
\ee
We see that
\be
\hat O \,  |\uparrow\rangle_b |\downarrow\rangle_c=|\downarrow\rangle_b |\uparrow\rangle_c \ , \quad \hat O \,  |\downarrow\rangle_b |\uparrow\rangle_c=-|\uparrow\rangle_b |\downarrow\rangle_c \ ,\quad \hat O \,  |\uparrow\rangle_b |\uparrow\rangle_c=0\ , \quad \hat O \,  |\downarrow\rangle_b |\downarrow\rangle_c=0 \ .
\ee
For small $\epsilon$, we find
\be
e^{\epsilon\hat O}|\psi\rangle_{pair}={1\over \sqrt{2}} \Big((1+\epsilon) |\uparrow\rangle_b |\downarrow\rangle_c- (1-\epsilon)|\downarrow\rangle_b |\uparrow\rangle_c\Big ) \ ,
\ee
so that the state $|\psi\rangle_{pair}$ rotates slightly towards the nonentangled state $|\widetilde{\psi}\rangle_{pair}$. Our model will use the nonlocal interaction generated by $\hat O $ between the hole and its radiation to bring the Page curve down at the end of Hawking evaporation.

Let the total number of pairs created by the black hole be $N$. Consider the $k$th pair which is created in the state
\be
|\psi\rangle^{(k)}_{pair}={1\over \sqrt{2}} \Big (  |\uparrow\rangle_{b_k} |\downarrow\rangle_{c_k}- |\downarrow\rangle_{b_k} |\uparrow\rangle_{c_k}\Big ) \ .
\ee
There are $N-k$ further steps in the evaporation process after the creation of this pair. In each of these steps we assume that the state of the $b_k, c_k$ is modified by the action of the operator
\be
e^{{1\over N-k}{\pi\over 4}\hat O_k}\quad \text{with} \quad \hat O_k=\sigma^-_{b_k}\sigma^+_{c_k}- \sigma^+_{b_k}\sigma^-_{c_k} \ .
\ee
At the end of the evaporation process, the state of the $b_k$ and $c_k$ pair is
\be
e^{{\pi\over 4}\hat O_k}|\psi\rangle^{(k)}_{pair}= |\uparrow\rangle_{b_k} |\downarrow\rangle_{c_k} \ ,
\ee
such that $b_k$ and $c_k$ are in a nonentangled state. 

The above expressions describe the effect of our nonlocal interaction on one entangled pair. Let us now write down the nonlocal interaction that acts on all pairs that are produced in the evaporation process. After step $n$ of the evaporation process we apply the following nonlocal operator to the quanta $\{b_i, c_i\}$ (with $i=1, \dots, n$)
\be
\hat P_n=\prod_{k=1}^n e^{{1\over N-k}{\pi\over 4}\hat O_k} \ .
\ee
We can see that as long as we are not near the endpoint of evaporation; i.e. that $N-n\gg 1$, the change to any quantum $b_k$ at infinity over the timescale of one Hawking emission is small. The overall change to all the quanta at infinity is not small in any one step, since each quantum $b_k$ suffers a change. But if we look only at a few $b$ quanta and only over a time that is short compared to the Hawking evaporation time, then we will find it hard to detect the change in the radiation quanta $b$. 

In this model we have maintained the semiclassical Hawking pair creation at the horizon (\ref{dnone}). We have still managed to bring the Page curve down to zero at the end of the evaporation process, by using nonlocal interactions between the hole and infinity. But we must note two things:

\begin{enumerate}[start=1,
    labelindent=\parindent,
    leftmargin =2\parindent,
    label = (\alph*)]
\item \label{(a7)}
The evaporation process is not like that of a piece of coal, since the state of the radiation quanta has been modified after the quanta have receded far from the hole and it is this modification which removed the entanglement. Thus we violate condition \ref{C1} of section~\ref{seccoal}.

\item \label{(b7)}
We do not know of any effects in string theory that will create nonlocal effects like the one we used in the above model.
\end{enumerate}

\subsubsection{Identifying bits between the hole and the radiation region}\label{identify}

We had seen in section~\ref{secintroidentify} that one class of wormhole models seeks to argue that bits are not independent between the region $r<10\,r_h$ and the distant radiation region. This argument then proceeds by noting that most models of black hole evaporation assume that bits in the black hole and at infinity are independent and so such models are not relevant to the actual physics of black holes. The argument may then continue by saying that in gravity the information is `in some way' nonlocally encoded and perhaps is already at infinity, so one should not worry about an information paradox (for some arguments of this kind, see \cite{Papadodimas:2012aq,Raju:2021lwh}).

But such arguments face some immediate difficulties. In the lab we can make two well separated qubits and, to the accuracy of our lab measurements, these two behave as independent degrees of freedom. One can certainly argue that quantum gravity effects can make some small change to this independence -- a change that is too small to have been observed in the lab so far. But then one has to answer the following questions: 

\begin{enumerate}[start=1,
    labelindent=\parindent,
    leftmargin =2\parindent,
    label = (\alph*)]
\item \label{(a8)}
What exactly is the small change that we should make to independent bits so that they are no longer independent and how will this change resolve the problem of growing entanglement? It does not help to just say that the bits in the black hole and infinity are not exactly independent and so all earlier arguments are incorrect. One has to show what the meaning is of having bits that are not independent and to also show how in a lab setting one does get approximately independent bits in agreement with experiments. 

\item \label{(b8)}
These models also seek to maintain semiclassical dynamics around the horizon. But this dynamics gives rise to entangled pairs $b,c$. Such a pair has $2\times 2=4$ independent states. When the $b$ quantum moves to infinity, it increases the entanglement of the radiation with the hole by $\log 2$. If we now wish to argue that quanta in the hole and at infinity are not independent, then we have to require that this identification happen after $c$ falls deep into the hole; if we identify $b$ and $c$ while they are in the horizon region, then they will not reproduce the $4$-dimensional Hilbert space that is required by the semiclassical approximation. Therefore, abstract arguments about bits not being independent do not help; one has to show what exactly happens such that one gets approximately independent bits at the horizon and yet avoids the monotonic growth of entanglement of the black hole with its radiation. 

\item \label{(c8)}
It also does not help to argue that in gravity the `information is somehow encoded at infinity'. The information paradox can be cast as a precise question about observations of low-energy quanta at infinity: we have described this experiment in section~\ref{experiment}. Qubits in the lab have all the effects of quantum gravity acting on them since we cannot switch off quantum gravity in the real world. Similarly, the radiation bits in the gedanken experiment of section~\ref{experiment} have all the effects of quantum gravity acting on them. Any argument that the `information is somehow at infinity' has to be explicit about what happens to the radiation quanta when they are passed through a Stern-Gerlach type apparatus at infinity and how the Page curve measured by such an experiment will come down.
\end{enumerate}
In what follows we will try to investigate the idea of `degrees of freedom not being independent between the hole and the radiation' by making simple bit models. In each case we find that the model runs into difficulties, either with loss of unitarity or with quanta at infinity not behaving in accordance with expected low-energy dynamics. It is certainly possible that the proponents of the idea of `non-independent bits' have different models in mind; in that case the discussion below should hopefully trigger an investigation of explicit simple models to clarify what the proposed ideas are.

We had noted in section~\ref{secintroidentify} that there were two ways in which we can seek to identify $c$ with $b$ after $c$ falls deep in the hole and $b$ moves off to infinity. In method \ref{(i)}, we simply identify the states of these two bits, which drops the $4$-dimensional Hilbert space to a $2$-dimensional Hilbert space via the identification
\be
|0\rangle_b|0\rangle_c\r |0\rangle_b|0\rangle_c \ , \quad |1\rangle_b|1\rangle_c\r |1\rangle_b|1\rangle_c\ ,\quad |1\rangle_b|0\rangle_c\r 0\ , \quad |0\rangle_b|1\rangle_c\r 0 \ .
\ee
This is a nonunitary evolution and we will not explore this model further. 

In method \ref{(ii)}, we keep all $4$ states of the $b,c$ pair, but introduce a nonlocal interaction which raises the energy of $3$ of the states. Then the low-energy space has a relation between the states of the $b,c$ bits. We now explain this model in more detail and note that the consequence of such an interaction is that the $b$ bit at infinity will not behave like a normal bit in experiments.

Consider again just one entangled pair, with the states $|\uparrow\rangle_b\,,\ |\downarrow\rangle_b$ for the radiation bit and $|\uparrow\rangle_c\,,\ |\downarrow\rangle_c$ for the bit in the black hole. There are $4$ states overall for this system. We can write these states as a singlet and a triplet of isospin
\begin{equation} \label{kkfive}
\begin{aligned}
J=0\ , \ \ M=0:& \qquad |0,0\rangle={1\over \sqrt{2}}\Big( |\uparrow\rangle_b|\downarrow\rangle_c - |\downarrow\rangle_b|\uparrow\rangle_c\Big) \ ,\\
J=1\ , \ \ M=1:& \qquad |1,1\rangle=|\uparrow\rangle_b|\uparrow\rangle_c \ ,\\
J=1\ , \ \ M=0:& \qquad |1,0\rangle={1\over \sqrt{2}}\Big( |\uparrow\rangle_b|\downarrow\rangle_c + |\downarrow\rangle_b|\uparrow\rangle_c\Big) \ ,\\
J=1\ , \ \ M=-1:& \qquad |1,-1\rangle=|\downarrow\rangle_b|\downarrow\rangle_c \ .
\end{aligned}
\end{equation}
Now suppose we add a nonlocal interaction between the $b$ and $c$ quanta such that the triplet state is raised to a very high energy. Then the low-energy physics can access only the singlet state and in this state the spin of the $b$ quantum is tied the spin of the $c$ quantum. We may therefore regard this as a situation where the radiation bit $b$ has been `identified' with the bit in the black hole $c$. As we will now see, however, imposing such an identification will affect the dynamics of the radiation bit $b$ in a way which will make it behave in a way that is different from what we would expect for normal bits at infinity; i.e. bits that have not emerged as members of Hawking pairs radiated by the black hole. 

We will consider Hamiltonians that are invariant under isospin rotations; such Hamiltonians on the $b,c$ pair of spins can be written as
\be
H=A I + B\,\vec \sigma^{(b)}\cdot \vec\sigma^{(c)} \ ,
\ee
where $A$ and $B$ are real-valued constants. We define the total isospin
\be
\vec \sigma^{(T)}=\vec \sigma^{(b)}+\vec \sigma^{(c)} \ ,
\label{kksix}
\ee
which gives the identity
\be
\vec\sigma^{(b)}\cdot \vec\sigma^{(c)}=\h \left ( (\vec \sigma^{(T)})^2   \right ) -3 I \ .
\label{kkseven}
\ee
Let $J$ denote the angular momentum quantum number of the total isospin; thus $J=0$ for the singlet and $J=1$ for the triplet. We have $(\vec \sigma^{(T)})^2= 4J(J+1)$, giving $(\vec \sigma^{(T)})^2=0$ for the singlet and $(\vec \sigma^{(T)})^2=8$ for the triplet. We can raise the energy of the triplet by assuming an interaction between the $b$ and $c$ quanta of the form
\be
H_{wormhole}=A \big(\vec \sigma^{(T)}\big)^2 \quad \text{with} \quad A>0 \ .
\label{kkeight}
\ee
In the limit where we take $A$ to be large, the triplet will be inaccessible to low-energy dynamics. The quanta $b,c$ can then be said to be `identified' since knowing the state of one of them gives the state of the other as determined by the form of the singlet $J=0$ in (\ref{kkfive}). Let us now look at the dynamics that we get by including such an interaction. The $b$ quantum is at $r\gg M$ near asymptotic infinity, while the $c$ quantum is in the black hole. We are now interested in the dynamics of a $b$ quantum which is connected to a $c$ quantum by a wormhole through an interaction of the form (\ref{kkeight}). In the absence of the wormhole interaction, let us assume that the dynamics of the $b$ quantum is given by a Hamiltonian $H^{(b)}_{normal}$. If $b$ is connected by a wormhole to $c$, then the total Hamiltonian is
\be
H_{total}=H^{(b)}_{normal} +H_{wormhole} \ .
\ee
As a concrete example, we take
\be
H^{(b)}_{normal}=B \sigma_z^{(b)} \ ,
\ee
and so
\be
H_{total}=A \big(\vec \sigma^{(T)}\big)^2+  B \sigma_z^{(b)} \ .
\ee
We wish to compute the action of the evolution operator
\be
U(t)=e^{-it H_{total}} \ ,
\ee
on the states (\ref{kkfive}). Note that
\be
\sigma_z^{(b)}|0,0\rangle=|1,0\rangle \ ,\quad \sigma_z^{(b)}|1,0\rangle=|0,0\rangle \ ,
\ee
and so the singlet state does not remain a singlet under the action of $\sigma_z^{(b)}$. For an infinitesimal time interval $dt$, we obtain a matrix in the 2-d space spanned by $\{|0,0\rangle,\, |1,0\rangle\}$. We have
\begin{align}
U(dt)|0,0\rangle &= \Big(1-idtA\big(\vec \sigma^{(T)}\big)^2\Big)\Big(1-idt B \sigma_z^{(b)}\Big)|0,0\rangle \nonumber\\
&= \Big(1-idtA\big(\vec \sigma^{(T)}\big)^2\Big)\big(|0,0\rangle-idtB |1,0\rangle\big) = |0,0\rangle -idt B |1,0\rangle \ ,\\
U(dt)|1,0\rangle &= \Big(1-idtA\big(\vec \sigma^{(T)}\big)^2\Big)\Big(1+idtB \sigma_z^{(b)}\Big) |1,0\rangle\nonumber\\
&= \Big(1-idtA\big(\vec \sigma^{(T)}\big)^2\Big)\big(|1,0\rangle-idtB |0,0\rangle\big) =|1,0\rangle -idt B |0,0\rangle -idt sA|1,0\rangle \ .
\end{align}
Thus the effective Hamiltonian in the 2-d space spanned by $\{|0,0\rangle,\, |1,0\rangle\}$ is
\be
H_{ef\!f}=\begin{pmatrix}0 & B \\ B & 8A \end{pmatrix} \ ,
\ee
with the eigenvalues
\be
\lambda=4A+\sqrt{16 A^2+B^2}\ , \quad \lambda=4A-\sqrt{16A^2+B^2} \ .
\ee
We now see that if there is a quantum near spatial infinity that does not have the wormhole interaction (\ref{kkeight}), then we would obtain expressions similar to the above with $A=0$. Since we need $A$ to be large in the wormhole interaction, we see that the eigenfrequencies obtained in the evolution operator $U$ will be very different for the cases with and without the wormhole interaction.

\section{The requirements for a bit model of the wormhole paradigm}\label{requirement}

We have investigated several bit models for the wormhole paradigm. In these models we have explicitly seen the nonlocalities required in the \emph{exact} theory between the region $r<10\,r_h$ and the region near infinity. It is important to explore these and other such models, since any nonlocalities claimed for the exact theory must ultimately be shown to exist as effects in string theory or whatever other theory of gravity one has in mind. 

In this section we will give explicitly the requirements for any bit model of the wormhole paradigm. As we have noted at the start of this article, it is very possible that our bit models do not capture what some of the proposed models are saying. But any such proposed model, when recast in its essential terms of a bit model, must satisfy the criteria listed below. Recasting the model in these terms will make manifest the requirement of nonlocality.

In appendix~\ref{appb} we give, as an example, a bit model for Hawking pair creation at the horizon. For the wormhole paradigm, we have effective fields instead of the semiclassical fields used by Hawking and so we have to replace the bits in Appendix~\ref{appb} with effective bits. The bit model must therefore have the following features:
\begin{enumerate}[start=1,
    labelindent=\parindent,
    leftmargin =2\parindent,
    label = (\arabic*)]

\item \label{(13)}
For the wormhole paradigm we require a smooth horizon in some effective variables. Consider a mode of the effective field straddling the horizon. When this mode has wavelength $r\lesssim r_h/10$, say, then this mode is in approximately the vacuum state. In the bit model of appendix~\ref{appb} we model this mode at this stage by two coupled harmonic oscillators. The state of these coupled oscillators must be the \emph{vacuum} state. Thus we need the analogue of (\ref{ftconditions}):
\be
\hat a^\dagger _{ef\!f,i}|0\rangle_{ef\!f,a} =0\ , \quad i=1, 2 \ .
\label{ftconditionsq}
\ee

\item \label{(23)}
The requirement of low-energy semiclassical dynamics at the horizon gives, say, an effective scalar field satisfying $\square\phi_{ef\!f}\approx 0$. With these dynamics, the field mode in \ref{(13)} above stretches as it evolves along the horizon. When the wavelength of the mode becomes $\gtrsim r_h$, a pair of on-shell quanta $b_{ef\!f},c_{ef\!f}$ emerge from this mode. In the bit model using two harmonic oscillators in \ref{(13)} above, the two oscillators are decoupled since they correspond to parts of the mode that are well separated. Following the steps in appendix~\ref{appb}, we must then get the effective field analogue of 
(\ref{ftaone})
\be
|0\rangle_{ef\!f,a}=Ce^{-{a\over 4\omega^2} \hat b_{ef\!f}^\dagger\hat c_{ef\!f}^\dagger}|0\rangle_{ef\!f,b}\otimes |0\rangle_{ef\!f,c} \ .
\label{ftaoneq}
\ee

\item \label{(33)}
The requirements \ref{(13)}, \ref{(23)} spell out in bit model terms what it means to have semiclassical dynamics (\ref{twopp}) at the horizon. We now add the other requirement of the wormhole paradigm: the bits at infinity have a Page curve that comes down at the end of the evaporation process.
\end{enumerate}
In our investigations, we have not found any bit model for the wormhole paradigm which involves the black hole radiating like a piece of coal as seen from outside; this is expected since the effective small corrections theorem of section~\ref{sec small} forbids this possibility. We have found models of the wormhole paradigm where there are nonlocal Hamiltonians between the region $r<10\, r_h$ and infinity, or a nonlocal transfer of bits from $r<10\, r_h$ to infinity through a wormhole. 

There has been a quite some confusion on what the wormhole paradigm is saying and so it is particularly important to have a picture in terms of the simple bit model described above. In particular, it is important to note that effective semiclassical dynamics at the horizon needs both conditions \ref{(13)} and \ref{(23)} above: \ref{(13)} tells us that how the modes should be entangled at the horizon if they are to generate a smooth manifold and, \ref{(23)} tells us the consequence of the dynamics $\square\phi_{ef\!f}\approx 0$. Thus \ref{(13)} and \ref{(23)} together incorporate Hawking's observation that a smooth horizon leads to the creation of entangled pairs. 

These observations are important for models of the following kind. Suppose we say that our effective description gives a smooth horizon in the region $r_{QES}<r<r_h$, where $r_{QES}$ is the radius of a quantum extremal surface. Suppose we also say that the region $r<r_{QES}$ is an `island' where the degrees of freedom are some combination of the radiation modes $\{ b\}$ at infinity.\footnote{The quanta $b_{ef\!f}$ must reduce to exact quanta $b$ when they reach infinity since we assume that physics at infinity is `normal'.} It may seem that with these two statements we have maintained a smooth horizon and also allowed ourselves a departure from semiclassical physics in the region $r<r_{QES}$ inside the hole. But here we will face a conflict with the semiclassical dynamics conditions \ref{(13)} and \ref{(23)} above. The smoothness of the horizon will give (\ref{ftconditionsq}) and then \ref{(23)} will give the entangled pairs (\ref{ftaoneq}) that move into the island. The crucial point is that the dynamics $\square \phi_{ef\!f}\approx 0$ forces the fact that the $c_{ef\!f}$ quanta falling into the island \emph{are made out of the same oscillator degrees of freedom that gave the horizon modes yielding} \ref{(13)}. This follows from the analogue of (\ref{inmaster2}) for the effective modes
\be
f_1 \hat a_{ef\!f,1} + f^*_1 \hat a^\dagger_{ef\!f,1} + f_2 \hat a_{ef\!f,2} + f^*_2 \hat a^\dagger_{ef\!f,2}= g_1 \hat b_{ef\!f} + g^*_1 \hat b^\dagger_{ef\!f} + g_2 \hat c_{ef\!f} + g^*_2 \hat c^\dagger_{ef\!f} \ .
\label{inmaster}
\ee
Thus we cannot make an arbitrary choice of what the $c_{ef\!f}$ quanta in the island are, or what they are entangled with; in particular we cannot say that these $c_{ef\!f}$ quanta are made of the bits describing the earlier radiation quanta $\{ b\}$. We need some explicit nonlocal interactions that transfer the entanglement from the $c_{ef\!f}$ falling into the island to the radiation quanta at infinity.

\section{Discussion}\label{discussion}

Let us summarize our observations. The information paradox stems from two observations:

\begin{enumerate}[start=1,
    labelindent=\parindent,
    leftmargin =2\parindent,
    label = (\roman*)]

\item \label{(i6)}
The no-hair results imply that the quantum state around the horizon is the local vacuum state $|0\rangle$.

\item \label{(ii6)}
Such a vacuum state $|0\rangle$ produces entangled pairs in the state
\be
|\psi\rangle_{pair}={1\over \sqrt{2}}\Big (|0\rangle_b|0\rangle_c+|1\rangle_b|1\rangle_c\Big ) \ ,
\label{pairm}
\ee
giving rise to a monotonically increasing entanglement entropy $S_{ent}$ between the radiation and the remaining black hole.
\end{enumerate}
These arguments can be made precise by adding the result of the small corrections theorem:

\begin{enumerate}[resume,
    labelindent=\parindent,
    leftmargin =2\parindent,
    label = (\roman*)]

\item \label{(iii6)}
Small corrections to the state of the created pairs
\be
|\psi\rangle_{pair}={1\over \sqrt{2}}\Big (|0\rangle_b|0\rangle_c+|1\rangle_b|1\rangle_c\Big ) +O(\epsilon) \ , \quad \epsilon\ll 1 \ ,
\label{pairmm}
\ee
cannot resolve the problem, since we get for the entanglement after $N$ steps of pair creation
\be
 S_{ent}(N+1)>S_{ent}(N)+\ln 2-2\epsilon \ .
 \label{threem}
 \ee
The only assumption in obtaining (\ref{threem}) is that the radiated quanta $b$ have no relevant interaction with the remaining hole after they move to distances $r\gg r_h$ from the hole; this assumption just mirrors the behavior of photons that are radiated from a piece of burning coal.
\end{enumerate}
The fuzzball paradigm resolves the information paradox by violating \ref{(i6)}. Explicit construction of brane bound states in string theory yield horizon-sized quantum `fuzzballs' which are microstates with no horizon. Thus we do not get the vacuum $|0\rangle$ around a horizon that was used in the Hawking computation and fuzzballs radiate from their surface like a normal body, not through pair creation. Thus the Page curve comes down to zero at the end of the evaporation process as it would for a piece of burning coal. 

The wormhole paradigm does not seek to resolve the information paradox as stated in \ref{(i6)}-\ref{(iii6)}. Instead it starts by \emph{assuming} that some hitherto unknown quantum gravity effects cause the Page curve in the exact theory come down to zero at the end of the evaporation process. The goal of the paradigm is to see how such a behavior of the Page curve can be made compatible with dynamics where semiclassical low-energy dynamics emerges in some approximation around the horizon: i.e. that effective pair production occurs in the state
\be
|\psi_{ef\!f}\rangle_{pair}={1\over \sqrt{2}}\Big (|0\rangle_{b,ef\!f}|0\rangle_{c,ef\!f}+|1\rangle_{b,ef\!f}|1\rangle_{c,ef\!f}\Big ) +O(\epsilon) \ .
\label{twoppm}
\ee
But such a goal runs into an immediate issue. Suppose we require that the effective quanta $b_{ef\!f}$ and $c_{ef\!f}$ are made from the degrees of freedom in the region around the hole (say $r<10\,r_h$) and also that there are no relevant nonlocal effects connecting the hole to the distant region (say $r>100\,r_h$). Then we can make a simple adaption of the small corrections theorem to obtain the `effective small corrections theorem' where the Hawking quanta $b$ and $c$ in (\ref{pairmm}) are replaced by effective quanta $b_{ef\!f}$ and $c_{ef\!f}$. One then finds that  the requirement of (\ref{twoppm}) implies that \emph{the Page curve of the {exact} theory must keep rising monotonically}. 

The wormhole paradigm seeks to get around this problem by using various kinds of nonlocal effects between the region around the hole $r<10\,r_h$ and the distant radiation. Different approaches have suggested different kinds of nonlocalities, so it is useful to keep track of the aspects \ref{A1}-\ref{A4} listed at the start of section~\ref{secintro}. 

It is sometimes said that in the wormhole paradigm we assume the `central dogma': that, as seen from outside, the black hole radiates like a piece of coal. But a piece of coal satisfies the properties \ref{C1}-\ref{C3} listed in section~\ref{seccoal} and these properties say that there are no relevant nonlocal interactions between the coal and its distant radiation. If we assume that there are no such interactions, then the only kind of nonlocality we are left with is where we use both variables at $r<10\,r_h$ as well as variables in the radiation region to make the low-energy effective variables near the horizon. But as we saw in section~\ref{secnonlocaldef}, with such a construction of effective variables we are forced to the following situation: if we manipulate the exact bits in the radiation at infinity, then we can alter the dynamics that is observed in experiments in the black hole region $r<10\,r_h$. Some authors have noted this behavior as a feature of the wormhole paradigm, but others have not noted this nonlocality explicitly.

Other approaches to the wormhole paradigm invoke nonlocal Hamiltonian interactions between the hole and its distant radiation. A bit model for such nonlocal interactions can be made where observations of only a few radiation quanta for a short time will make it hard to see the nonlocality, while the Page curve still comes down to zero at the end of the evaporation process (section~\ref{secnonlocal}). We do not, however, know of any such nonlocal effects in string theory. It should be also noted that such models do not satisfy the `central dogma' since the region outside a piece of coal does not have any relevant nonlocal effects with the coal. 

We also investigate several other models which have been proposed. In each case we either find nonunitarity of evolution, or the fact that radiation quanta at infinity will behave differently in experiments from radiation quanta obtained from a piece of coal. 

There have been suggestions that simple semiclassical computations with gravity can tell us that the Page curve will come down like the Page curve of a normal body. We have argued that such is not the case: we did not find any way to obtain a decreasing Page curve from gravity without inputting this decrease via some feature of the exact quantum gravity theory. In our investigations of (1+1)-dimensional quantum gravity, we found that the possibility of topology change in gravity does \emph{not} imply that there should be a wormhole connecting different replica copies. Rather, the prescription (\ref{mfive}) that is used in the Page curve argument was an independent postulate that replaces the R\'{e}nyi entropy by a \emph{different} quantity. It is the curve stemming from this different quantity that comes down, not the Page curve deduced from the original R\'{e}nyi entropies. Thus we note a difference between the semiclassical Page curve arguments and the Gibbons-Hawking computation of black hole entropy: in the Gibbons-Hawking computations the starting point is a path integral that gives the entropy for \emph{any} quantum system, while our investigations so far indicate that this starting point is itself altered in the semiclassical Page curve arguments.

Thus we have to go further and ask for the origin of a prescription like (\ref{mfive}). We have noted that Euclidean path integrals can be used as a `trick' to generate entangled states, but that these tricks should be distinguished from true interactions that we may introduce in the theory. The essential constraint arises from the map $g_{ef\!f}=F[g_{exact}]$ (eq.(\ref{dthreeqq})) between the exact variables of the gravity theory and the effective semiclassical variables; this map then determines the dynamics of the effective theory (\ref{dfourqq}) as well as the form of quantities like the R\'{e}nyi entropy through a relation like (\ref{dfivepreqq}). The wormhole paradigm does not seek to give the map $g_{ef\!f}=F[g_{exact}]$, but any modification to definition of the R\'{e}nyi entropies that we have for the effective theory must stem from some feature of the exact theory via the map (\ref{dthreeqq}). We noted that due to the effective small corrections theorem, this map cannot be one where the effective degrees of freedom are obtained from some combination of the degrees of freedom in the region $r<10\,r_h$ (and no nonlocal interactions are introduced between the hole and its radiation). Thus some fundamental nonlocalities between the hole and its radiation need to be introduced to get a prescription like (\ref{mfive}) used in the Page curve argument. We do not have an understanding of what nonlocalities can give (\ref{mfive}), but we explored different possibilites that have been suggested in the wormhole literature and noted that they have nontrivial consequences for observations on the radiation at infinity.

We note that there have been quantum mechanical models made to explain the possibility of having traversable wormholes \cite{Gao:2016bin,Maldacena:2017axo}. In our understanding, such models seek to find a semiclassical gravity model having a wormhole, where this wormhole gives an effective description of quantum teleportation between two entangled quantum systems (which have a classical communication channel also present between them). In the discussion of teleportation in \cite{Maldacena:2017axo} one adds a bilocal operator between the two regions of the form
\be
e^{i\t g O_L O_R} \ ,
\ee
where $O_L$ and $O_R$ are operators on the two different systems and $\t g$ is a constant. The corresponding effect in the wormhole paradigm for black hole evaporation would be a bilocal operator that connects the region $r<10\,r_h$ to the region near infinity. Such an operator would be like the bilocal operator that we had in the model discussed in section~\ref{secnonlocal}, where interactions between the hole and infinity were invoked bring the Page curve down. In the model of section~\ref{secnonlocal} we had an interaction
\be
\hat O=\sigma^-_b\sigma^+_c- \sigma^+_b\sigma^-_c \ ,
\ee
where $b$ was an exact quantum near infinity and $c$ was an exact quantum in the region $r<10\,r_h$. Thus if we could extend the teleportation model to the problem of black hole evaporation, then we would be saying: (i) the \emph{exact} theory has a nonlocal interaction between the region $r<10\,r_h$ and infinity and, (ii) there is an effective semiclassical description of this exact nonlocal interaction where the horizon is smooth and the information appears to escape to infinity through a semiclassical wormhole. Note that the nonlocal interaction in the exact theory makes the hole different from a piece of coal: the nonlocal interaction violates condition \ref{C1} of section~\ref{seccoal}. Note that the following is impossible by the effective small corrections theorem: (i') the hole evaporates like a piece of coal (no nonlocal interactions in the exact theory between the region $r<10\,r_h$ and infinity and, (ii') in an effective description we get a wormhole transporting information out to infinity through a wormhole. In establishing that this combination of requirements (i') and (ii') is impossible, it is important to note that at infinity the semiclassical description of quanta has to agree with the exact description, since we define quanta at infinity by experiments done at infinity. 

It has been suggested that the nonlocalities of the wormhole paradigm are somehow automatically present in string theory or perhaps in any theory of quantum gravity. We do not believe such is the case. We hope to discuss this issue in a separate article. Here we just note that in the fuzzball paradigm there are no such nonlocalities between the fuzzball and its radiation: the fuzzball  indeed radiates just like a piece of coal and there are no effective variables where we get the semiclassical dynamics (\ref{twoppm}). Thus the fuzzball paradigm gives a natural and conceptually simple resolution of the information paradox.

\section*{Acknowledgements}

We would like to thank for helpful comments Vijay Balasubramanian, Iosif Bena, Davide Bufalini, Bidisha Chakrabarty, Ben Craps, Sumit Das, Marine Alexandra De Clerck, Philip Hacker, Daniel Harlow, Surbhi Khetrapal, Maria Knysh, Juan Maldacena, Emil Martinec, Henry Maxfield, Sameer Murthy, Rob Myers, Dominik Neuenfeld, Kevin Nguyen, Maxim Pavlov, Charles Rabideau, Mukund Rangamani, Sami Rawash, Allic Sivaramakrishnan, David Turton, Shreya Vardhan and Nicholas Warner. This work is supported in part by DOE grant DE-SC0011726.

\appendix

\section{Some details of the fuzzball paradigm}\label{appA}

In this appendix we review some aspects of the fuzzball paradigm.

\subsection{How fuzzballs differ from the traditional hole}

Let us first recall the set-up used in Hawking's original computation which yielded the information paradox \cite{Hawking:1975vcx}. Classically, the black hole spacetime is a vacuum outside of the central singularity. At the quantum level, one assumes that physics is semiclassical outside a Planck radius of the singularity $r=0$, so away from this singularity we need to consider only small fluctuations $h_{\mu\nu}$ around the classical background $\bar g_{\mu\nu}$
\be
g_{\mu\nu}=\bar g_{\mu\nu}+h_{\mu\nu}\ , \quad |h_{\mu\nu}|\ll 1 \ .
\label{metric fluctuation}
\ee
Consider for concreteness a solar-mass black hole; then the horizon radius is $r_h\approx 3\, \rm{km}$. For a ball-shaped region around the horizon with radius $r_b\ll r_h$, say $r_b=100 \, \rm{m}$, the state of our quantum fields in this ball ($|\psi\rangle$) is close to the local vacuum state $|0\rangle$. Thus we write
\be
\langle 0 | \psi\rangle =1-\delta_1\ , \quad |\delta_1|\ll 1 \ .
\label{q2onea}
\ee
With this state, the natural evolution of quantum fields on curved space gives the creation of entangled pairs in the state
 \be
|\psi\rangle_{pair}={1\over \sqrt{2}}\Big (|0\rangle_b|0\rangle_c+|1\rangle_b|1\rangle_c\Big ) +O(\epsilon) \ .
\label{pairpa}
\ee
This pair creation leads to the information paradox.

A fuzzball is in principle no different from a normal body like a planet or star; one may call it a `string star'. Thus we do not have (\ref{metric fluctuation}) and in particular we do not have a vacuum region around $r=r_h$. If we draw a ball-shaped region of radius $r_b\ll r_h$ and consider the state $|\psi\rangle$ in this region, then this $|\psi\rangle$ will not be at all close to the vacuum. Thus we write
\be
\langle 0 | \psi\rangle =1-\delta_2 \ , \quad |\delta_2|\sim 1 \ .
\label{dtwoa}
\ee
Further, we will not have the evolution giving the pair creation (\ref{pairpa}); the emission from the fuzzball will depend on the details of the particular fuzzball state that we have taken.

\subsection{Construction of fuzzball microstates}\label{commentfuzzball}

The relation (\ref{dtwoa}) and the consequent absence of (\ref{pairpa}) characterize the fuzzball paradigm. But this characterization just defines what a fuzzball \emph{is}. To establish that the fuzzball paradigm resolves the information puzzle one has to show that in string theory the microstates of black holes actually have this fuzzball nature. In particular, one has to understand how states in string theory bypass the traditional no-hair theorems and constraints like the Buchdahl theorem. Constructing examples of fuzzballs and understanding their nature has been a key task of the fuzzball program. 

Interestingly, this `fuzzball construction' requires one to use all the detailed properties of string theory. If one makes a numerical error in the tension of a brane or in the relation between the string coupling $g$ and Newton's constant $G_N$, the fuzzball structure can collapse, develop closed time like curves, or have singularities that do not correspond to valid brane sources in string theory. But when all computations are done correctly for any microstate, then one has always found a fuzzball rather than a black hole with horizon. 

Roughly speaking, one can think of the fuzzball as a region where the compact directions are not trivially tensored with the noncompact ones. In \cite{Gibbons:2013tqa} it was shown how the structure of fuzzballs bypasses the assumptions of the traditional no-hair theorems. In \cite{Mathur:2016ffb} a toy model was used to observe how Buchdahl-type theorems are bypassed.

It is sometimes claimed that the fuzzball paradigm is not fully established because the fraction of microstates that have been explicitly constructed is small. But this is an incorrect argument. For the simplest black hole -- the 2-charge extremal hole -- all microstates have been constructed and are found to be fuzzballs \cite{Lunin:2001jy,Lunin:2002iz,Kanitscheider:2007wq}. For more complicated holes, one cannot currently find all microstates in closed form. Instead, one constructs specific examples of fuzzballs corresponding to different corners of configuration space \cite{Mathur:2005zp,Bena:2007kg,Chowdhury:2010ct,Bena:2015bea,Bena:2016agb,Bena:2016ypk}. By now so many of these corners have been explored that it does not look possible for any state in string theory to \emph{not} be a fuzzball. 
 
More precisely, we construct fuzzball states the way we would construct states of black-body radiation. For black-body radiation, we can explicitly write down the quantum wavefunction for the case where we have a few photons and we can also explicitly write a classical electromagnetic field to describe the limit where we have many many photons but distributed only over a few Fourier modes. The qualitative nature of the generic case (where we have many photons with occupation number $\sim 1$ per mode) can be inferred from these limits. Similarly, we can construct fuzzballs which have a few quantum string excitations around a classical geometry (see for example \cite{Lunin:2002fw,Gomis:2002qi,Gava:2002xb}) and also the limit where we have a large number of  excitations in the same state (see for example \cite{Hampton:2018ygz} where it was found that a large number of strings in the same state yield a new classical geometry). In both limits the microstates are fuzzballs with no horizon. We construct these limits as a demonstration that the no-hair theorem is broken in string theory. Note that a fuzzball is a quantum state of the full string theory; it is not a classical geometry, though for special states which are close to being coherent states one might be able to describe the fuzzball by giving the geometry that corresponds to the expectation value of the supergravity fields in the coherent state.  
    
The argument for the fuzzball paradigm is completed by using the (effective) small corrections theorem: if any microstate did have the horizon behavior (\ref{dtwoa}), then the information paradox cannot be resolved without an order unity violation of causality in string theory (i.e. without an introduction of nonlocal interactions between the hole and its radiation).

\subsection{Understanding the fuzzball resolution to the information paradox}

There has existed a certain misconception about the information paradox itself. This misconception has prevented people from understanding how the fuzzball construction program has resolved the information paradox.

Consider the following two connected beliefs: (i) we can assume, without working to demonstrate it, that some complicated string interactions will change the semiclassical black hole to a complicated quantum gravitational mess that will radiate like a piece of coal and, (ii) the goal of the information paradox is therefore to explain how effective semiclassical behavior will emerge from this quantum mess. Neither of these two beliefs are correct.

The information paradox arose because the no-hair results indicated that the black hole horizon has the vacuum state $|0\rangle$ in its vicinity. This vacuum gives the pair creation (\ref{pairpa}) and leads to the problem of monotonically growing entanglement entropy. Perturbative string theory does not help in resolving this problem; a string loop falls through the horizon like any other object would fall and the horizon returns to the state $|0\rangle$. \emph{Non-perurbative} string theory might change this situation, but the whole issue is to find out if and how this happens. The fuzzball construction accomplishes precisely this task, showing that microstates in string theory break the no-hair theorem by having sources of extended objects, a non-product structure of compact and noncompact directions, etc., with different mechanisms appearing in different duality frames. This demonstration that the no-hair theorem is violated in string theory by the fuzzball construction resolves the information paradox. But as noted in section \ref{commentfuzzball}, the fuzzball structure utilizes all the precise details of string theory. Thus the belief (i) is incorrect and misses the whole point of what we need to do to resolve the paradox.

As noted above, people with the above two beliefs asked for more fuzzball constructions before they would accept the fuzzball resolution. Why was that? After all, we do not construct all possible states of planets to agree that planets do not have an information paradox. Once we have understood how the no-hair theorems are broken in string theory, we should accept that the paradox is over. One may certainly study more and more fuzzballs to learn about the beautiful physics of these objects, but removing the \emph{paradox} itself needs only a set of examples demonstrating that the central assumption -- the vacuum $|0\rangle$ at the horizon -- need not hold in string theory.

The reason why people with the beliefs (i) and (ii) wanted to see more and more fuzzballs constructed was the following. They noted that the examples of fuzzballs which had been constructed behaved just like pieces of coal with no horizon. But by belief (ii), they expected that when more general fuzzballs were constructed, the generic fuzzball would exhibit the effective semiclassical behavior (\ref{dseven}). But the effective small corrections theorem tells us that this is not even possible! If the effective behavior (\ref{dseven}) emerges for some microstates, then in the exact theory these microstates cannot radiate like a piece of coal. Thus the belief (ii) is incorrect and has led to people not understanding that the fuzzball constructions have already told us how the information paradox is resolved in string theory.

\subsection{Fuzzball complementarity}

A common question about fuzzballs is the following. Given that the entire interior of the hole has been replaced by a fuzzball structure and that the semiclassical approximation (\ref{dseven}) cannot be obtained in any effective variables, is the traditional black hole geometry of any relevance at all? 

The effective semiclassical modes (\ref{dseven}) refer to the dynamics of modes that have energy $E\sim T$, where $T$ is the temperature of the hole; this is because the typical Hawking quantum will have an energy of the order of the temperature of the hole. The effective small corrections theorem says that the low-energy approximation (\ref{dfive}) cannot arise as an effective description of the fuzzball; else the Page curve will not come down in the exact string theory description of the fuzzball. But this still leaves open the possibility that there could be simple effective dynamics for infalling objects with energy $E\gg T$, and that this effective dynamics reflects the geometry of the traditional hole. Here one does not mean that infalling objects with energy $E\gg T$ will fail to see the fuzzball structure and travel on the traditional metric of the hole. Rather, an infalling object with $E\gg T$ excites collective modes on the space of all fuzzball configurations, and it may be possible to interpret this using the effective geometry of the traditional hole. This possibility is called \emph{fuzzball complementarity} \cite{Mathur:2017fnw,Mathur:2017wxv}.

We do not know if the conjecture of fuzzball complementarity is true. It was shown in \cite{Mathur:2013gua} that this conjecture is not ruled out by the firewall argument \cite{Almheiri:2012rt}. In \cite{Mathur:2015nra} a bit model was given to show how the evolution in the space of fuzzball states could be mapped to effective radial infall. A physical picture of collective modes on the space of fuzzballs emerges from the `vecro' hypothesis \cite{Mathur:2020ely}. The investigation of fuzzball complementarity is an interesting and important direction to pursue, however, it is crucial to note that the dynamics of $E\gg T$ modes has no bearing on the information paradox. The rising Page curve describes the entanglement of the quanta radiated by the hole and these are necessarily quanta with $E\sim T$. Whether a simple effective dynamics emerges when $E\gg T$ quanta fall into the hole is a \emph{different} question (it is sometimes called the infall problem). 

The wormhole paradigm is concerned with the Page curve and thus is concerned with the Hawking quanta which have $E\sim T$. The effective small corrections theorem says that if the semiclassical behavior (\ref{dfive}) and (\ref{dseven}) emerges in some code subspace, then in the exact theory the black hole cannot radiate like a piece of coal as seen from outside; i.e. one will need nonlocal effects between the hole and infinity.

\section{A bit model for Hawking pair creation at the horizon}\label{appb}

\subsection{The divergence of trajectories at the horizon}

Let us begin by describing the basic physics that leads to pair creation at the horizon of a black hole. We will first see how geodesics on the two sides of the horizon diverge away from each other. We will then make a toy model for the quantum vacuum and see how this divergence of trajectories leads to the creation of entangled particle pairs. Recall the Schwarzschild metric
\be
ds^2=-\Big(1-{2M\over r}\Big)dt^2+\Big(1-{2M\over r}\Big)^{-1} dr^2+ r^2\big(d\theta^2+\sin^2\theta d\phi^2\big) \ ,
\label{9.1}
\ee
where in this appendix we set Newton's constant to one. We wish to consider particles that are trying to escape from the hole.
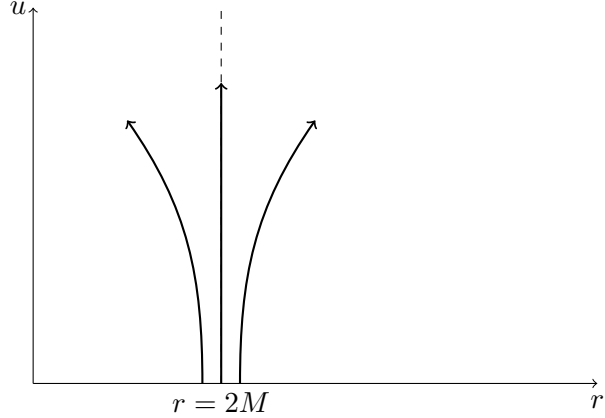
\begin{figure}[h]
\centering
\begin{tikzpicture}[scale=.5];

\def\xx{.5}

\draw [->] (0, 0) -- ( 0,10);
\node at ( -.4, 10) {$u$};

\draw [->] (0, 0) -- ( 15, 0);
\node at (15, -.5) {$r$};

\draw[dashed] (5,0) --(5,10);
\draw [->, thick] (5,0) -- (5,8);

\draw[->] [ thick] (5-\xx,0) to [out=90,in=125+180] (5-\xx-2,7);
\draw[->] [ thick] (5+\xx,0) to [out=90,in=55+180] (5+\xx+2,7);

\node at (5, -.5) {$r=2M$};

\end{tikzpicture}
\caption{The divergence of null geodesics either side of the black hole horizon at $r=2M$. Precisely along the horizon, the geodesic is radially stable. Here $u$ is the Eddington-Finkelstein coordinate defined in \eqref{EddFink}.}
\label{figintro.1}
\end{figure}
The particles that can escape most easily are massless particles, moving outwards radially at the speed of light. For such trajectories the only nonzero displacements are $dt,\,dr$ and we must have $ds^2=0$. There are three cases to consider:

\begin{enumerate}[start=1,
    labelindent=\parindent,
    leftmargin =2\parindent,
    label = (\roman*)]

\item \label{(i7)}
Suppose the particle starts a little outside the horizon, at $r=2M+\epsilon$. Then we have from $ds^2=0$
\be
{dr\over dt}=\pm \left( { 1-{2M\over r}}\right )  \ .
\ee
To have the particle go outwards, we take the positive sign. Let us ask for the time it takes for this particle to escape to a location $r_f$ that is away from the horizon
\be
\int_{2M+\epsilon}^{r_f} {dr\over 1-{2M\over r}}=\int _0^Tdt \ .
\ee
This gives
\be
T\approx  2M\log {1\over \epsilon} \ ,
\ee
and so the particle does ultimately escape, but the time to escape becomes large as $\epsilon$ goes to zero.

\item \label{(ii7)}
Suppose the particle starts a little \emph{inside} the horizon, at a position $r=2M-\epsilon$. We then have
\be
{dr\over dt}=\pm\left({ 1-{2M\over r}}\right ) =\mp\left ({{2M\over r}-1}\right ) \ .
\ee 
Let us compute the time for the particle to reach a position $r_f$ that is away from the horizon $0<r_f<2M$. This needs the negative sign in the above relation and we find
\be
\int_{2M+\epsilon}^{r_f} {dr\over {2M\over r}-1}=\int _0^Tdt \ ,
\ee
giving
\be
T\approx  2M\log {1\over \epsilon} \ .
\ee
Thus the particle escapes the vicinity of the horizon, but again the time to escape becomes large as $\epsilon$ goes to zero. 
\end{enumerate}
We see that a small region straddling the horizon gets stretched to a \emph{large region} after we wait for a sufficiently long time. In fact we can start with an arbitrarily small region
\be
|r-2M|<\epsilon \ ,
\ee
and see that after a time the region will stretch to a size $\sim M$ which describes the length scale of the Schwarzschild geometry. This persistent stretching at the horizon is what will lead to the evaporation of the hole. Quantum fields on spacetime can be modeled by a set of coupled harmonic oscillators. When a slice stretches, the distance between neighboring points increases. This makes the coupling between the corresponding oscillators weaker. This change of coupling can convert a vacuum state of the oscillators to a state that contains pairs of excitations. But excitations of the oscillators describing the quantum field correspond to particles. Thus we will find that the stretching of slices seen above will lead to the creation of particle pairs from the vacuum. We can make a bit model of pair creation using just two oscillators as follows. 

We consider two oscillators, one on each side of the horizon. These oscillators will be coupled to each other, the way neighboring oscillators are coupled in quantum field theory, and we will let the initial state of the system be the ground state of the coupled oscillator pair. We have seen above that geodesics on the two sides of the horizon separate away from each other. We will model this effect by removing the coupling between the oscillators at some time $t=0$.\footnote{Of course in the black hole the coupling changes over a Kruskal time of order the horizon radius, but replacing this by a sudden change of coupling captures the physics with only changes of factors of order unity.} We will find that the two oscillators will now have pairs of excitations and the overall  states will be \emph{entangled} between the two oscillators. This state has all the features of the full quantum problem that will be relevant to the information paradox, so this is a useful toy model.

Let the oscillator $\phi_L$ denote a wavemode just inside the horizon and the oscillator $\phi_R$ a mode just outside the horizon. On the `earlier' time slice the wavemodes are close to each other and their corresponding oscillators should be coupled. At late times, the modes are far from each other, and the corresponding oscillators should be almost decoupled. We let the oscillators be coupled for $t<0$ and decoupled for $t>0$. Thus the Lagrangian is
\bea
{\cal L}&=& \h \dot \phi_L^2 +\h \dot \phi_R^2 -\h \omega^2 \phi_L^2 -\h \omega^2 \phi_R^2 - a\phi_R\phi_L\ \ , \quad t<0\nn
&=& \h \dot \phi_L^2 +\h \dot \phi_R^2 -\h \omega^2 \phi_L^2 -\h \omega^2 \phi_R^2\ \ , \quad t>0 \ .
\eea

\subsection{The state for $t<0$}

We can decouple these two oscillators by going to a new basis
\be
\phi_1={1\over \sqrt{2}} \big( \phi_L+\phi_R\big) \ , \quad\phi_2={1\over \sqrt{2}} \big( \phi_L-\phi_R\big) \ ,
\ee
which gives two uncoupled oscillators with frequencies
\be
\phi_1: ~~~\omega_1=\sqrt{\omega^2+a\,}\ , \quad \omega_2=\sqrt{\omega^2-a\,} \ \ .
\ee
The oscillator with variable $\phi_1$ has creation and annihilation operators $\hat a_1,\, \hat a^\dagger_1$ and the oscillator with variable $\phi_1$ has creation and annihilation operators  $\hat a_2,\, \hat a^\dagger_2$. We wish match our notation as closely as possible to the notation of quantum field operators on curved space. On a (1+1)-dimensional spacetime we would have an infinite line of points where a scalar field $\phi$ would be defined. In place of this, in our toy model, we now just have two points. In place of the field modes $f(t,x)$ at time $t$ on this line $x$, we now have a function of $t$ defined on two points. We write functions on this 2-point space using a  2-component vector $(a,b)$, with $a$ corresponding to $\phi_L$ and $b$ corresponding to $\phi_R$. We therefore define two-component functions
\be
f_1={1\over \sqrt{2\omega_1}}e^{-i\omega_1 t}  {1\over \sqrt{2}}(1,1), ~~~~ f_2={1\over \sqrt{2\omega_2}}e^{-i\omega_2 t}  {1\over \sqrt{2}}(1,-1) \ .
\label{inmodes}
\ee
The inner product between modes $f$ and $g$ is
\be
(f,g)=i[f^* \cdot \p_t g -\p_t f^* \cdot  g] \ ,
\ee
and they are normalized as
\be
(f_i,f_j) = \delta_{ij}\ , \quad(f^*_i, f^*_j)=-\delta_{ij}\ , \quad(f^*_i, f_j)=0 \ .
\ee
Now consider the `field operator'
\be
\hat \phi = \big( \hat \phi_1, \hat\phi_2\big) \ .
\ee
Since the oscillators have been decoupled in the $\phi_1, \phi_2$ basis, we can expand the field operator just the way we did for a single oscillator 
\be
\hat \phi = f_1 \hat a_1 + f^*_1 \hat a^\dagger_1 + f_2 \hat a_2 + f^*_2 \hat a^\dagger_2 \ .
\label{appaq}
\ee
We start with the vacuum state for these coupled oscillators
\be
\hat a^\dagger _i|0\rangle_a =0\ , \quad i=1, 2 \ .
\label{ftconditions}
\ee

\subsection{Evolution for $t>0$ }

At the late time slice the field modes on the two sides of the horizon are well separated and the coupling between them is weak. We have modeled this by letting the oscillators corresponding to these modes be decoupled for $t>0$. The analogue of the modes (\ref{inmodes}) is
\be
g_1={1\over \sqrt{2\omega}}e^{-i\omega t} (1,0)\ , \quad g_2={1\over \sqrt{2\omega}}e^{-i\omega t}  (0,1) \ .
\label{inmodesq}
\ee
Note that we also have
\be
(g_i,g_j) = \delta_{ij} \ , \quad (g^*_i, g^*_j)=-\delta_{ij} \ , \quad (g^*_i, g_j)=0 \ .
\ee
The same field operator $\hat \phi$ can be written as
\be
\hat \phi = g_1 \hat b + g^*_1 \hat b^\dagger + g_2 \hat c + g^*_2 \hat c^\dagger \ ,
\label{appaqq}
\ee
and so we have
\be
f_1 \hat a_1 + f^*_1 \hat a^\dagger_1 + f_2 \hat a_2 + f^*_2 \hat a^\dagger_2= g_1 \hat b + g^*_1 \hat b^\dagger + g_2 \hat c + g^*_2 \hat c^\dagger \ .
\label{inmaster2}
\ee

\subsection{Matching at $t=0$}

As we did in the case of the single oscillator in the Heisenberg picture, we wish to express the conditions (\ref{ftconditions}) (which define our state $|0\rangle_a$) as conditions involving the oscillators $\{\hat b, \hat b^\dagger, \hat c, \hat c^\dagger\}$. This will then allow us to express the state $|0\rangle_a$ in terms of $\hat b^\dagger, \hat c^\dagger $ excitations. We take the inner product $(g_1 , \cdot\,)$ on both sides of (\ref{inmaster2}). This gives
\bea
\hat b &=& ( g_1, f_1) \hat a_1 + (g_1, f_1^*)\hat a^\dagger_1 + ( g_1, f_2) \hat a_2 + (g_1, f_2^*)\hat a^\dagger_2\nn
&=&{\omega+\omega_1\over 2\sqrt{2}\sqrt{\omega\omega_1}} \hat a_1 + {\omega-\omega_1\over 2\sqrt{2}\sqrt{\omega\omega_1}}\hat a^\dagger_1 +{\omega+\omega_2\over 2\sqrt{2}\sqrt{\omega\omega_2}}\hat a_2 + {\omega-\omega_2\over 2\sqrt{2}\sqrt{\omega\omega_2}}\hat a^\dagger_2 \ ,\nn
\hat c &=& ( g_2, f_1) \hat a_1 + (g_2, f_1^*)\hat a^\dagger_1 + ( g_2, f_2) \hat a_2 + (g_2, f_2^*)\hat a^\dagger_2\nn
&=&{\omega+\omega_1\over 2\sqrt{2}\sqrt{\omega\omega_1}} \hat a_1 + {\omega-\omega_1\over 2\sqrt{2}\sqrt{\omega\omega_1}}\hat a^\dagger_1 -{\omega+\omega_2\over 2\sqrt{2}\sqrt{\omega\omega_2}}\hat a_2 - {\omega-\omega_2\over 2\sqrt{2}\sqrt{\omega\omega_2}}\hat a^\dagger_2\ .
\eea
While we can find the state $|0\rangle_a$ in terms of $\hat b^\dagger$ and $\hat c^\dagger $ for any value of the coupling $a$, the algebra is a little simpler for $a\ll \omega^2$. In this limit we have, keeping the leading order expression for each term,
\be
\omega_1\approx \omega + {a\over 2\omega}, ~~~ \omega_2\approx \omega - {a\over 2\omega} \ .
\ee
Then we have for the operators and their conjugates
\bea
\hat b &\approx & {1\over \sqrt{2}}\hat a_1 - {a\over 4 \sqrt{2}\omega^2}\hat a^\dagger_1 +  
{1\over \sqrt{2}}\hat a_2 + {a\over 4 \sqrt{2}\omega^2}\hat a^\dagger_2 \ ,\nn
\hat c &\approx & {1\over \sqrt{2}}\hat a_1 - {a\over 4 \sqrt{2}\omega^2}\hat a^\dagger_1 -  
{1\over \sqrt{2}}\hat a_2 - {a\over 4 \sqrt{2}\omega^2}\hat a^\dagger_2 \ ,\nn
\hat b^\dagger &\approx & {1\over \sqrt{2}}\hat a^\dagger_1 - {a\over 4 \sqrt{2}\omega^2}\hat a_1 +  
{1\over \sqrt{2}}\hat a^\dagger_2 + {a\over 4 \sqrt{2}\omega^2}\hat a_2 \ ,\nn
\hat c^\dagger &\approx & {1\over \sqrt{2}}\hat a^\dagger_1 - {a\over 4 \sqrt{2}\omega^2}\hat a_1 -  
{1\over \sqrt{2}}\hat a^\dagger_2 - {a\over 4 \sqrt{2}\omega^2}\hat a_2 \ .
\eea
Now we note that the combination
\be
\hat b + {a\over 4\omega^2} \hat c^\dagger \ ,
\ee
has only annihilation operators $\hat a_1$ and $\hat a_2$. Thus
\be
\left ( \hat b + {a\over 4\omega^2} \hat c^\dagger \right ) |0\rangle_a=0 \ ,
\ee
which has the solution
\be
|0\rangle_a=Ce^{-{a\over 4\omega^2} \hat b^\dagger\hat c^\dagger}|0\rangle_b\otimes |0\rangle_c \ .
\label{ftaone}
\ee
We see that if we have two oscillators with the same frequency, weakly coupled together and then we remove the coupling suddenly, the ground state of the initial system becomes an entangled state of the two uncoupled oscillators.

\subsection{The entangled nature of the final state}

We can now see the entangled nature of the state (\ref{ftaone}). We can expand the exponential in (\ref{ftaone}) to find
\be
|0\rangle_a=C\left [ ~ |0\rangle_b\otimes |0\rangle_c - \left ( {a\over 4\omega^2}\right )  |1\rangle_b\otimes |1\rangle_c +\left ( {a\over 4\omega^2}\right ) ^2 |2\rangle_b\otimes |2\rangle_c +\dots \right] \ . 
\label{ftaoneqq}
\ee
Thus the above model with two oscillators gives a bit model for Hawking radiation. When the wavelength of the mode is small compared to the horizon radius (say $\lambda\sim r_h/10$), the parts of the mode that are just inside and just outside the horizon are strongly coupled and such coupled modes are described by the oscillators (\ref{appaq}). When the wavelength of the mode is large ($\lambda\gtrsim r_h$) then the inside and outside parts are weakly coupled and are described by the mode expansion (\ref{appaqq}). The relation (\ref{inmaster2}) relates these two mode expansions and encodes the essence of the Hawking pair creation process.

\bibliographystyle{utphys}
\bibliography{fuzzball_wormholev2.bib}

\end{document}